\documentclass[aps, prx, reprint, superscriptaddress, longbibliography, floatfix]{revtex4-2}
\usepackage{graphicx}
\usepackage{dcolumn}
\usepackage{bm}
\usepackage{physics}
\DeclareMathAlphabet{\pazocal}{OMS}{zplm}{m}{n}
\usepackage{amsmath}
\usepackage{amsfonts}
\usepackage{mathrsfs}
\usepackage{subfigure}
\usepackage{color}
\usepackage{xcolor}
\usepackage[colorlinks=true,linkcolor=blue,anchorcolor=red,citecolor=blue, urlcolor=blue]{hyperref}

\begin{document}
\preprint{APS/123-QED}
 
\title{Kinetics of Peierls dimerization transition: Machine learning force-field approach}


\author {Ho Jang}
\affiliation{Department of Physics, University of Virginia, Charlottesville, Virginia, 22904, USA}

\author {Yang Yang}
\affiliation{Department of Physics, University of Virginia, Charlottesville, Virginia, 22904, USA}

\author {Gia-Wei Chern}
\affiliation{Department of Physics, University of Virginia, Charlottesville, Virginia, 22904, USA}

\begin{abstract}
We present a machine learning (ML) force-field framework for simulating the non-equilibrium dynamics of charge-density-wave (CDW) order driven by the Peierls instability. Since the Peierls distortion arises from the coupling between lattice displacements and itinerant electrons, evaluating the adiabatic forces during time evolution is computationally intensive, particularly for large systems. To overcome this bottleneck, we develop a generalized Behler-Parrinello neural-network architecture -- originally formulated for ab initio molecular dynamics -- to accurately and efficiently predict forces from local structural environments. Using the locality of electronic responses, the resulting ML force field achieves linear scaling efficiency while maintaining quantitative accuracy. Large-scale dynamical simulations using this framework uncover a two-stage coarsening behavior of CDW domains: an early-time regime characterized by a power-law growth $L \sim t^{\alpha}$ with an effective exponent $\alpha \approx 0.7$, followed by a crossover to the Allen-Cahn scaling $L \sim \sqrt{t}$ at late times. The enhanced early-time coarsening is attributed to anisotropic domain-wall motion arising from electron-mediated directional interactions. This work demonstrates the promise of ML-based force fields for multiscale dynamical modeling of condensed-matter lattice models. 
\end{abstract}

\date{\today}
\maketitle

\section{Introduction}\label{sec:intro}

Charge-density waves (CDWs) occupy a central position in condensed matter physics as prototypical examples of spontaneous electronic ordering that intertwines lattice, charge, and other internal degrees of freedom~\cite{tranquada95,fradkin15,kivelson03}. Their significance lies both in their fundamental theoretical role -- as manifestations of broken symmetry and collective quantum behavior -- and in their ubiquity across a wide range of materials, from quasi-one-dimensional conductors to layered transition-metal dichalcogenides, cuprate superconductors, and kagome metals~\cite{gruner1988,Grner1994DensityWI,thorne_charge-density-wave_1996,Monceau2012}. Although charge modulation can, in principle, emerge without an accompanying lattice distortion~\cite{kivelson03,tranquada95}, most CDW phases are stabilized by a cooperative lattice distortion arising from strong electron–phonon coupling~\cite{gruner1988,Grner1994DensityWI}. The entanglement between electronic and lattice degrees of freedom gives rise to a host of unconventional properties, including nonlinear transport, giant dielectric susceptibility, and metastable conduction states~\cite{Monceau2012,anderson1973,balandin2021}. In recent years, the rich phenomenology of CDWs in low-dimensional systems -- particularly transition-metal dichalcogenides -- together with their potential for next-generation electronic and optoelectronic applications, has sparked renewed interest in the nonequilibrium dynamics and control of CDW order~\cite{porer2014,samnakay2015,cho2016,chen_charge_2015,zhao_moire_2022}.

A minimal theoretical framework for investigating CDW phases is provided by the Holstein model, in which itinerant electrons couple locally to dispersionless optical phonons via a deformation-potential interaction~\cite{holstein1959,scalettar1989,noack91}. Near half filling, strong electron–phonon coupling drives a transition to a commensurate checkerboard charge order on the square lattice, breaking the sublattice symmetry of the underlying bipartite structure. Moreover, the Holstein model can be simulated without encountering the fermion sign problem in determinant quantum Monte Carlo~\cite{EYLoh1990,Troyer2005}, making it a well-established platform for systematic and unbiased numerical studies of electron-phonon coupled systems. 

Beyond the Holstein framework, a more realistic microscopic description of CDW formation is provided by models incorporating the Peierls mechanism, in which itinerant electrons couple to bond distortions that modulate the electronic hopping amplitudes~\cite{peierls1955}. The Su–Schrieffer–Heeger (SSH) model offers a paradigmatic realization of this instability, where electron–lattice coupling enters through bond-dependent hopping integrals. In one dimension, the Peierls instability renders the metallic Fermi surface unstable to a lattice distortion with wave vector $2k_F$, resulting in bond dimerization, the opening of a single-particle energy gap at the Fermi level, and the emergence of a CDW ground state. The SSH model further predicts topological solitons and domain-wall excitations, which play a central role in the electronic properties of conducting polymers~\cite{SSH}. In higher dimensions, the Peierls distortion becomes increasingly intricate due to multiple competing nesting vectors and complex elastic couplings, resulting in a rich variety of ordering patterns~\cite{Yam2020,Xing2021}.

While extensive studies have elucidated the thermodynamic and transport properties of CDW order, the dynamical evolution of CDW domains--particularly those arising from the Peierls instability--remains far less understood. Even the fundamental phase-ordering kinetics following the emergence of local CDW order have received limited theoretical attention. A central open question is whether CDW coarsening follows the conventional power-law growth characteristic of classical order-parameter dynamics, or instead exhibits anomalous scaling behavior stemming from the coupled electron–lattice degrees of freedom. Understanding these coarsening pathways is essential for clarifying how lattice geometry, elastic anisotropy, and electron-mediated interactions govern the pattern formation and morphology of Peierls-driven CDW states~\cite{tsen2015}. 

Microscopic simulations of Peierls-type dynamics, however, remain computationally prohibitive due to the need for repeated electronic structure evaluations. Consequently, large-scale studies of CDW evolution have typically relied on empirical time-dependent Ginzburg–Landau approaches, which neglect the microscopic feedback between electronic and lattice degrees of freedom. Quantum simulations, on the other hand, are often restricted to small system sizes or to mean-field approximations that overlook spatial fluctuations and inhomogeneity. These limitations highlight the need for a machine-learning (ML) force-field framework capable of retaining microscopic fidelity while enabling large-scale, nonequilibrium simulations of CDW coarsening dynamics in the adiabatic limit.

The remarkable efficiency of ML force-field methods, as demonstrated in numerous ML-based {\em ab initio} molecular dynamics simulations~\cite{behler07,bartok10,Kermode2015,smith17,zhang18,behler16,deringer19,Mueller2020,mcgibbon17,suwa19,chmiela17,chmiela18,sauceda20,iftimie_ab_2005}, stems from their linear-scaling performance. As emphasized by Kohn, such scalability arises from the locality principle, or the nearsightedness of electronic matter, which asserts that the electronic response at a given site depends primarily on its immediate atomic environment~\cite{kohn1996,prodan2005}. The Behler–Parrinello (BP) architecture~\cite{behler07} provides a practical realization of this concept by decomposing the total energy into atomic contributions determined from local symmetry-invariant descriptors, with forces obtained as analytical energy gradients. This framework naturally enforces locality, symmetry preservation, and energy conservation, making it ideally suited for developing scalable ML force fields for Peierls-type electron–lattice dynamics.

In this work, we extend the BP force-field framework to simulate the adiabatic dynamics of electron–lattice systems exhibiting the Peierls instability. A central element of our formulation is a symmetry-aware descriptor that ensures the ML model faithfully preserves the symmetries of the underlying electronic Hamiltonian. In lattice-based systems, the structure of the relevant symmetry groups differs fundamentally from that for molecular dynamics: the continuous Euclidean symmetry $E(3)$ is replaced by discrete lattice translations combined with local point-group operations. Furthermore, the transformation properties of on-site lattice displacements are inherently coupled to the discrete rotational symmetries of the lattice. To capture these features, we construct an efficient descriptor based on group-theoretical bispectrum coefficients, adapted to the lattice symmetry environment. Large-scale simulations using this framework reveal anomalous coarsening dynamics of Peierls-induced CDW domains, arising from complex electron-mediated lattice interactions.

The remainder of this paper is organized as follows. Section~\ref{sec:peierls} introduces the microscopic Hamiltonian and the adiabatic dynamics of Peierls-induced CDW order. Section~\ref{sec:MLFFmodel} outlines the general ML force-field framework employed in this study and presents benchmark results on force-prediction accuracy and correlation functions for small lattices. The results of large-scale ML-driven simulations of CDW domain coarsening are presented in Section~\ref{sec:largeML}. In Section~\ref{sec:preferDW&coarsening}, we analyze the microscopic origin of the anomalous growth dynamics and preferential domain-wall orientations observed in the simulations. Finally, Section~\ref{sec:conclusion} summarizes the main results and discusses potential extensions of this work.

\section{Adiabatic dynamics of Peierls electron-lattice systems}

\label{sec:peierls}

\subsection{CDW order in Su-Schrieffer-Heeger model}

We consider a modified Su-Schrieffer-Heeger (SSH) model with spinless fermions on a square lattice as a representative system for charge-density-wave (CDW) phases arising from the Peierls instability. The total Hamiltonian consists of electronic and lattice parts, $\hat{\mathcal{H}} = \hat{\mathcal{H}}_e(\{\mathbf u_i\}) + \mathcal{V}_L(\{\mathbf u_i\})$, where $\hat{\mathcal{H}}_e$ denotes the tight-binding electronic Hamiltonian parametrized by lattice displacements, and $\mathcal{V}_L$ represents the elastic energy of the lattice. The electronic Hamiltonian is given by
\begin{eqnarray}
    & & \hat{\mathcal{H}}_e=-\sum_{\langle ij \rangle} [t_{\rm nn} - g \hat{\mathbf n}_{ij} \cdot(\mathbf u_j - \mathbf u_i)] (\hat{c}_i^\dagger \hat{c}_j + \mathrm{h.c.}) \nonumber \\ 
    & & \qquad\quad -t_{\rm nnn}\sum_{\langle\langle ij \rangle\rangle}(\hat{c}_i^\dagger \hat{c}_j + \mathrm{h.c.}),
    \label{eqn:electronicHamiltonian}
\end{eqnarray}
where $\hat{c}_i^\dagger$ ($\hat{c}_i$) creates (annihilates) a spinless electron at site $i$, $\mathbf{u}_i$ is the two-dimensional lattice displacement at site $i$, and $\mathbf{p}_i$ its conjugate momentum. The first term describes nearest-neighbor (NN) hopping modulated by the bond distortion, with $t_{\mathrm{nn}}$ the hopping amplitude in the undistorted lattice, $g$ the electron–phonon coupling strength, and $\hat{\mathbf{n}}_{ij}$ the unit vector pointing from site $i$ to~$j$. The second term accounts for next-nearest-neighbor (NNN) hopping with amplitude $t_{\mathrm{nnn}}$. The classical elastic energy takes the form
\begin{eqnarray}
        \mathcal{V}_L = \sum_i \left(\frac{\mathbf{p}_i^2}{2m}+\frac{k\mathbf{u}_i^2}{2}\right)+\kappa \sum_{\langle ij \rangle} \mathbf{u}_i\cdot\mathbf{u}_j,
        \label{eqn:classicalHamiltonian}
\end{eqnarray}
where the first term represents the harmonic lattice energy characterized by ion mass $m$ and spring constant~$k$, while the second introduces a quadratic coupling $\kappa$ between neighboring displacements, favoring correlated distortions.

\begin{figure}
    \centering
    \includegraphics[width=\linewidth]{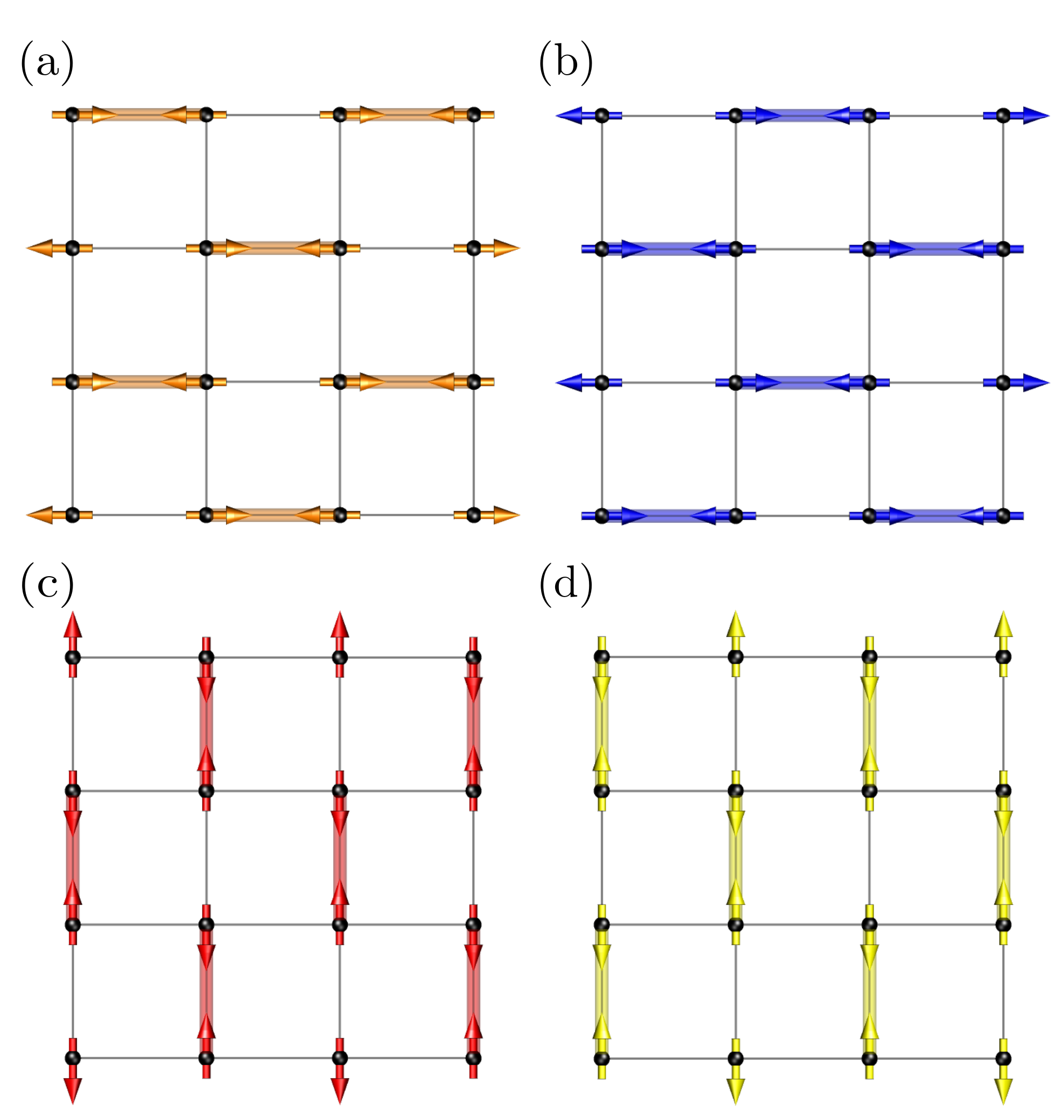}
    \caption{Schematic illustration of the four degenerate charge-density-wave (CDW) ground states of the Su–Schrieffer–Heeger (SSH) model on a square lattice. Colored edges denote the short, dimerized bonds corresponding to regions of enhanced charge density. The arrow colors indicate distinct domain types that emerge during the coarsening process, as shown in Fig.~\ref{fig:MLlargesimulation}.}
    \label{fig:4pipi}
\end{figure}

Due to perfect Fermi-surface nesting in the nearest-neighbor tight-binding model on a square lattice at half filling, lattice distortions with wave vector $\mathbf{Q} = (\pi, \pi)$ are energetically favored. Depending on the details of the lattice elastic energy, however, several distinct CDW patterns may emerge. In this work, we focus on a CDW state characterized by a staggered arrangement of dimerized bonds, as illustrated in Fig.~\ref{fig:4pipi}. This ordering simultaneously breaks the tetragonal symmetry (the equivalence between the $x$ and $y$ directions) and the sublattice symmetry of the square lattice, leading to a fourfold degenerate CDW configuration.

Phenomenologically, the CDW order can be described by a vector order-parameter field $\bm{\phi}(\mathbf r)$ defined through
\begin{equation}
\mathbf{u}_i = \bm{\phi}(\mathbf r_i)\,e^{i \mathbf{Q}\cdot \mathbf{r}_i},
\end{equation}
where $\mathbf{u}_i$ denotes the lattice displacement at site $i$. The ideal CDW configurations shown in Fig.~\ref{fig:4pipi} correspond to uniform order parameters $\bm{\phi} = \pm A \hat{\mathbf{x}}$ and $\pm A \hat{\mathbf{y}}$, with $A$ the amplitude of the ground-state CDW order. The corresponding Ginzburg–Landau free energy exhibiting this fourfold degeneracy can be expressed as an XY-type model with a two-dimensional cubic anisotropy,
\begin{eqnarray}\label{eqn:TDGL}
 \mathcal{F} = \int d^2\mathbf{r}\Bigl[ \tfrac{1}{2}\sum_{\alpha = x, y}(\nabla \phi_\alpha)^2  - \tfrac{a}{2}\bm{\phi}^2 + \tfrac{b}{4}\bm{\phi}^4 
+ \frac{c}{4} \,\phi_x^2 \phi_y^2 \Bigr],\quad
\end{eqnarray}
where $a, b, c > 0$ are phenomenological parameters. The last, anisotropic term with $c > 0$ represents an effective repulsion between the $\phi_x$ and $\phi_y$ components, favoring ground states in which the vector $\bm{\phi}$ aligns along either the $x$ or $y$ direction.

On symmetry grounds, the same free-energy functional also serves as a continuum field-theory of the four-state clock model. The Hamiltonian of a general $q$-state clock model is given by $\mathcal{H}_{\rm clock} = -J \sum_{\langle ij \rangle} \cos(\theta_i - \theta_j)$, where the angular variable $\theta_i = 2\pi m_i / q$ and $m_i$ is an integer between $0$ and $q - 1$. Both the thermal phase transitions and phase-ordering dynamics of $q$-state clock models have been extensively studied~\cite{Puri2009,Bray1994,chatterjee2018,chatterjee2020,Corberi06}. The thermal transition of the $q = 4$ clock model has long been recognized as subtle, as it lies at the boundary between the single Ising-like transition for small $q$ and the emergence of two Kosterlitz–Thouless (KT) transitions for $q \ge 5$~\cite{jose1977}. Subsequent studies established that the $q = 4$ model exhibits a single continuous phase transition in the two-dimensional Ising universality class. This behavior reflects the fact that the discrete $Z_4$ symmetry can be mapped onto two coupled Ising variables, thereby precluding the KT-type transitions and intermediate quasi–long-range ordered phase characteristic of larger $q$.

Systematic studies of phase-ordering kinetics in $q$-state clock models have been conducted in recent years. It is well established that the relaxation process is governed by the dynamics of topological defects associated with symmetry breaking. In the $q$-state clock models, two primary types of defects emerge: interfaces separating distinct ordered domains and point-like defects corresponding to the intersections of multiple domains. The latter behave analogously to vortices in the XY model, which represents the $q \to \infty$ limit of the clock model. Kinetic Monte Carlo simulations with nonconserved spin dynamics have revealed that the coarsening process in $q$-state clock models follows the Allen–Cahn power-law growth, $L \sim \sqrt{t}$, originally derived for domain coarsening in systems with nonconserved Ising order. These findings underscore the dominant role of interface motion in controlling the coarsening dynamics of the $q$-state clock model.

\subsection{Adiabatic dynamics of SSH model}

Our objective is to investigate the phase-ordering dynamics of the $Z_4$ CDW phases arising from the Peierls instability and to compare their behavior with the dynamical universality class of the four-state clock model. Fully quantum simulations of phase-transition dynamics—particularly in open or dissipative systems—are notoriously challenging, as the system remains far from equilibrium throughout the ordering process. Conventional QMC methods are restricted to equilibrium sampling and thus cannot capture the real-time evolution of the order parameter. To make the problem computationally tractable, we employ a semiclassical approximation in which the lattice degrees of freedom are treated as classical dynamical variables. The validity of this approach has been well established in previous studies, which demonstrated that the dimerized CDW ground state and its associated dynamics are accurately reproduced within the semiclassical approximation~\cite{Tang1988,Ono2000}.

However, the treatment of phonons as classical variables in the conventional SSH model with “optical” phonons—described by an elastic energy $\frac{k}{2}[(\mathbf{u}_j - \mathbf{u}_i) \cdot \hat{\mathbf{n}}_{ij}]^2$ for each bond $(ij)$—inadvertently introduces accidental ground-state degeneracies. In addition to the four $Z_4$ dimerized configurations shown in Fig.~\ref{fig:4pipi}, other charge-density-wave (CDW) states characterized by $(\pi,0)$ and $(0,\pi)$ ordering vectors also become energetically degenerate with the $(\pi,\pi)$ ground states~\cite{Ono2000,Watanabe2007}. To confine the system to the desired $Z_4$ CDW manifold, we incorporate two modifications to the original SSH model: a second-neighbor hopping $t_{\rm nnn}$, which breaks the perfect Fermi-surface nesting at half-filling~\cite{Yuan01,yuan02}, and a nearest-neighbor antiferro-distortive coupling $\kappa$, which explicitly favors $(\pi,\pi)$ distortions.

Within the semiclassical framework, the lattice (phonon) degrees of freedom are treated as classical dynamical variables coupled to a thermal bath to emulate a quench into a nonequilibrium state. The ensuing dissipative dynamics are naturally formulated in terms of stochastic Langevin equations, which incorporate both damping and thermal noise in accordance with the fluctuation–dissipation theorem~\cite{ermak80,hohenberg77,glauber_time-dependent_1963,hanggi1982}. The equation of motion for the classical displacement field is then given by
\begin{equation}\label{eqn:LangevinGeneral}
    m\frac{d^2\mathbf{u}_i}{dt^2}\,=\,-\frac{\partial E }{\partial \mathbf{u}_i}\,-\gamma\frac{d\mathbf{u}_i}{dt} + \boldsymbol{\eta} _i(t).
\end{equation}
where $E$ is the effective energy of the system, $\gamma$ is a damping constant and $\boldsymbol{\eta}_i(t)$ is a two-dimensional Gaussian noise satisfying
\begin{equation}\label{eqn:langevin}
    \begin{split}
        \langle\boldsymbol{\eta}_i(t)\rangle\,&=\,0\\
        \langle\eta^a_{i}(t)\eta^b_{j}(t')\rangle\,&=\,2\gamma m k_BT\delta_{ij}\delta_{ab}\delta(t-t'),
    \end{split}
\end{equation}
where $a,b$ label Cartesian components ($x$ or $y$), $T$ denotes the temperature of the thermal bath, and $k_B$ is the Boltzmann constant.  The effective energy has contributions from both the classical lattice elasticity and the electron–lattice coupling
\begin{eqnarray}
    E = \langle \hat{\mathcal{H}}_e \rangle + \mathcal{V}_L.
\end{eqnarray}
The resultant driving forces $\mathbf F_i = - \partial E / \partial \mathbf u_i = \mathbf F^{e}_i + \mathbf F^{L}_i$ can also be separated into two terms. The elastic contribution is given by
\begin{equation}\label{eqn:elasticforce}
    \mathbf{F}_i^{L} = -\frac{\partial \mathcal{V}_L}{\partial\mathbf{u}_i}
    = -k \mathbf{u}_i - \kappa \sum_{j \in \mathcal{N}(i)} \mathbf{u}_j,
\end{equation}
where $\mathcal{N}(i)$ denotes the nearest neighbors of site-$i$. The first term represents the local harmonic restoring force, while the second term describes the nearest-neighbor antiferro-distortive coupling. The electron-mediated force is obtained from the Hellmann–Feynman theorem~\cite{diventra00,todorov10,lu12,dundas09}, $\partial \langle \hat{\mathcal{H}}_e \rangle / \partial \mathbf{u}_i = \langle \partial \hat{\mathcal{H}}_e / \partial \mathbf{u}_i \rangle$, which yields
\begin{eqnarray}
    \label{eqn:electronforce}
    \mathbf F^{e}_i = -\biggl\langle\frac{\partial \hat{\mathcal{H}}_e}{\partial \mathbf u_i} \biggr\rangle = g \sum_{j \in \mathcal{N}(i)} \hat{\mathbf n}_{ij} 
    \left(\langle c_{i}^\dagger c^{\,}_{j}\rangle + \mathrm{h.c.}\right).
\end{eqnarray}
Here $\hat{\mathbf n}_{ij}$ again denotes the unit vector pointing from site-$i$ to $j$. This term captures the net imbalance of electronic bond correlations adjacent to site-$i$, which acts as the microscopic driving force for lattice distortions.

Because electronic relaxation occurs on a timescale much shorter than that of domain growth, the evolution of the CDW state can be accurately described within the adiabatic limit. In analogy with quantum molecular dynamics, we adopt the Born–Oppenheimer approximation~\cite{marx2009}, whereby the electronic degrees of freedom are assumed to instantaneously relax to their ground state corresponding to the instantaneous lattice configuration. Consequently, the electronic subsystem remains in quasi-equilibrium throughout the lattice evolution, and the bond correlators in Eq.~(\ref{eqn:electronforce}) can be evaluated by solving the electron Hamiltonian as
\begin{eqnarray}
    & & \langle c_{i}^\dagger c^{\,}_{j}\rangle 
    = \frac{1}{\mathcal{Z}_e}{\rm Tr}\left( e^{-\hat{\mathcal{H}}_e/k_B T} c_{i}^\dagger c^{\,}_{j} \right) \nonumber \\
    & & \qquad \,\,\,\, =\sum_{m} f_{\rm FD}(\epsilon_m-\mu) U^{(m)*}_i U^{(m)}_j,
\end{eqnarray}
where $\mathcal{Z}_e$ denotes the partition function of the quasi-equilibrium electronic subsystem, $f_{\rm FD}(x) = 1/(1 + e^{x/T})$ is the Fermi–Dirac distribution, and $U^{(m)}_i$ are the eigenvectors of the tight-binding Hamiltonian in the site representation. Even within the adiabatic approximation, however, simulating the dynamical evolution of the Peierls system remains computationally demanding: the electronic forces must be evaluated at every time step, and the $\mathcal{O}(N^3)$ scaling of exact diagonalization (ED) rapidly becomes prohibitive for large system sizes.

In contrast, the machine-learning (ML) framework enables the evaluation of lattice forces with linear computational complexity, $\mathcal{O}(N)$, at each time step. This dramatic reduction in cost makes it possible to perform large-scale simulations of nonequilibrium phenomena such as CDW domain coarsening. Although other linear-scaling methods exist--most notably the kernel polynomial method (KPM)~\cite{weiler18,wang2018}--these techniques are typically limited to noninteracting (quadratic) fermionic Hamiltonians. By comparison, once trained, the ML model can be readily extended to more complex strongly correlated systems which explicitly incorporates electron–electron interactions, such as the Hubbard repulsion~\cite{berger96,johnston2013,costa2020}. The ML approach therefore provides an efficient and broadly applicable alternative for exploring the nonequilibrium dynamics of strongly correlated electron–phonon systems at substantially reduced computational cost.

\begin{figure*}[t]
\centering
\includegraphics[width=0.96 \linewidth]{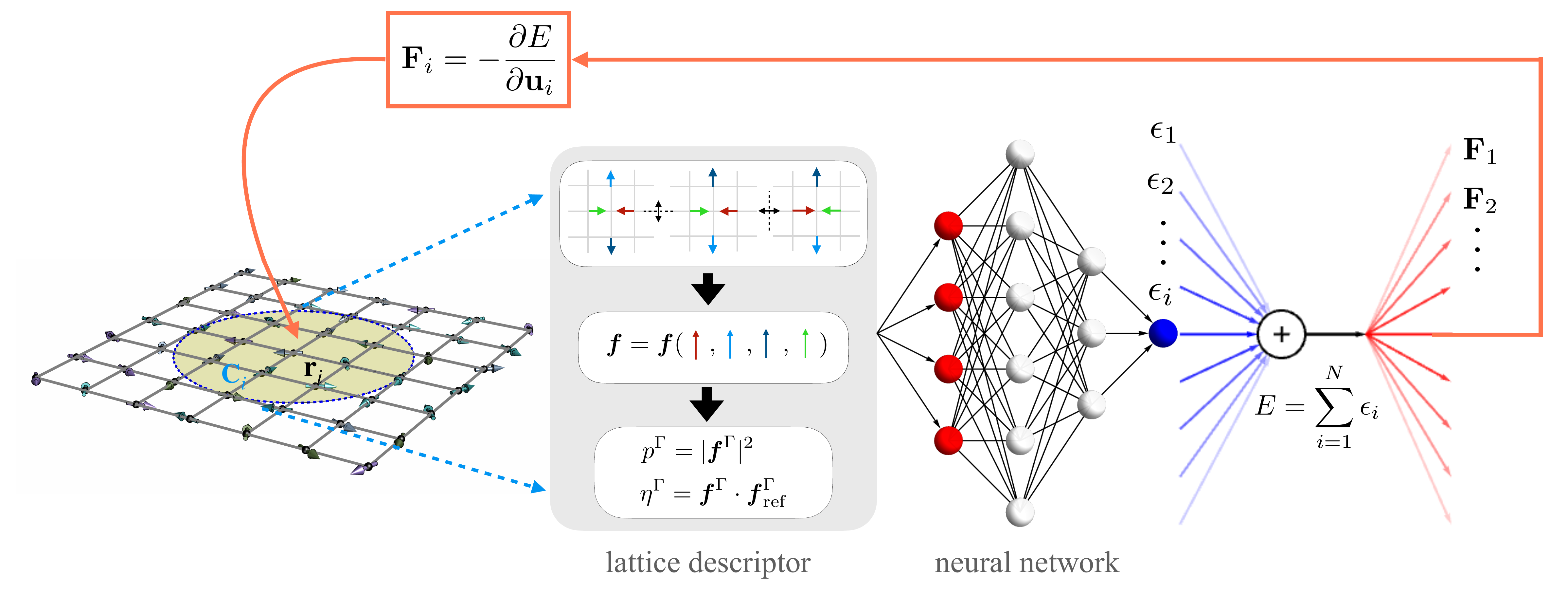}
\caption{Schematic diagram of the Behler-Parrinello architecture for ML force-field model of the Peierls system. The model comprises two main components: (i) a lattice descriptor and (ii) a multilayer neural network. The descriptor preserves the $D_4$ point-group symmetry of the lattice, with symmetry-adapted features $\{\eta_i^{(\Gamma,r)}\}$ constructed from the irreducible representations (IRs) of the local lattice environment $\mathcal{C}_i$. These features are input to the neural network to predict the local energy $\epsilon_i$, and the total energy $E$ is obtained by summing over all lattice sites. Atomic forces $\mathbf F _i$ are then computed as derivatives of the total energy with respect to the local lattice displacements $\mathbf{u}_i$.
}
    \label{fig:ML-schematic}
\end{figure*}

\section{machine learning force-field model and benchmarks}

\label{sec:MLFFmodel}

As discussed in Sec.~\ref{sec:intro}, the development of linear-scaling electronic-structure methods is founded on the principle of locality (or nearsightedness) in many-electron systems~\cite{kohn1996,prodan2005}. This principle asserts that the physical properties associated with a given atom depend primarily on its immediate local environment, while contributions from distant electronic degrees of freedom decay rapidly with distance. Modern machine-learning (ML) techniques provide a natural and efficient framework for incorporating this locality principle into $\mathcal{O}(N)$ electronic-structure methodologies.

A particularly successful realization of this idea is the class of ML-based interatomic potential and force-field models for ab initio molecular dynamics (MD) simulations~\cite{behler07,bartok10,Kermode2015,smith17,zhang18,behler16,deringer19,Mueller2020,mcgibbon17,suwa19,chmiela17,chmiela18,sauceda20,iftimie_ab_2005}. In conventional quantum MD, atomic forces are obtained by explicitly solving the many-electron Schrödinger equation—within a chosen electronic-structure approximation—at each integration step, a procedure that quickly becomes computationally prohibitive for large systems. To overcome this limitation, ML frameworks grounded in the locality principle have been developed to learn and reproduce these forces with orders-of-magnitude reduction in computational cost.

Among these approaches, energy-based ML models -- pioneered by Behler and Parrinello~\cite{behler07} and later extended by Bart\'ok \textit{et al.}~\cite{bartok10} -- represent a particularly powerful paradigm. In this framework, the total energy of a system is decomposed into a sum of local atomic contributions, each determined by the atom’s local environment. Crucially, the Behler–Parrinello (BP) scheme employs symmetry-preserving descriptors that encode the geometric and chemical environments in a manner invariant under translations, rotations, and permutations of identical atoms. The explicit enforcement of these fundamental symmetries ensures that the learned potential yields physically consistent energies and forces, while greatly enhancing transferability and data efficiency. As a result, BP-type models achieve both physical fidelity and computational scalability, providing a robust foundation for constructing accurate, many-body ML potentials applicable to systems of arbitrary size and composition. Notably, the BP-type force-field framework has recently been generalized to model adiabatic dynamics in a variety of condensed-matter lattice systems, including  itinerant electron magnets lattices, Holstein, and Jahn-Teller models~\cite{chen23_CDW,Supriyo2024,ma19,liu22,zhang2022,zhang2020,zhang2021,zhang2023,Yunhao2024,puhan22, sheng22, puhan21, puhan23}.

\subsection{Behler-Parrinello architecture}

The central objective of a BP-type force-field model is to predict the total energy of the system, from which the forces are obtained as its gradient. It is worth noting that this effective energy, obtained by integrating out the electronic degrees of freedom within the adiabatic approximation, serves as an effective potential energy for the phonons. A useful practical approach is to approximate this potential energy as a Taylor series expansion of displacement vector field:
\begin{eqnarray}
    E = E_0 + \frac{1}{2}\sum_{ij} \Phi^{(2)}_{ia, jb} u^a_i u^b_j + \sum_{ijk} \Phi^{(3)}_{ia, jb, kc}u^a_i u^b_j u^c_k + \cdots, \nonumber \\
\end{eqnarray}
where the coupling coefficients $\Phi^{(k)}$, known as the interatomic force constants (IFCs), characterize the $k$-body interactions in the expansion, and repeated $a, b, c$ indices indicate sum over $x, y, z$ components. These coefficients are subject to symmetry constraints arising from discrete lattice translations and point-group operations. Once the independent parameters are identified, their values can be obtained through linear or nonlinear regression methods. However, the number of independent parameters increases rapidly with the interaction order~$k$, posing a significant challenge for conventional fitting approaches.

Machine-learning (ML)–based force-field methods offer a more automated and efficient means of constructing such energy models. Here, we present a scalable ML force-field framework based on the BP architecture, which integrates a symmetry-aware descriptor with a neural-network regression model. A schematic illustration of the framework is shown in Fig.~\ref{fig:ML-schematic}. The ML model maps the lattice displacements ${\mathbf{u}_i}$ of the entire system onto the total energy $E$. Crucially, to ensure scalability, the key idea of this approach is to decompose the total energy, $E = \langle \hat{\mathcal{H}}_e \rangle + \mathcal{V}_L$, into a sum of local energy contributions $\epsilon_i$ associated with individual lattice sites:
\begin{equation}
    E = \sum_i\epsilon_i = \sum_i \varepsilon(C_i).
\end{equation}
As indicated by the second equality above, and consistent with the principle of locality, each local energy $\epsilon_i$ depends only on the immediate lattice environment $\mathcal{C}_i$ of site~$i$, through a function $\varepsilon(\cdot)$ that is intrinsic to the specific SSH model under consideration. Specifically, the local lattice environment is defined by the displacement vectors of neighboring sites within a cutoff radius $r_c$:
\begin{equation}
    \mathcal{C}_i = \left\{\mathbf{u}_j\,\big| \,|\mathbf{r}_j-\mathbf{r}_i| \leq \mathit{r}_c\right\},
\end{equation}
The cutoff distance is determined by the intrinsic locality of the interatomic forces. Crucially, the local energy function $\varepsilon(\mathcal{C}_i)$, which encodes the complex dependence of the site energy on the neighborhood displacements, is approximated by a deep neural network. The total energy $E$ is then obtained by applying the same local model uniformly over all lattice sites, effectively constructing a super-neural-network that automatically preserves translational symmetry. The driving forces, given by the energy gradients $\mathbf{F}_i = -\partial E / \partial \mathbf{u}_i$, can be efficiently computed through automatic differentiation across this super-network.


\subsection{Lattice descriptor}

Despite the remarkable expressive power of deep neural networks, as guaranteed by the universal approximation theorem, the intrinsic symmetries of the underlying electronic Hamiltonian are not automatically enforced in the resulting ML model. In computer science, techniques such as data augmentation are often employed to help neural networks learn the symmetries present in the training data. However, owing to the statistical nature of ML optimization, such symmetries can only be captured approximately. To remedy this limitation, symmetry-aware descriptors have been introduced to explicitly encode the exact symmetry constraints into ML force-field models, ensuring that the learned potentials rigorously respect the fundamental invariances of the physical system.

The importance of symmetry-aware descriptors was already emphasized in the original work of Behler and Parrinello~\cite{behler07}. To encode local atomic environments while rigorously preserving the symmetries of the potential energy function, they introduced a family of feature variables known as atom-centered symmetry functions (ACSFs)~\cite{behler07,behler11}. Constructed from interatomic distances and angular correlations among neighboring atoms, ACSFs ensure explicit invariance under the SO(3) rotational and reflection symmetry group~\cite{behler11}. Following this seminal development, a wide range of alternative atomic descriptors have been proposed and implemented in ML force-field frameworks~\cite{bartok13,ghiringhelli15,himanen20,Rupp2012,behler11,shapeev16,drautz19,Hansen2015,Faber2015,huo22}. Among these, the group-theoretical bispectrum method provides a particularly systematic formulation by representing local atomic environments through rotationally invariant combinations of spherical harmonics~\cite{bartok10,bartok13}. More recently, many of these descriptors have been recognized as specific instances of the Atomic Cluster Expansion (ACE) formalism~\cite{drautz19}, which offers a unified and hierarchical framework for constructing complete, symmetry-adapted representations of atomic environments.

In the context of condensed-matter lattice systems, the construction of symmetry-aware descriptors differs fundamentally from that for atomic or molecular systems. The key distinction lies in the structure of the underlying symmetry groups: the continuous Euclidean symmetry $E(3)$ governing atomic configurations is replaced by discrete lattice translations combined with site-specific point-group operations. Moreover, additional internal symmetries often arise from the nature of the dynamical variables themselves—such as local magnetic moments, charge densities, or lattice distortions. A general group-theoretical framework for developing symmetry-invariant descriptors under these combined spatial and internal symmetries has recently been formulated, with several concrete implementations demonstrated in recent works.

For the square-lattice SSH model, the relevant point-group symmetry is $D_4$. The symmetry operations of this group act simultaneously on the real-space lattice sites and on the corresponding displacement vectors. To describe these combined transformations, consider the local neighborhood $\mathcal{C}_i$ centered at site $i$, and let $\hat{g}$ denote a discrete rotation or reflection in the point group that maps site $j$ to site $k$. Let $O(\hat{g})$ be the orthogonal matrix representation of $\hat{g}$. The action of $\hat{g}$ on the local environment $\mathcal{C}_i$ is then given by
\begin{eqnarray}
& & \mathbf r_k - \mathbf r_i = O(\hat{g}) \cdot (\mathbf r_j - \mathbf r_i), \nonumber \\
& & \mathbf u_k \,\to \, \mathbf u'_k = O(\hat{g}) \cdot \mathbf u_j.
\end{eqnarray}
This relation captures the coupled transformation of lattice geometry and displacement fields under the point-group operations of $D_4$.

A proper representation of the local neighborhood $\mathcal{C}_i$ must remain invariant under the coupled symmetry operations acting on both lattice coordinates and displacement vectors. A systematic framework for constructing such invariants is provided by the group-theoretical bispectrum method~\cite{kondor07}.
Descriptors derived from bispectrum coefficients have been widely applied in conjunction with Gaussian process regression for quantum molecular dynamics simulations~\cite{bartok10,bartok13}.
More recently, this group-theoretical approach has been extended to electronic lattice systems in condensed-matter physics, providing a general theoretical foundation for constructing symmetry-adapted descriptors~\cite{puhan22,ma19}.

Here we outline the group-theoretical procedure for deriving invariant feature variables for the SSH model; further details are provided in Appendix~\ref{appendix:A}. First, we note that the set of displacement vectors in the neighborhood, denoted by $\mathcal{C}_i$, constitutes a high-dimensional representation of the point group. This representation can be systematically decomposed into the fundamental irreducible representations (IRs) of the group. The decomposition is greatly simplified by the fact that the original representation matrix is naturally block-diagonal, with each block corresponding to distortions at a fixed distance from the central site. Standard group-theoretical techniques can then be applied independently to each block to obtain the full decomposition~\cite{hamermesh_group_1989}.

As an illustrative example, consider the displacement vectors of the four nearest neighbors, denoted as $\mathbf u_A, \mathbf u_B, \mathbf u_C,$ and $\mathbf u_D$. Together they form an eight-dimensional reducible representation of the $D_4$ group, which can be decomposed as: $8 = A_1 \oplus A_2 \oplus B_1 \oplus B_2 \oplus 2E$. Based on this decomposition, we construct the corresponding symmetry-adapted linear combinations (SALCs), expressed in vector form as
\[
    \bm f^{(\Gamma, r)} = \left(f^{(\Gamma, r)}_1, f^{(\Gamma, r)}_2, \cdots, f^{(\Gamma, r)}_{n_\Gamma} \right)
\]
where $\Gamma$ labels the IR type, $r$ indexes its multiplicity, and $n_\Gamma$ is the dimension of the IR. For instance, the totally symmetric combination is $f^{A_1} = u_A^y + u_B^x + u_C^y + u_D^x$, which reflects the coupling between lattice and displacement symmetries, leading to the mixing of $x$ and $y$ components.
Similarly, one of the doublet combinations transforming as the $E$ representation is given by 
\[
    \bm f^{(E, 1)} = (u_A^x - u_B^x + u_C^x - u_D^x, 
    -u_A^y + u_B^y - u_C^y + u_D^y).
\]
These symmetry-adapted linear combinations $\bm{f}^{(\Gamma, r)}$ serve as the building blocks to derive invariant feature variables. First, the squared magnitude of each irreducible component, $p^{\Gamma}_r = \left|\bm f^{\Gamma}_r\right|^2$, is by construction invariant under all point-group operations. The complete set of these quantities, ${p^{\Gamma}_r}$, forms the power spectrum, which serves as a basic set of symmetry-invariant feature variables. The power spectrum is a special case of a more general class of invariants known as bispectrum coefficients~\cite{kondor07,bartok13}. A bispectrum coefficient involves the product of three IR components combined with the Clebsch–Gordon coefficients that couple different IRs. These quantities encode not only the invariant magnitudes but also the relative phase relationships among distinct IR channels, providing a comprehensive and symmetry-faithful characterization of the local environment.

For most point groups, the dimensionality $n_\Gamma$ of each IR is small, while the multiplicity index $r$ can be large. As a result, the number of possible bispectrum coefficients grows rapidly and often includes significant redundancy. To obtain a more compact representation, a refined construction based on reference IR coefficients $\bm f^{\Gamma}_{\text{ref}}$ has been introduced~\cite{puhan21}. A reference coefficient is defined once for each IR type and derived using the same decomposition procedure, but evaluated on coarse-grained displacement vectors $\overline{\mathbf u}$. The reference IR provides a natural means to define a phase variable for each IR component, $\exp({\phi^\Gamma_r}) \equiv \bm f^\Gamma_r \cdot \bm f^\Gamma_{\rm ref} / |\bm f^\Gamma_r |\, |\bm f^\Gamma_{\rm ref}| = \pm 1$, from which the relative phases between equivalent IRs can be directly inferred. The relative phases between distinct IR types are, in turn, encoded by bispectrum coefficients constructed from the reference IR alone.
By combining the invariant amplitudes with their corresponding phases, one obtains a set of complex invariant feature variables, $G^\Gamma_r = p^\Gamma_r \,\exp(i \phi^\Gamma_r)$ which are further supplemented by the bispectrum coefficients $B^{\Gamma_1,\Gamma_2,\Gamma_3}_{\text{ref}}$ derived from the reference IR. 

As illustrated in Fig.~\ref{fig:ML-schematic}, rather than directly feeding the raw displacement configuration $\mathcal{C}_i$ into the neural network, we employ the symmetry-invariant feature variables ${ G^{(\Gamma, r)} }$ as inputs to the model. This means that the system energy depends on the lattice displacements through such feature variables
\begin{eqnarray}
\label{eq:E_G}
E = \sum_i \varepsilon_{\bm \theta}\bigl[ G^{(\Gamma, r)}(\mathcal{C}_i) \bigr],
\end{eqnarray}
where $\bm \theta = \{ w, b \}$ denotes the set of trainable parameters (weights and biases) of the neural-network representation of the energy function. Since the feature variables are constructed to be invariant under all symmetry operations of the point group, Eq.~(\ref{eq:E_G}) ensures that both the local energy $\varepsilon_{\bm \theta}$ and the total system energy $E$ are symmetry-invariant by design.

\subsection{Benchmarks of the ML model}

\begin{figure}[t]
    \centering
    \includegraphics[width=0.99\linewidth]{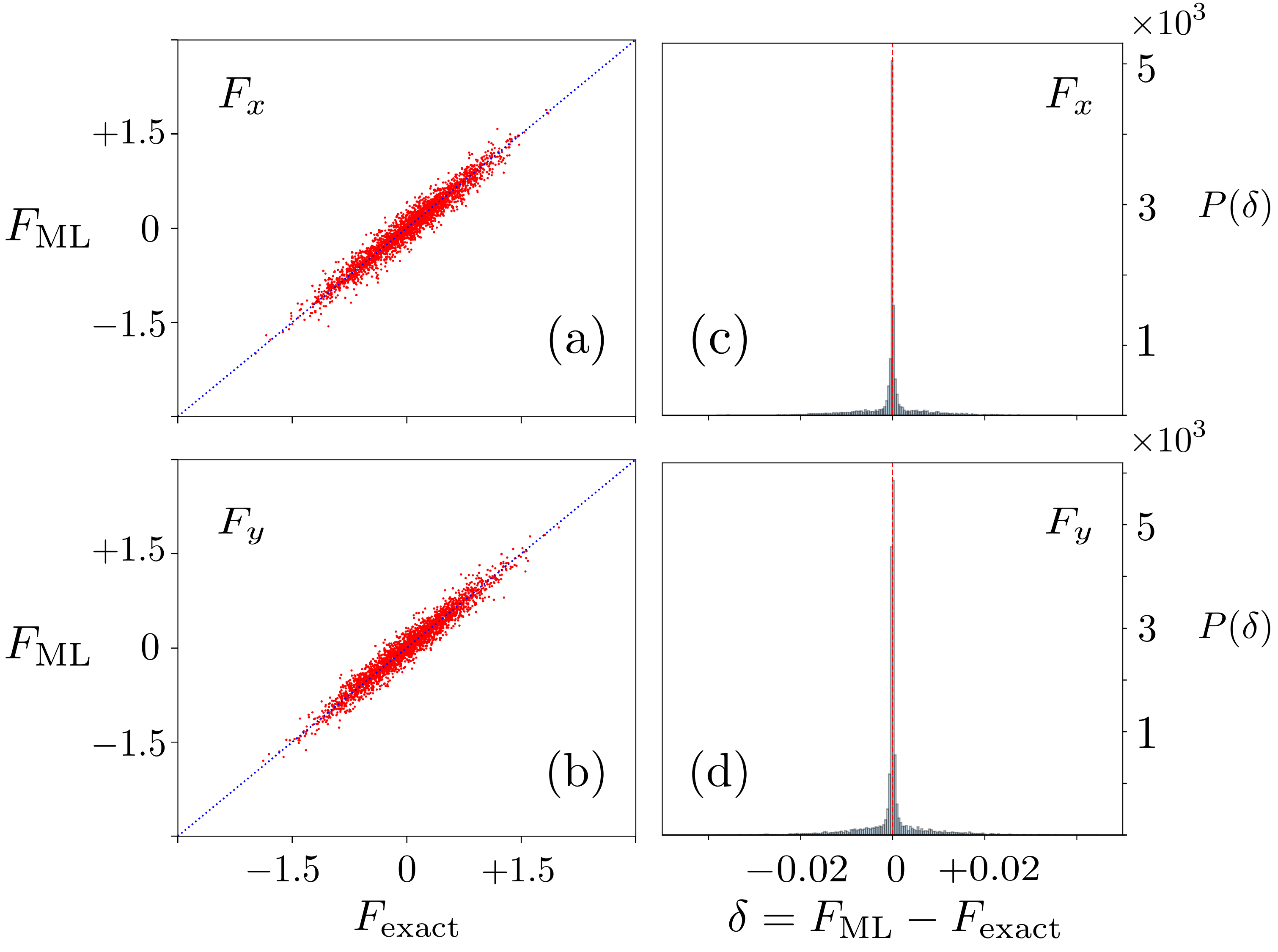}
    \caption{Benchmark of the ML force-field model against the exact force on the lattice. Panels (a) and (b) compare the ML-predicted forces $F_{\mathrm{ML}}$ with the exact forces $F_{\mathrm{eexact}}$ along the $x$ and $y$ directions, respectively. Panels (c) and (d) show the corresponding error distributions $\delta = F_{\mathrm{ML}} - F_{\mathrm{exact}}$ in each direction. The standard deviations of the prediction errors are $\sigma_x = 0.002$ and $\sigma_y = 0.003$.}
    \label{fig:benchmark-force}
\end{figure}

The neural-network (NN) model used in this work was implemented and trained using the PyTorch framework~\cite{paszke19,Nair10,barron2017,Paszke2017,kingma2017}. The architecture consists of a seven-layer fully connected feed-forward network, designed to approximate the nonlinear mapping between the symmetry-invariant feature variables ${ G^{(\Gamma, r)} }$ and the corresponding local energy contributions. The network employs rectified linear unit (ReLU) activation functions for all hidden layers and a linear activation at the output layer. Optimization of the trainable parameters $\bm{\theta} = \{ w, b \}$ was performed using the Adam stochastic gradient descent algorithm with an adaptive learning rate schedule. Further details of the network architecture, hyperparameters, and training procedure are provided in Appendix~\ref{appendix:B}.

The supervised training dataset was generated from ED-based Langevin simulations of the SSH lattice model on a $50 \times 50$ system. The lattice dynamics were computed within the adiabatic approximation, wherein the electronic subsystem is assumed to instantaneously follow the classical lattice displacements. As described in Appendix~\ref{appendix:C}, the SSH model and its associated equations of motion can be expressed in terms of a small set of dimensionless parameters, whose specific values are also listed there. Both the ED and ML-based Langevin simulations were carried out using this dimensionless formulation of the dynamical equations. The time evolution of the lattice variables was integrated using the Velocity–Verlet algorithm, a symplectic scheme that preserves time reversibility and ensures long-term numerical stability~\cite{verlet67,vyas2010}.

A total of 2000 snapshots of $\{\mathbf{u}_i, \mathbf{F}_i \}$ pairs were collected at uniform time intervals along the ED–Langevin trajectories. The dataset includes configurations drawn from three representative dynamical regimes: (i) random initial states, (ii) intermediate coarsening processes, and (iii) late-time configurations near the ground state. This diverse sampling strategy ensures that the neural network is trained over a wide range of local environments, enabling it to accurately capture both transient and near-equilibrium lattice dynamics.

The predictive accuracy of the ML force-field model was benchmarked against the exact forces obtained from the ED simulations, as shown in Fig.~\ref{fig:benchmark-force}. Panels (a) and (b) compare the ML-predicted forces $F_{\mathrm{ML}}$ with the exact forces $F_{\mathrm{exact}}$ along the $x$ and $y$ directions, respectively, while panels (c) and (d) display the corresponding prediction error distributions $\delta = F_{\mathrm{ML}} - F_{\mathrm{exact}}$. The resulting standard deviations, $\sigma_x = 0.002$ and $\sigma_y = 0.003$, demonstrate that the NN achieves quantitatively accurate force predictions within the numerical precision of the ED data. The force units and normalization conventions are specified in Appendix~\ref{appendix:C}.

\begin{figure}[t]
    \centering
    \includegraphics[width=0.99\linewidth]{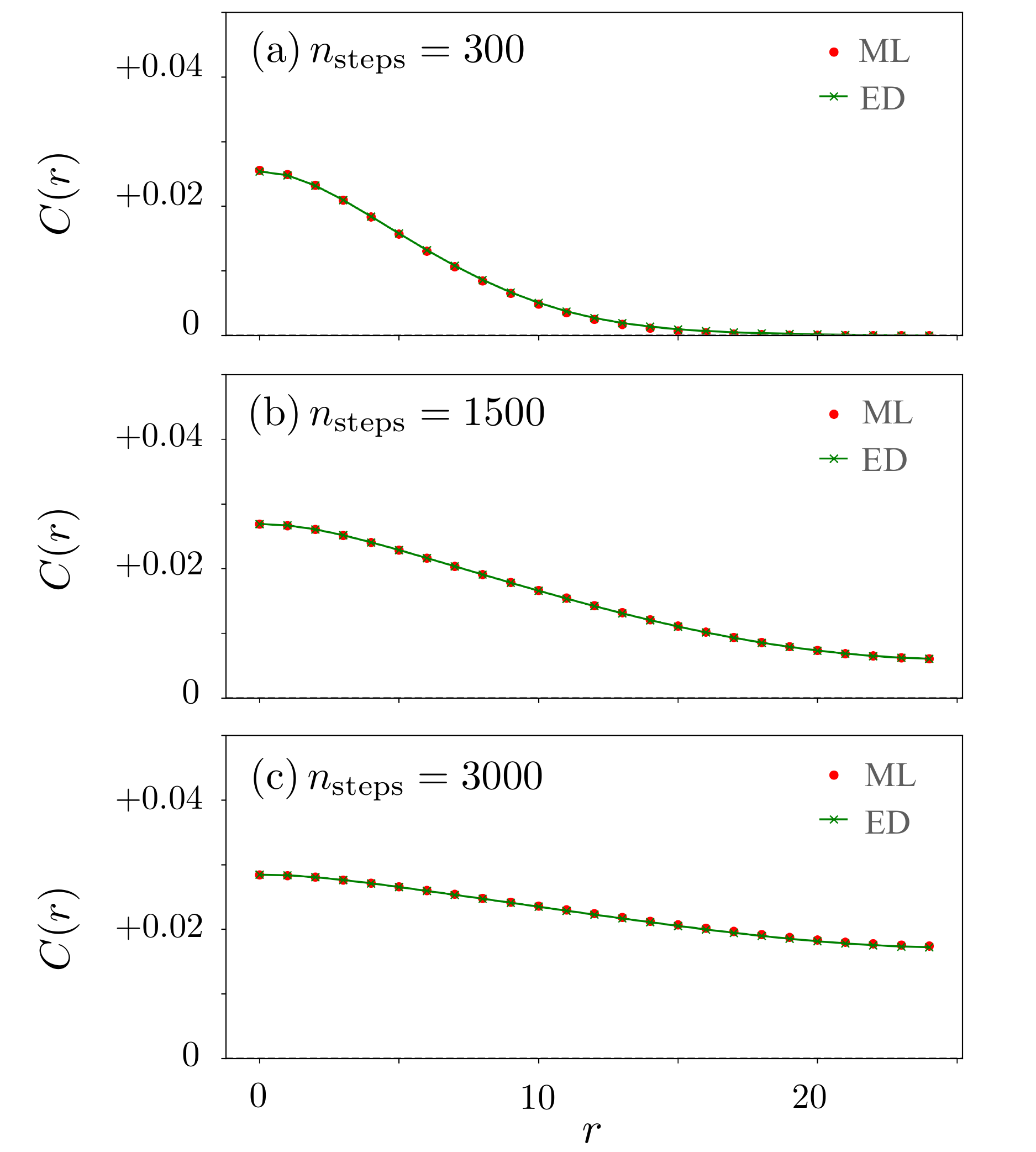}
    \caption{Comparison of time-dependent correlation functions $C(r,t)$ obtained from Langevin simulations using the ML force-field model and the exact-diagonalization (ED) method. The correlation function is evaluated as the average of $C^{xx}_{ij}$ and $C^{yy}_{ij}$ for site pairs $(i,j)$ separated by distance $r = r_{ij}$ along the diagonal direction. Each curve represents an ensemble average over 100 independent simulations on a $50\times50$ lattice with randomized initial conditions.}
    \label{fig:benchmark-CF}
\end{figure}

In addition to the force-prediction benchmark, we further performed dynamical validation to demonstrate that the ML force-field model can faithfully reproduce the real-time evolution of the SSH model. To this end, the trained neural network was incorporated into the Langevin dynamics framework, and thermal quench simulations were carried out using this hybrid scheme (hereafter referred to as ML–Langevin simulations). These simulations were designed to test not only the instantaneous force accuracy but also the long-time dynamical consistency of the ML model relative to ED–based dynamics.

\begin{figure*}[t]
    \centering
    \includegraphics[width=\linewidth]{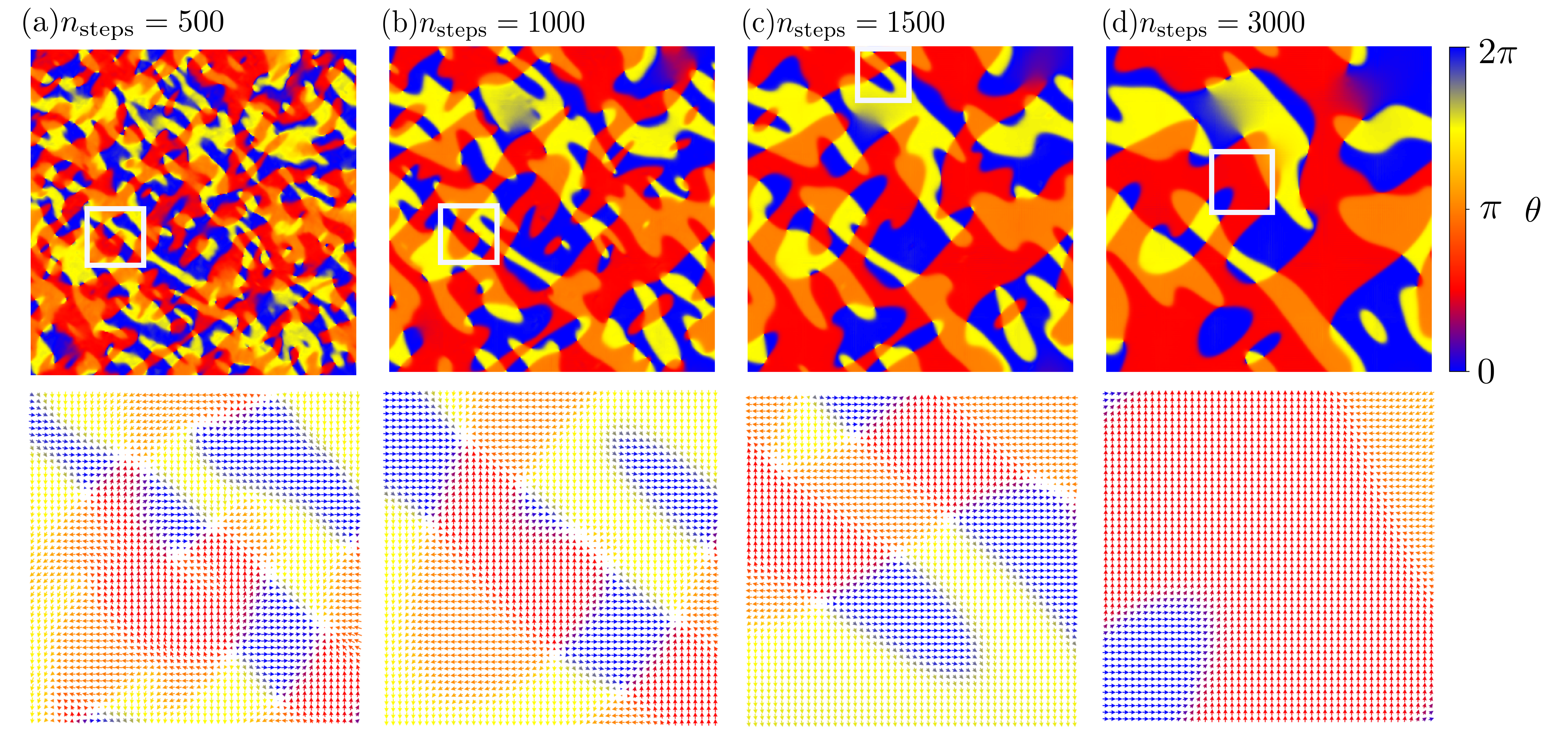}
    \caption{Snapshots from ML-Langevin simulations on a $300\times300$ lattice showing the temporal evolution of the phase field $\theta_i$, defined via the local CDW order parameter $\bm{\phi}_i = (\phi^x_i, \phi^y_i)$ [Eq.~(\ref{eq:local_CDW_order})]. The phase $\theta_i = \arctan(\phi^y_i / \phi^x_i)$ varies continuously from $0$ to $2\pi$, with $\theta_i = 0, \pi/2, \pi$, and $3\pi/2$ corresponding to the four degenerate ground states illustrated in Fig.~\ref{fig:4pipi}. The bottom panels display vector plots of $\bm{\phi}_i$ within the white boxed regions of the corresponding top panels. At early times, the system exhibits diagonal domain structures that give rise to anomalous coarsening behavior. At later stages, the evolution crosses over to the conventional Allen-Cahn coarsening regime characterized by curvature-driven domain growth.}
    \label{fig:MLlargesimulation}
\end{figure*}

To quantitatively compare the dynamical evolution, we evaluated the equal-time two-point correlation function of the lattice displacement field,
\begin{equation}\label{eqn:corrftn}
C^{ab}_{ij}(t) = \langle u^a_i(t) u^b_j(t) \rangle - \langle u^a_i(t) \rangle \langle u^b_j(t) \rangle,
\end{equation}
where $a,b$ denote Cartesian components. The averaging $\langle \cdots \rangle$ was performed over all lattice sites and over an ensemble of independent initial configurations to ensure statistical convergence. Fig.~\ref{fig:benchmark-CF} presents a direct comparison of the correlation functions obtained from the ML–Langevin and ED–Langevin simulations. Ensemble averages were taken over 100 independent runs to suppress stochastic fluctuations. The two sets of results exhibit excellent quantitative agreement across all timescales, confirming that the ML model accurately captures both the short-time relaxation and the long-time coarsening dynamics.

\section{Machine learning dynamical simulation of CDW pordering}

\label{sec:largeML}

As discussed in Sec.~\ref{sec:MLFFmodel}, the linear scalability of the ML force-field model offers a substantial computational advantage for large-scale dynamical simulations. Whereas the ED approach scales as $\mathcal{O}(N^3)$ with the number of lattice sites $N$, the ML–Langevin framework achieves $\mathcal{O}(N)$ scaling owing to its strictly local energy decomposition. This reduction in computational complexity enables access to long-time dynamical regimes and system sizes that are otherwise computationally intractable within the conventional ED-based approach.

The linear scaling of the ML–Langevin framework enables simulations of system sizes far exceeding those accessible by ED–based approaches, providing a powerful platform for investigating emergent mesoscale phenomena such as charge-density-wave (CDW) phase ordering in Peierls systems. Using the same quench temperature and dynamical parameters as in the ED–Langevin benchmarks, we performed large-scale ML–Langevin simulations on a $300 \times 300$ lattice. Representative snapshots from a single dynamical run at different time steps following the thermal quench are shown in Fig.~\ref{fig:MLlargesimulation}. To quantify the evolution of the inhomogeneous CDW textures during relaxation, we introduce a vector order parameter $\bm \phi_i = (\phi_i^x, \phi_i^y)$ to characterize the local CDW order:
\begin{equation}
	\label{eq:local_CDW_order}
 	\bm \phi_i = \biggl( \mathbf u_i - \tfrac{1}{4}\sum_{j \in \mathcal{N}(i)} \mathbf u_j \biggr)\exp(i\mathbf{Q}\cdot \mathbf{r}_i),
\end{equation}
where  $\mathcal{N}(i)$ denotes the nearest neighbors of site-$i$, and the prime on the summation denotes the nearest neighbors of site $i$. This definition incorporates the sublattice phase factor into the relative lattice distortions, such that within a single domain the lattice modulation is represented by a uniform order parameter. From the local order parameters, we define the corresponding phase angle $\theta_i = \mathrm{arctan}(\phi_i^y / \phi_i^x)$. The four degenerate CDW ground states illustrated in Fig.~\ref{fig:4pipi} correspond to $\theta_i = 0, \pi/2, \pi,$ and $3\pi/2$, respectively. The same color scheme is used to represent the angular variable $\theta_i$ and the CDW domain coloring in Fig.~\ref{fig:4pipi}.

At early times following the thermal quench [Fig.~\ref{fig:MLlargesimulation}(a)], numerous small CDW domains of different colors emerge from the initially disordered state, signaling rapid local ordering driven by the Peierls instability. As the system evolves [Fig.~\ref{fig:MLlargesimulation}(b)], these nascent domains begin to merge and compete, forming a dense network of domain walls (interfaces) that separate regions of distinct CDW phases. At this stage, point-like defects—junctions where four domain walls meet and the four degenerate CDW variants intersect—are abundant. These defects are analogous to vortices in the XY model, but with distinct energetics owing to the reduction of the continuous $O(2)$ rotational symmetry to the discrete $Z_4$ subgroup. At intermediate times [Fig.~\ref{fig:MLlargesimulation}(c)], both the domain-wall density and the defect number decrease systematically as smaller domains vanish and neighboring walls annihilate via curvature-driven motion. At late times [Fig.~\ref{fig:MLlargesimulation}(d)], the system approaches a nearly ordered configuration consisting of a few large CDW domains separated by straight or gently curved interfaces, with most point-like defects annihilated. The remaining domain walls preferentially align along the diagonal directions of the square lattice, consistent with the electron-mediated anisotropy of the domain-wall energy discussed below. Overall, this time sequence vividly captures the nonequilibrium coarsening dynamics of a four-state CDW order parameter, including the intertwined evolution of extended interfaces and localized topological defects.

\begin{figure}[t]
    \centering
    \includegraphics[width=0.92\linewidth]{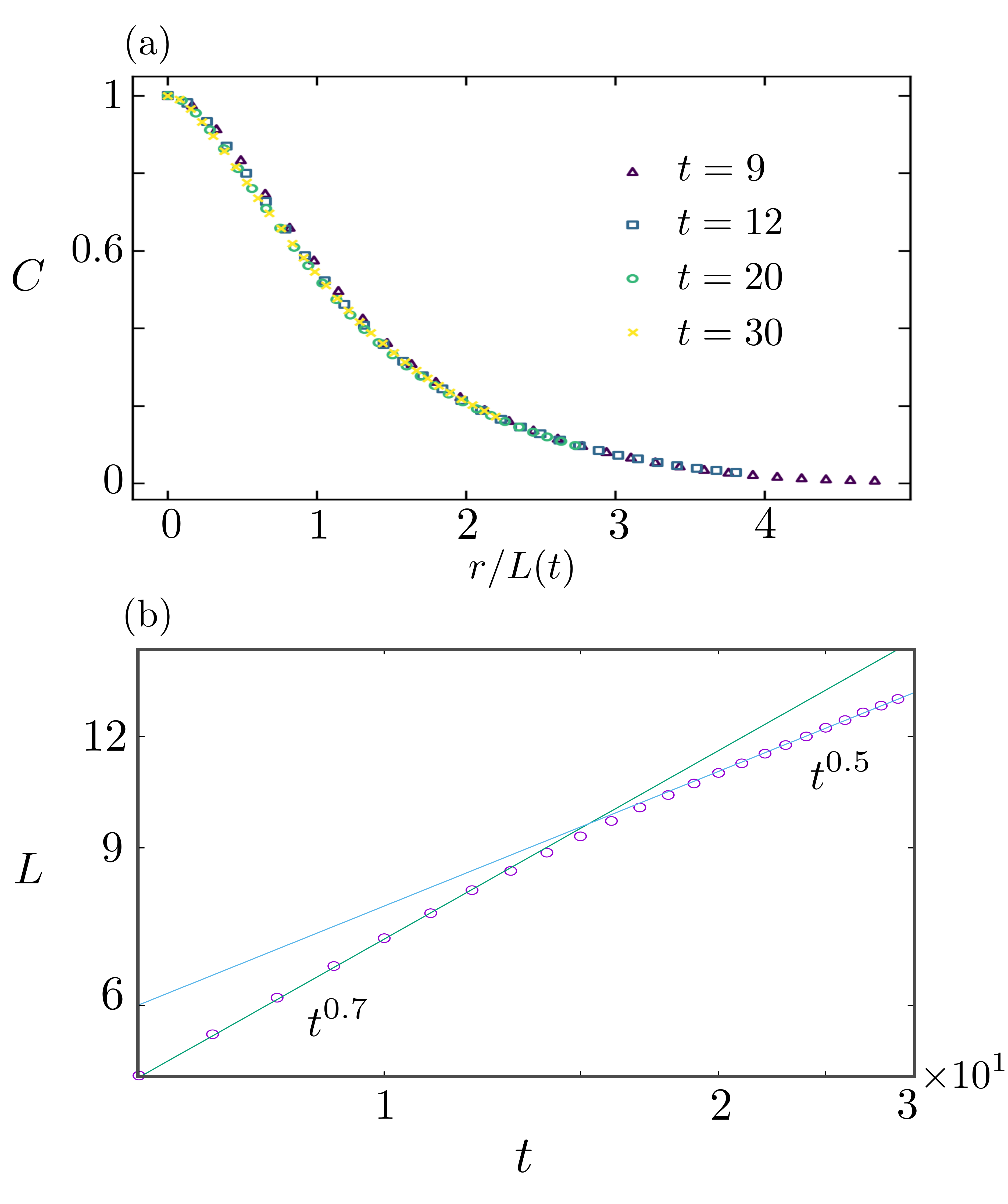}
    \caption{Dynamical scaling and correlation-length growth obtained from ML–Langevin simulations on a $300\times300$ lattice. Panel (a) demonstrates excellent collapse of the correlation functions with the rescaled distance $r/L(t)$. Panel (b) shows correlation length $L(t)$ as a function of time, which initially exhibits enhanced growth with an exponent $\alpha \approx 0.7$ before crossing over to the Allen–Cahn regime with the expected exponent $\alpha=0.5$.}
    \label{fig:Large_dynL}
\end{figure}

To quantitatively characterize the evolving domain morphology during phase ordering, we compute the time-dependent displacement correlation function defined in Eq.~(\ref{eqn:corrftn}), from which a time-dependent correlation length~$L(t)$ is extracted. The correlation length is determined using the half-maximum condition, $C(L(t),t) = \tfrac{1}{2}C(0,t)$, which provides a robust measure of the typical CDW domain size. When the correlation functions at different times are plotted as a function of the rescaled distance $r/L(t)$, the data points collapse onto a single universal curve, as shown in Fig.~\ref{fig:Large_dynL}(a). This collapse demonstrates the emergence of dynamical scaling symmetry~\cite{Bray1994,Puri2009} in the coarsening process,
\begin{equation}
C(\mathbf{r},t) = g(|\mathbf{r}| / L(t)),
\end{equation}
where $g(x)$ is a time-independent scaling function that characterizes the universal spatial correlations of the CDW domains. The existence of such scaling implies statistical self-similarity in the domain morphology during coarsening.

Interestingly, the extracted time dependence of the characteristic domain size~$L(t)$, shown in Fig.~\ref{fig:Large_dynL}(b), reveals two distinct dynamical regimes, both exhibiting power-law growth,
\begin{eqnarray}
L(t) \sim t^{\alpha}.
\end{eqnarray}
At late times, the coarsening follows the Allen–Cahn growth law with an exponent $\alpha = 1/2$, characteristic of curvature-driven domain-wall motion in systems with a nonconserved order parameter. In contrast, during an extended intermediate regime, a markedly faster coarsening with an effective exponent $\alpha \approx 0.7$ is observed. This enhanced coarsening is attributed to anisotropic domain-wall motion, where the walls preferentially propagate along certain crystallographic directions favored by the underlying lattice symmetry. Such directionally biased coarsening behavior, and its microscopic origin in the Peierls lattice, will be discussed in detail in the next section.

\section{Anomalous CDW coarsening}

\label{sec:preferDW&coarsening}

The coarsening dynamics following a symmetry-breaking transition are primarily governed by the motion and annihilation of topological defects that emerge spontaneously during a rapid quench from the disordered phase. These defects--such as domain walls, vortices, and disclinations--are localized singularities of the order-parameter field that dictate the kinetics of phase ordering. In the $q$-state clock model, which interpolates between the discrete Ising ($q=2$) and continuous XY ($q \to \infty$) limits, symmetry breaking gives rise to two dominant defect types: domain walls separating neighboring ordered states and point-like vortices at the junctions of multiple domains.

For $q=2$, corresponding to the Ising universality class, the nonconserved scalar order parameter obeys the Allen–Cahn growth law with exponent $\alpha = 1/2$, independent of lattice geometry~\cite{Bray1994,Puri2009}. This scaling arises from curvature-driven interface motion governed by the Allen–Cahn equation~\cite{Puri2009}, 
\begin{eqnarray}
	\label{eq:AC-eq}
	v_{\rm dw} = - c\, \kappa_{\rm dw}, 
\end{eqnarray}
where $v_{\rm dw}$ is the domain-wall velocity along the local normal direction, $\kappa_{\rm DW}$ denotes the local curvature of the interface, and $c$ is a constant. For domains with a characteristic size $L$, the velocity can be approximated as $v_{\rm dw} \sim dL/dt$, while the typical curvature scales as $\kappa_{\rm dw} \sim 1/L$. Substituting these relations into the above Allen-Cahn equation yields the asymptotic growth law $L(t) \sim t^{1/2}$. 

In the asymptotic limit $q \to \infty$, the model approaches the well-known two-dimensional XY system, whose elementary topological excitations are vortices and antivortices. In this regime, phase ordering proceeds predominantly through vortex–antivortex annihilation, and the characteristic coarsening length scale $L$ can be identified with the mean separation between these point defects. Assuming dissipative Brownian dynamics for the vortices, their motion satisfies $v_{\rm vtx} \sim F / \gamma$, where $v_{\rm vtx}$ denotes the terminal velocity, $\gamma$ is the viscous damping coefficient, and $F$ represents the average mutual force between a vortex and an antivortex. Within the 2D XY framework, vortices form a Coulomb gas with a logarithmic interaction potential, leading to a force scaling as $F \sim -1/L$. Approximating the defect velocity by $v_{\rm vtx} \sim dL/dt$, one obtains the equation of motion $dL/dt \sim -1/L$, which yields a square-root temporal growth law. A more refined treatment, however, reveals a logarithmic correction to this scaling, resulting in $L(t) \sim (t/\ln t)^{1/2}$~\cite{Yurke93,Bray00}. 

For finite $q > 2$, extensive numerical studies have shown that the late-time coarsening dynamics generally revert to the Allen–Cahn scaling~\cite{chatterjee2018,chatterjee2020,Corberi06,Kaski83,Kaski85,Kaski85b,Grest84,Enomoto90}, reflecting the dominant role of curvature-driven interface motion. The role of vortices in the coarsening process of the $q$-state clock model, however, is more subtle. In this case, vortices correspond to point-like junctions where multiple domain walls intersect. Owing to the absence of a continuous $O(2)$ symmetry, these vortices--although they may carry discrete analogs of topological charge similar to those in the XY model--are not genuinely deconfined topological defects. Consequently, the effective interaction potential between a vortex and an antivortex is proportional to the total length of the domain walls connecting them. The motion of vortices is therefore primarily governed by the reduction of domain-wall lengths, which again proceeds through the curvature-driven mechanism.

Our large-scale ML-Langevin simulations, as shown in Fig.~\ref{fig:Large_dynL}, exhibit Allen-Cahn-type domain growth at late times, consistent with the scenario in which coarsening is governed predominantly by curvature-driven interface motion, while vortex-antivortex annihilation plays only a secondary role. Rather than moving independently, these point-like defects tend to be bound to domain-wall intersections. Their annihilation typically occurs as a byproduct of domain-wall coalescence or the collapse of isolated domains. This observation supports the picture that the phase-ordering kinetics in the Peierls system are predominantly governed by curvature-driven domain-wall motion, with point defects acting as passive spectators in the late-time coarsening regime.

\begin{figure}[t]
    \centering
    \includegraphics[width=\linewidth]{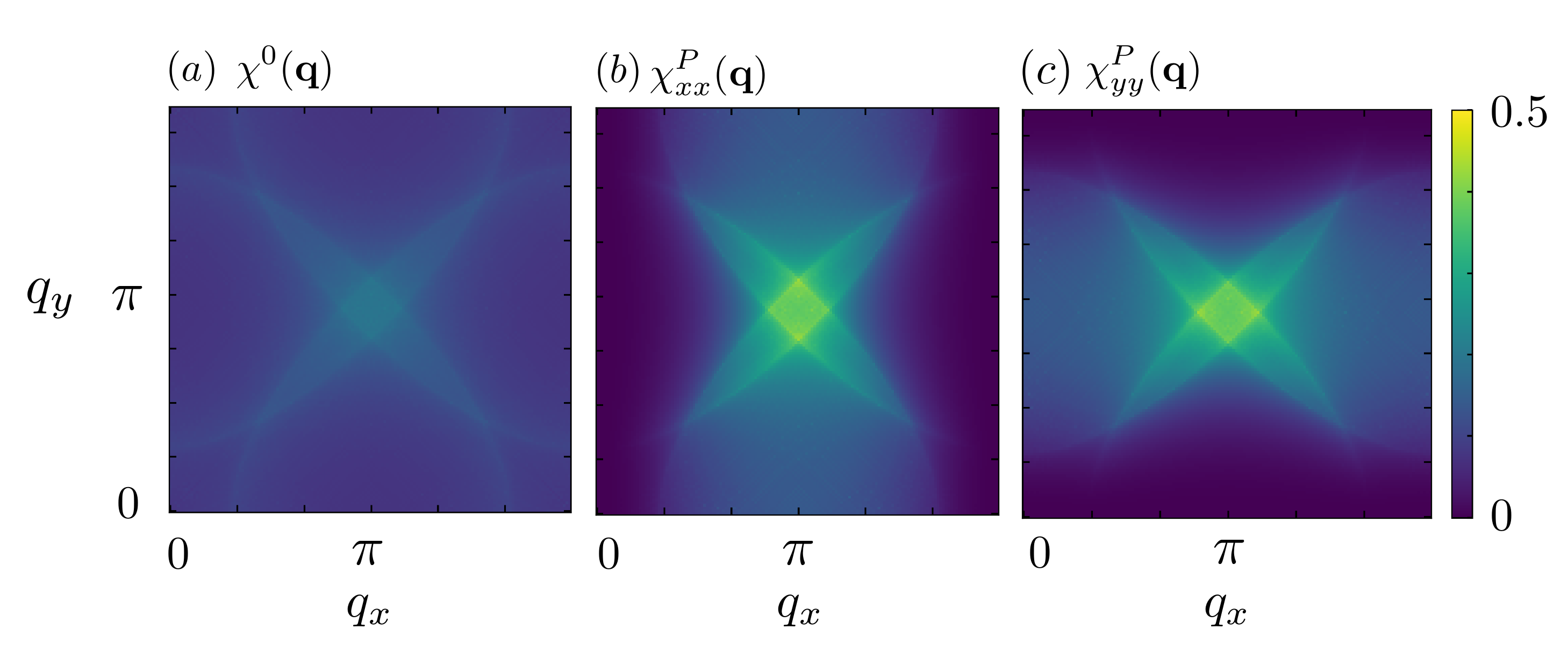}
    \caption{
(a) Lindhard susceptibility $\chi^L(\mathbf{q})$ of the square-lattice tight-binding model with both NN and NNN hopping.
(b,c) Electronic lattice susceptibilities $\chi^{P}_{xx}(\mathbf{q})$ and $\chi^{P}_{yy}(\mathbf{q})$, respectively, of the SSH model. In contrast to the Lindhard response, the lattice susceptibilities exhibit enhanced overall magnitude. After including the quadratic lattice-coupling energy, the second derivative of $\chi^{P}_{\mu\mu}(\mathbf{q})$ ($\mu = x, y$) attains a minimum along the diagonal direction near $\mathbf{Q} = (\pi,\pi)$, reflecting the preferred orientation of domain-wall formation.}
    \label{fig:peierlssus}
\end{figure}

The enhanced growth exponent $\alpha \approx 0.7$ observed at the early stages of coarsening can be attributed to the distinctive domain morphology that rapidly develops following the quench in the SSH model. As illustrated in Fig.~\ref{fig:MLlargesimulation}, the CDW domains display pronounced directional anisotropy along the two diagonal axes of the square lattice, particularly during the early stages of the relaxation process. The origin of this directional anisotropy can be understood from a perturbative analysis of the SSH model in the weak-coupling regime. By treating the electron-lattice coupling as a perturbation to the square-lattice tight-binding model, we obtain the following expression for the effective electronic energy,
\begin{equation}
    \label{eq:effect_EL}
    \mathcal{E}_e = \langle \hat{\mathcal{H}_e} \rangle 
    = -4g^2 \sum_{\mathbf{q}}\sum_{a,b=x,y} \chi_{ab}^P(\mathbf{q}) {u}^*_{a}(\mathbf{q})u^{\,}_b(\mathbf q),
\end{equation}
where  $\mathbf{q}$ is the momentum within the first Brillouin zone, $\mathbf{u}(\mathbf{q})$ is the Fourier transform of the lattice displacement field, and $\chi_{ab}^P(\mathbf{q})$ represents the electronic lattice susceptibility. The detailed derivation of Eq.~(\ref{eq:effect_EL}) is provided in Appendix~\ref{appendix:D}.
The resulting susceptibility, shown in Fig.~\ref{fig:peierlssus}, exhibits a pronounced peak at $\mathbf{Q} = (\pi, \pi)$. Notably, $\chi_{ab}^P(\mathbf{q})$ decays most slowly along the diagonal directions near the staggered wave vector $\mathbf{Q}$, reflecting enhanced electronic nesting along these momentum paths. In fact, such anisotropy can already be seen even in the Lindhard susceptibility of the square-lattice tight-binding model, as shown in Fig.~\ref{fig:peierlssus}(a).  When the normal of a domain wall is oriented parallel to these high-susceptibility directions, the energy cost associated with spatial variations of the order parameter is minimized. As a result, the domain walls preferentially align along the diagonals of the square lattice, consistent with the anisotropic domain morphology observed in our simulations.


\begin{figure}[t]
    \centering
    \includegraphics[width=0.95\linewidth]{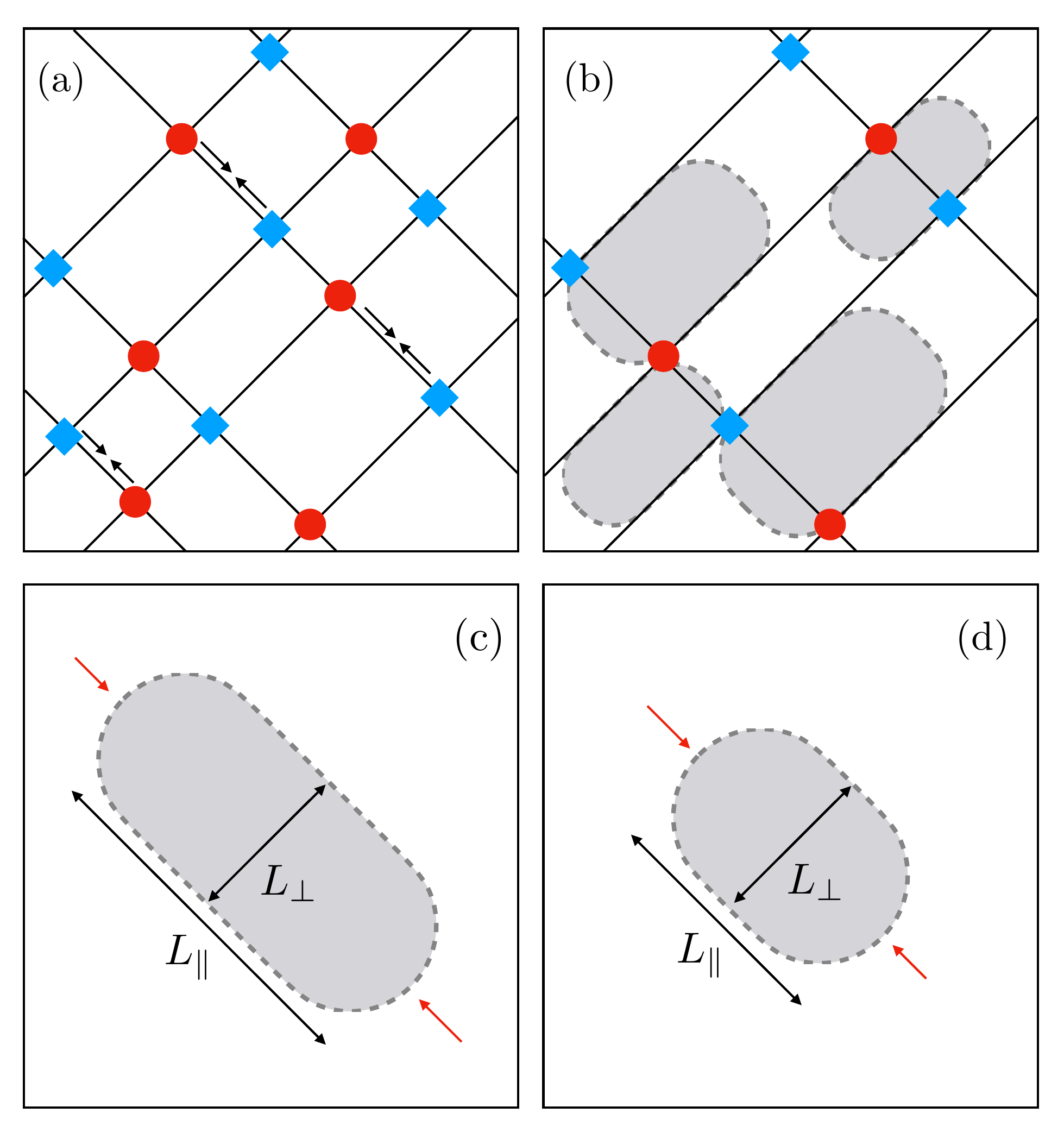}
    \caption{Schematic illustration of the domain evolution process at intermediate stages.
(a) At early times after the quench, a network of vortices interconnected by diagonally oriented domain walls forms an irregular cross-hatch pattern.
(b) Annihilation of vortex-antivortex pairs reduces the defect density and leads to the emergence of elongated CDW domains aligned along the diagonal directions.
(c,d) The subsequent coarsening is characterized by anisotropic contraction, where the long axis of each elongated domain shrinks approximately linearly in time while the short axis remains nearly constant.}
    \label{fig:domainstructure}
\end{figure}

The directional anisotropy gives rise to a dense network of elongated domain walls and an anomalous coarsening behavior during the early stages of relaxation. This process is schematically illustrated in Fig.~\ref{fig:domainstructure}. Immediately following the thermal quench, numerous topological defects (vortices and antivortices) are nucleated throughout the system. Owing to the underlying anisotropy, these initial vortices tend to organize into a network interconnected by diagonally oriented domain walls, as depicted in Fig.~\ref{fig:domainstructure}(a). The resulting morphology resembles an irregular cross-hatch pattern composed of intersecting diagonal filaments. As the system continues to evolve, the subsequent annihilation of vortex-antivortex pairs--driven by positional fluctuations and the inherent irregularity of the network--gives rise to coarsened domains that are preferentially elongated along the diagonal axes, as shown in Fig.~\ref{fig:domainstructure}(b).

The early-stage coarsening proceeds primarily through the shrinkage and eventual disappearance of these elongated domains. As illustrated in Fig.~\ref{fig:domainstructure}(c), such anisotropic domains can be characterized by two distinct length scales: $L_{\parallel}$ (along the wall) and $L_{\perp}$ (perpendicular to the wall). Because the interface curvature along the long axis is nearly zero, the motion in the perpendicular direction is strongly suppressed, i.e., $v_{\perp} = dL_{\perp}/dt \approx 0$, in accordance with the Allen–Cahn relation [Eq.~(\ref{eq:AC-eq})]. In contrast, the two short ends of an elongated domain possess a large interface curvature, $\kappa_{\rm dw} \sim 1/L_{\perp}$, which drives the contraction of the long axis. The corresponding equation of motion can be written as $dL_{\parallel}/dt \sim -1/L_{\perp}$, which remains approximately constant since $L_{\perp}$ varies little over time. Consequently, the domain length decreases linearly in time, $L_{\parallel} \sim L_0 - (c/L_{\perp}) t$, where $L_0$ is the initial domain length and $c$ is a proportionality constant. This linear-in-time contraction of elongated domains thus contribute to the enhanced apparent growth exponent observed in the ML-Langevin simulations.

It is also worth emphasizing that, although elongated domains with their long axes oriented along the diagonal directions may be energetically more favorable (depending on the aspect ratio of $L_{\parallel} / L_{\perp}$), the curvature-driven motion of domain walls--being intrinsically local--does not induce a global deformation that converts an initially isotropic (circular) domain into an anisotropic (elongated) one. Within the Allen-Cahn framework, the interface velocity is determined solely by the local curvature and thus acts to reduce interfacial area without introducing large-scale directional anisotropy. As a result, once the small elongated domains that emerge at the early stage of the relaxation process have vanished, the subsequent coarsening of the remaining, larger CDW domains proceeds through isotropic curvature-driven dynamics, recovering the conventional Allen-Cahn growth law.


\section{Conclusion and Outlook}
\label{sec:conclusion}

In this work, we investigated the coarsening dynamics of Peierls systems through large-scale ML-Langevin simulations enabled by a scalable machine-learning force-field framework. The substantial computational acceleration relative to exact diagonalization allows access to mesoscale coarsening phenomena otherwise beyond reach. Within the adiabatic approximation, the coupled dynamics of the lattice and quasi-equilibrated electrons govern the time evolution of the CDW state. To this end, we extended the Behler–Parrinello ML force-field model—originally developed for ab initio molecular dynamics—to the Peierls system. The neural network is trained to predict the lattice force field from atomic distortions, with symmetry-preserving descriptors constructed via a group-theoretical bispectrum formulation.

Training datasets were generated from exact-diagonalization calculations on $50\times50$ lattices. When integrated with Langevin dynamics, the ML force field accurately reproduces both microscopic lattice forces and the overall temporal evolution of the Peierls system. By replacing repeated electronic-structure evaluations with ML inference, the computational cost is reduced by several orders of magnitude, enabling simulations of large systems necessary for studying domain coarsening and topological defect kinetics. These large-scale simulations reveal a two-stage coarsening process, both regimes exhibiting power-law domain growth. While the late-stage evolution follows the conventional Allen-Cahn growth law, consistent with the expected $q=4$ clock-model dynamics, an enhanced growth exponent $\alpha \approx 0.7$ emerges at intermediate times. This anomalous behavior can be attributed to a pronounced directional anisotropy of the domain walls that develops during the early stage of the relaxation process. 

In real materials, the coarsening dynamics of CDWs are expected to depend sensitively on the microscopic details of the electron-lattice coupling and lattice geometry. Therefore, multi-scale modeling that integrates microscopic electronic calculations with mesoscopic dynamical simulations is crucial for capturing the realistic evolution of ordered states. However, performing such first-principles-based dynamical simulations is extremely demanding due to the necessity of resolving intricate electron-lattice interactions over large systems and long timescales. In this context, scalable ML approaches provide a powerful and efficient alternative, capable of encoding the essential electron-lattice interactions while drastically reducing computational overhead. The present ML-Langevin framework thus establishes a promising route toward quantitative, large-scale simulations of nonequilibrium dynamics in complex correlated systems, with potential extensions to phenomena such charge-spin coupled order and photoinduced CDW phase transitions.

\begin{acknowledgments}
The authors thank Sheng Zhang for useful discussions. H.J. acknowledges the support from the Owens Family Foundation. Y.Y. and G.W.C. are supported by the US Department of Energy Basic Energy Sciences under Contract No.~DE-SC0020330. The authors acknowledge Research Computing at The University of Virginia for providing computational resources and technical support.
\end{acknowledgments}

\appendix

\section{Descriptor of lattice displacements}
\label{appendix:A}
\begin{figure*}[t]
    \centering
    \includegraphics[width=0.92 \linewidth]{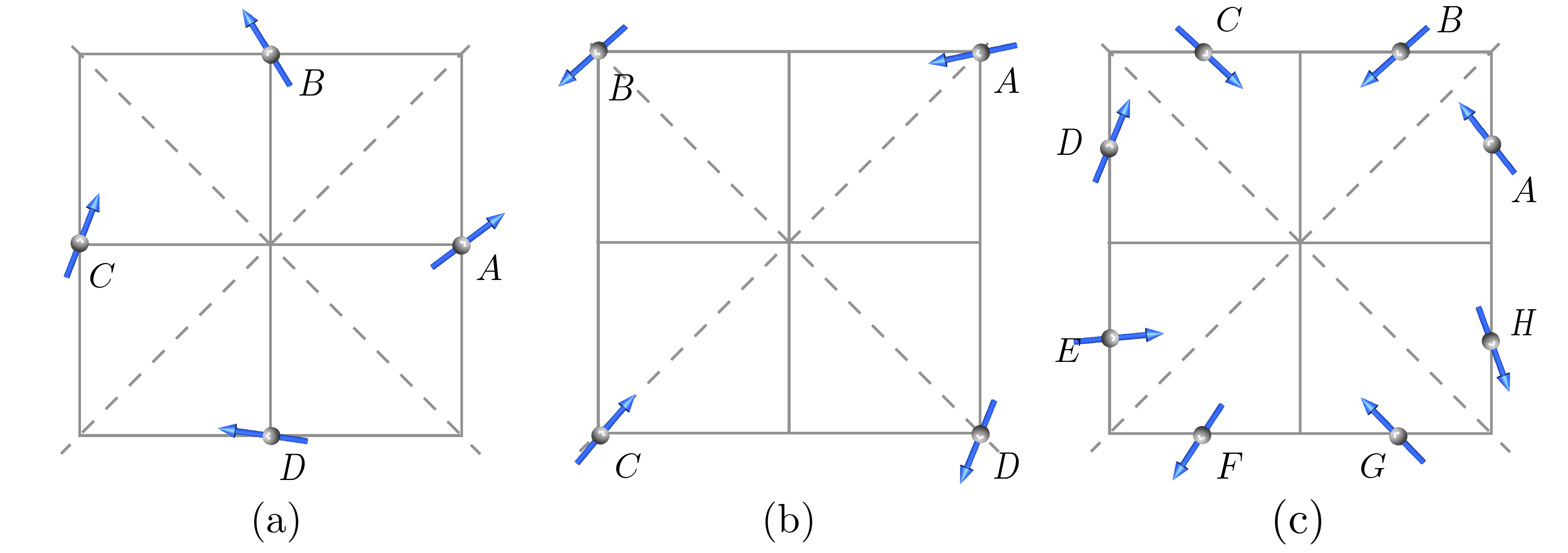}
    \caption{Schematic diagrams of the three types of neighbor blocks: 
    (a) type-I with four neighbor sites along the $x$ or $y$ axes, 
    (b) type-II with four neighbors along the $y = \pm x$ diagonals, and 
    (c) type-III with eight neighbors. The arrows represent the lattice displacement $\mathbf{u} = (u^x, u^y)$.
    }
        \label{fig:appendixA-descriptor}
\end{figure*}

Here, we present details about the calculations of IR coefficients and constructions of invariant feature variables, $\{G^{(\Gamma,r)}\}$. As introduced in Sec.~\ref{sec:MLFFmodel}, the descriptor is designed to preserve the discrete lattice symmetry of the underlying Hamiltonian, defined in terms of local displacements ${\mathbf{u}_i}$. 
Since the distance from the central site $i$ remains invariant under all $D_4$ symmetry operations, the neighborhood representation $C_i$ becomes block-diagonal, with each block determined by the distance from $i$. This neighborhood representation can be systematically decomposed into irreducible representations (IRs) \cite{hamermesh_group_1989}. These invariant blocks can be classified into three types, as illustrated in Figs.~\ref{fig:appendixA-descriptor}(a),(b), and (c). 

First, each of the type-I neighbors, labeled as $A$, $B$, $C$, and $D$ in Fig.~\ref{fig:appendixA-descriptor}(a), possesses both $x$ and $y$ components of the real lattice displacement vector. Consequently, these four neighbors collectively form an eight-dimensional representation of the $D_4$ point group. Under the symmetry operations of $D_4$, both the spatial coordinates and the corresponding displacement vectors at each site transform simultaneously. The eight-dimensional representation decomposes as    
\[8 = A_1 \oplus A_2\oplus B_1\oplus B_2 \oplus 2E,
\] with 
\begin{equation*}
\begin{split}
f^{A_1} &= u_A^x + u_B^y - u_C^x - u_D^y\\
    f^{A_2} &= u_A^y - u_B^x - u_C^y + u_D^x\\
    f^{B_1} &= u_A^x - u_B^y - u_C^x + u_D^y\\
    f^{B_2} &= u_A^y + u_B^x - u_C^y - u_D^x\\
    \bm{f}^{(E, 1)} &= (u_A^x - u_B^x + u_C^x - u_D^x,\; -u_A^y + u_B^y - u_C^y + u_D^y)\\
    \bm{f}^{(E, 2)} &= (u_A^x - u_B^x + u_C^x - u_D^x,\; -u_A^y + u_B^y - u_C^y + u_D^y).
\end{split}
\end{equation*}\\

The type-II blocks correspond to neighbors on the two diagonal directions at angles of $\pm \pi/4$, as illustrated in Fig.~\ref{fig:appendixA-descriptor}(b). Similarly, the total eight-dimensional representation decomposes as
\[8 = A_1 \oplus A_2\oplus B_1\oplus B_2 \oplus 2E,\] with 
\begin{equation*}
\begin{split}
    f^{A_1} &= u_A^x - u_A^y + u_B^x + u_B^y - u_C^x + u_C^y - u_D^x - u_D^y\\
    f^{A_2} &= u_A^x + u_A^y - u_B^x + u_B^y - u_C^x - u_C^y + u_D^x - u_D^y\\
    f^{B_1} &= u_A^x - u_A^y - u_B^x - u_B^y - u_C^x + u_C^y + u_D^x + u_D^y\\
    f^{B_2} &= u_A^x + u_A^y + u_B^x - u_B^y - u_C^x - u_C^y + u_D^x + u_D^y\\
    \bm{f}^{(E, 1)} &= (u_A^y - u_B^y + u_C^y - u_D^y,\; +u_A^x - u_B^x + u_C^x - u_D^x)\\
    \bm{f}^{(E, 2)} &= (u_A^y + u_B^y + u_C^y + u_D^y,\; u_A^x + u_B^x + u_C^x + u_D^x).
\end{split}
\end{equation*}\\

Finally, the type-III blocks correspond to eight neighbors, as shown in Fig.~\ref{fig:appendixA-descriptor}(c). The total sixteen-dimensional representation decomposes as 
\[16 = A_1 \oplus A_2\oplus B_1\oplus B_2 \oplus 2E,\] with 
\begin{equation*}
\begin{split}
    f^{(A_1,1)} &= u_A^y + u_B^x - u_C^x + u_D^y - u_E^y - u_F^x + u_G^x - u_H^y\\
    f^{(A_1,2)} &= u_A^x + u_B^y + u_C^y - u_D^x - u_E^x - u_F^y - u_G^y + u_H^x\\
    f^{(A_2,1)} &= u_A^y - u_B^x - u_C^x - u_D^y - u_E^y + u_F^x + u_G^x + u_H^y\\
    f^{(A_2, 2)} &= u_A^x - u_B^y + u_C^y + u_D^x - u_E^x + u_F^y - u_G^y - u_H^x\\
    f^{(B_1, 1)} &= u_A^y - u_B^x + u_C^x + u_D^y - u_E^y + u_F^x - u_G^x - u_H^y\\
    f^{(B_1, 2)} &= u_A^x - u_B^y - u_C^y - u_D^x - u_E^x + u_F^y + u_G^y + u_H^x\\
    f^{(B_2, 1)} &= u_A^y + u_B^x + u_C^x - u_D^y - u_E^y - u_F^x - u_G^x + u_H^y\\
    f^{(B_2, 2)} &= u_A^x + u_B^y - u_C^y + u_D^x - u_E^x - u_F^y + u_G^y - u_H^x\\
    \bm{f}^{(E, 1)} &= (u_B^y - u_C^y + u_F^y - u_G^y,\; u_A^x - u_D^x + u_E^x - u_H^x)\\
    \bm{f}^{(E, 2)} &= (u_A^x + u_D^x + u_E^x + u_H^x,\; u_B^y + u_C^y + u_F^y + u_G^y)\\
    \bm{f}^{(E, 3)} &= (u_A^y - u_D^y + u_E^y - u_H^y,\; u_B^x - u_C^x + u_F^x - u_G^x)\\
    \bm{f}^{(E, 4)} &= (u_B^x + u_C^x + u_F^x + u_G^x,\; u_A^y + u_D^y + u_E^y + u_H^y).\\
\end{split}
\end{equation*}
Using these basis IR functional forms, each block can be reconstructed into IRs systematically depending on its type. For instance, the block including site $i$ can be reconstructed as,
\begin{eqnarray}
    \bm f^{(\Gamma, r)} = \left(f^{(\Gamma, r)}_1, f^{(\Gamma, r)}_2, \cdots, f^{(\Gamma, r)}_{n_\Gamma} \right)
\end{eqnarray}
where $\Gamma$ is one of different IR types $A_1, A_2, B_1, B_2,$ or $E$, $r$ denotes its multiplicity, and $n_\Gamma$ is the dimension of the block that site $i$ belongs. One possible approach to construct a set of invariant variables from these IRs is through the power spectrum, 
\begin{eqnarray}
    p^{(\Gamma, r)} = \left|\bm{f}^{(\Gamma, r)} \right|^2,
\end{eqnarray}
which is obtained from the magnitudes of the IRs. However, the power spectrum exhibits spurious symmetry, as configurations that are not related by point-group operations still yield identical magnitudes. This additional symmetry arises from the invariance of the relative angle between the two IRs of the same type $\Gamma$, given by $\cos\theta_{12} = (\bm{f}_1^{(\Gamma,r)} \cdot \bm{f}_2^{(\Gamma,r)}/(|\bm{f}_1^{(\Gamma,r)}||\bm{f}_2^{(\Gamma,r)}|)$.

The bispectrum coefficients can incorporate all relevant invariants and relative phases to fully encode symmetry~\cite{bartok13,kondor07}. The coefficients are,
\begin{eqnarray}
B^{\Gamma,\Gamma_1,\Gamma_2}_{r,r_1,r_2} = \sum_{m,n,l} C_{m,n,l}^{\Gamma,\Gamma_1,\Gamma_2} f_{m}^{(\Gamma,r) *}f_{n}^{(\Gamma_1, r_1)}f_{l}^{(\Gamma_2, r_2)},
\end{eqnarray}
where $C_{m,n,l}^{\Gamma;\Gamma_1,\Gamma_2}$ are the Clebsh-Gordan coefficients of the point group to describe the different transformation properties of the three IRs. Even though the bispectrum coefficients are complete symmetry-invariant, its high redundancy limits its practicality. 

To balance completeness and efficiency, we introduce a reference frame, $\bm{f}_{\mathrm{ref}}^{(\Gamma,r)}$. The IRs in this reference frame are constructed from the averaged displacements over eight angular partitions of the lattice. Each partition corresponds to a sector spanning an angle of $\pi/4$, containing all sites whose positions fall within that region. The averaged displacement in partition $K$ is defined as 
$\bar{\mathbf{u}}_K = \frac{1}{M} \sum_{j\in K} \mathbf{u}_j$, 
where M is the number of sites in the partition. Using the average displacements in eight partitions, $\bm{f}_{\mathrm{ref}}^{(\Gamma,r)}$ is formulated in the same manner as the type-III blocks.
The implementation based on $\bm{f}_{\mathrm{ref}}^{(\Gamma,r)}$ reduces sensitivity to small variations in the local environment~\cite{puhan22}. The final symmetry-invariant variables with phase information are then defined as $\eta^{(\Gamma,r)} = (\bm{f}^{(\Gamma,r)} \cdot \bm{f}_{\mathrm{ref}}^{(\Gamma,r)})/(|\bm{f}^{(\Gamma,r)}||\bm{f}^{(\Gamma,r)}_{\mathrm{ref}}|)$, and the full set of $\{\eta^{(\Gamma,r)}\}$ serve as direct input to the neural network model.
\section{Neural network architecture and Loss function}
\label{appendix:B}

In this study, the neural network model is a fully connected architecture implemented in PyTorch \cite{paszke19,Nair10,barron2017,Paszke2017,kingma2017}. The input dimension is determined by the cutoff radius $r_c$, and the network consists of seven layers with sizes $322 \times 4096 \times 2048 \times 1024 \times 256 \times 128 \times 1$. The output layer is a single neuron that predicts the local energy. The loss function is the mean-square error (MSE) of the total force, readily evaluated using PyTorch’s automatic differentiation,
\begin{equation}
L = \frac{1}{N}\sum_{i=1}^N \left| \mathbf F_i^{\mathrm{ML}} - \mathbf F_i^{\mathrm{exact}}\right|^2 ,
\end{equation}
with a batch size of 16. Training was performed using 2000 snapshots ranging from random initial states to final configurations, with corresponding forces as labels. The model was optimized over 2000 epochs using the Adam algorithm with a learning rate of 0.0001.

\section{Dimensional analysis}
\label{appendix:C}

Here, we perform a dimensional analysis of the SSH model and derive a dimensionless form of both the electronic Hamiltonian and the lattice equation of motion, which serve as the foundation for our numerical simulations. From the classical elastic energy [Eq.~(\ref{eqn:classicalHamiltonian})], the natural oscillator frequency associated with the on-site lattice displacement is given by
\begin{eqnarray}
\omega_0 = \sqrt{\frac{k}{m}}.
\end{eqnarray}
The corresponding fundamental period, $t_0 = 2\pi / \omega_0$, defines the characteristic timescale of the lattice dynamics and is adopted as the time unit in our simulations. A dimensionless time is then defined as
\begin{eqnarray}
    \tilde{t} = t/t_0 = \omega_0 t.
\end{eqnarray}
Next, we identify a characteristic length scale $u_0$ for the lattice displacements. A finite distortion is energetically favored through the Peierls coupling term $-g B u_0$, where $B$ denotes the average bond correlator $\langle c_i^\dagger c_j^{\phantom{\dagger}} \rangle$. This distortion is counteracted by the elastic restoring energy $k u_0^2 / 2$. Assuming the bond correlator is of order unity, $B \sim 1$, the balance between these two competing energy scales yields the characteristic displacement amplitude
\begin{eqnarray}
u_0 = \frac{g}{k}.
\end{eqnarray}
The characteristic momentum scale $p_0$ can be obtained from the definition of momentum, $du_0/dt = p_0/m$, which gives the relation $\omega_0 u_0 \sim p_0 / m$. Accordingly, we define
\begin{eqnarray}
p_0 = m \omega_0 u_0 = \frac{m \omega_0 g}{k}.
\end{eqnarray}
Using the above characteristic scales, we introduce the dimensionless displacement and momentum variables, 
\begin{eqnarray}
\tilde{\mathbf{u}}_i = \frac{\mathbf{u}_i}{u_0}, \quad\tilde{\mathbf{p}}_i = \frac{\mathbf{p}_i}{p_0},
\end{eqnarray}
to derive the dimensionless lattice equation of motion.
A characteristic electronic energy scale is naturally provided by the nearest-neighbor hopping $t_{\rm nn}$. In terms of dimensionless displacements, the electron Hamiltonian can be expressed as 
\begin{eqnarray}
\frac{\hat{\mathcal{H}}_e}{t_{\mathrm{nn}}}
&=& 
\sum_{\langle ij \rangle}
\!\left[
-1 + \lambda\,
\hat{\mathbf{n}}_{ij} \!\cdot\!
(\tilde{\mathbf{u}}_j - \tilde{\mathbf{u}}_i)
\right]
(\hat{c}_i^{\dagger} \hat{c}_j + \mathrm{h.c.})
\nonumber \\[4pt]
&&
+\, \tilde{t}_{\mathrm{nnn}}
\sum_{\langle\!\langle ij \rangle\!\rangle}
(\hat{c}_i^{\dagger} \hat{c}_j + \mathrm{h.c.})
\label{eqn:He_tilde}
\end{eqnarray}
where the electron-lattice coupling is now characterized by a dimensionless parameter
\begin{eqnarray}
    \lambda = \frac{g^2}{k t_{\mathrm{nn}}}.    
\end{eqnarray}
The ratio $\tilde{t}_{\mathrm{nnn}} = t_{\mathrm{nnn}} / t_{\mathrm{nn}}$ describes the amplitude of next-nearest-neighbor hopping relative to $t_{\rm nn}$. The dynamical equation for the displacement vectors can now be cast in a dimensionless form:
\begin{eqnarray}
    & & \frac{d^2 \tilde{\mathbf{u}}_i}{d\tilde{t}^2} = 
    \sum_{j\in \mathcal{N}(i)} \hat{\mathbf n}_{ij} \left(\langle c^\dagger_i c^{\,}_j \rangle + \mbox{h.c.}\right) \nonumber \\
    & & \quad \qquad - \tilde{\mathbf{u}}_i - \tilde{\kappa} \sum_{j \in \mathcal{N}(i)} \tilde{\mathbf{u}}_j - \tilde{\gamma}
\frac{d \tilde{\mathbf{u}}_i}{d\tilde{t}}
+ \tilde{\boldsymbol{\eta}}_i(t),
\end{eqnarray}
where $\tilde{\kappa} = \kappa / k$ denotes the dimensionless quadratic lattice coupling,
$\tilde{\gamma} = \gamma / (m \omega_0)$ is the dimensionless damping constant, and
$\tilde{\boldsymbol{\eta}}$ represents the normalized dimensionless thermal noise.
The latter is modeled as Gaussian white noise satisfying
\begin{eqnarray}
\langle \tilde{\boldsymbol{\eta}}_i(t) \rangle 
&=& 0, \nonumber\\[4pt]
\langle \tilde{\eta}_{i a}(t)\, \tilde{\eta}_{j b}(t') \rangle 
&=& \tilde{D}\,
\delta_{ij}\, \delta_{ab}\, \delta(t - t'),
\end{eqnarray}
where $\tilde{D} = 2 \gamma m k_{\mathrm{B}} T / g^2$ is the dimensionless diffusion constant.
The system is therefore fully specified by five dimensionless parameters:
$\lambda$, $\tilde{t}_{\mathrm{nnn}}$, $\tilde{\kappa}$, $\tilde{\gamma}$, and $\tilde{D}$. In the simulations, these parameters are set to
$\lambda \approx 0.21$, $\tilde{t}_{\mathrm{nnn}} = 0.2$, $\tilde{\kappa} \approx 0.13$,
$\tilde{\gamma} \approx 0.18$, and $\tilde{D} \approx 10^{-4}$.



\section{Susceptibility of Peierls instability}
\label{appendix:D}

Here, we evaluate the electronic susceptibility of a square-lattice tight-binding model in response to a Peierls lattice distortion within the framework of standard many-body perturbation theory. We begin with the conventional tight-binding Hamiltonian that includes both nearest-neighbor (NN) and next-nearest-neighbor (NNN) hopping processes, which serves as the unperturbed Hamiltonian
\begin{eqnarray}
\hat{\mathcal{H}}_0
= -t_{\mathrm{nn}}\sum_{\langle ij\rangle}(\hat{c}_i^\dagger\hat{c}_j+\mathrm{h.c.})
- t_{\mathrm{nnn}}\sum_{\langle\!\langle ij\rangle\!\rangle}(\hat{c}_i^\dagger\hat{c}_j+\mathrm{h.c.}) . \qquad
\end{eqnarray}
This Hamiltonian can be readily diagonalized using Fourier transform
\[
\hat{c}_{i}^\dagger = \frac{1}{\sqrt{N}}\sum_{\mathbf{k}} e^{i\mathbf{k}\cdot \mathbf{r}_i} \hat{c}_{\mathbf{k}}^\dagger.
\]  
The Hamiltonian becomes $\hat{\mathcal{H}}_0 = \sum_{\mathbf{k}} \epsilon(\mathbf k) \hat{c}_{\mathbf{k}}^\dagger \hat{c}_{\mathbf{k}}$, where the single-particle energy is 
\begin{eqnarray}
\epsilon(\mathbf{k}) =  -2t_{\rm nn}(\cos k_x + \cos k_y) - 4t_{\rm nnn} \cos k_x \cos k_y , \qquad\,\,
\end{eqnarray}  
where $\mathbf{k}$ lies in the first Brillouin zone.  The resultant unperturbed many-body ground state is given by a Slater determinant of filled quasiparticles up to the Fermi level:
\begin{eqnarray}\label{eqn:appendixD_groundstate}
    |\Psi_0\rangle = \prod_{\epsilon(\mathbf k) \le \epsilon_F} \hat{c}^\dagger_{\mathbf k} |{\rm vacuum}\rangle.
\end{eqnarray}

Next the electron-lattice coupling is treated as a small perturbation. For convenience, we separate it to two parts associated with the $x$ and $y$ components of the displacement vectors
\begin{eqnarray}
\begin{split}
    \hat{\mathcal{H}}_{1} &= \hat{\mathcal{H}}_{1}^x + \hat{\mathcal{H}}_{1}^y,\\
    \hat{\mathcal{H}}_{1}^x &= g\sum_{\langle ij\rangle} \left(u_{j}^x-u_i^x\right) (\hat{c}_{i}^\dagger \hat{c}^{\,}_{j} + \mathrm{h.c.}),\\
    \hat{\mathcal{H}}_{1}^y &= g \sum_{\langle ij\rangle} \left(u_{j}^y-u_i^y\right) (\hat{c}_{i}^\dagger \hat{c}^{\,}_{j} + \mathrm{h.c.}),
\end{split}
\end{eqnarray}
Introducing the Fourier transform of the displacement field:
\begin{equation}
    \mathbf{u}(\mathbf{q}) = \frac{1}{\sqrt{N}} \sum_j \mathbf{u}_j e^{i \mathbf{q} \cdot \mathbf{r}_j},
\end{equation}
the perturbation Hamiltonian can be expressed as
\[
\begin{split}
    \hat{\mathcal{H}}_1^x &= \frac{2ig}{\sqrt{N}} \sum_{\mathbf{k}_1,\mathbf{k}_2} 
    \hat{c}_{\mathbf{k}_2 - \mathbf{k}_1}^\dagger \hat{c}^{\,}_{\mathbf{k}_2} 
    \left[\sin(k_1^x - k_2^x) + \sin(k_2^x)\right] u_{\mathbf{k}_1}^x,\\
    \hat{\mathcal{H}}_1^y &= \frac{2ig}{\sqrt{N}} \sum_{\mathbf{k}_1,\mathbf{k}_2} 
    \hat{c}_{\mathbf{k}_2 - \mathbf{k}_1}^\dagger \hat{c}^{\,}_{\mathbf{k}_2} 
    \left[\sin(k_1^y - k_2^y) + \sin(k_2^y)\right] u_{\mathbf{k}_1}^y.
\end{split}
\]  
It is straightforward to see that the first-order energy correction vanishes $\Delta E^{(1)} = \langle \Psi_0 | \hat{\mathcal{H}}_1 | \Psi_0 \rangle = 0 $. The second-order energy correction, which serves as effective energy due to electron-lattice coupling Eq.~(\ref{eq:effect_EL}) in the main text, is given by  
\begin{eqnarray}
	\label{eq:E_eff1}
    \mathcal{E}_e = 
    \sum_{m > 0} \frac{
        \langle \Psi_0|\hat{\mathcal{H}}_1|\Psi_m\rangle
        \langle \Psi_m|\hat{\mathcal{H}}_1|\Psi_0\rangle
    }{E_{0} - E_m},
\end{eqnarray}  
where $|\Psi_m\rangle$ with $m > 1$ denotes the many-body excited states of $\hat{\mathcal{H}}_1$. For a perturbation Hamiltonian $\hat{\mathcal{H}}_1$ with biquadratic fermion operators, the relevant excited states correspond to electron-hole pairs above the filled Fermi sea. This gives the following matrix elements
\[
\begin{aligned}
\langle \Psi_m|
\hat{c}_{\mathbf{q}_1}^\dagger \hat{c}_{\mathbf{q}_2}
|\Psi_0\rangle
&= f_{\rm FD}(\epsilon_f - \epsilon_{\mathbf{q}_2})\, f_{\rm FD}(\epsilon_{\mathbf{q}_1} - \epsilon_f) \\
&- f_{\rm FD}(\epsilon_f - \epsilon_{\mathbf{q}_1})\, f_{\rm FD}(\epsilon_{\mathbf{q}_2} - \epsilon_f),
\end{aligned}
\]
where $f_{\rm FD}(x) = [1+\exp(x/T)]^{-1}$ is the Fermi function. Substituting these matrix elements into Eq.~(\ref{eq:E_eff1}), the electron energy can then be conveniently expressed as
\begin{eqnarray}
    \mathcal{E}_e  
    = -4g^2 \sum_{\mathbf{q}}\sum_{a,b=x,y} \chi_{ab}^P(\mathbf{q}) {u}^*_{a}(\mathbf{q})u^{\,}_b(\mathbf q),
\end{eqnarray}
where the susceptibility is 
\begin{equation}
\chi_{ab}^{\mathrm{P}}(\mathbf{q}) 
= -\frac{1}{N} \sum_{\mathbf{k}} 
\frac{f_{\rm FD}(\epsilon_{\mathbf{k}+\mathbf{q}}) - f_{\rm FD}(\epsilon_{\mathbf{k}})}
     {\epsilon_{\mathbf{k}+\mathbf{q}} - \epsilon_{\mathbf{k}}}
\, g_{ab}(\mathbf{k}, \mathbf{q}),
\end{equation}
with the $g_{ab}$ function given by  
\[
g_{ab}(\mathbf{k},\mathbf{q}) 
= \,
  \big(\sin(q_a+k_a) - \sin k_a\big)
  \big(\sin(q_b+k_b) - \sin k_b\big),
\]  
where $a,b$ are $x,y$ direction.  The effective lattice interaction energy $\mathcal{E}_e$ represents the lattice–lattice interaction mediated by electronic degrees of freedom. When $g_{ab}$ is unity and the next-nearest-neighbor hopping is absent, $\chi^{P}_{ ab}(\mathbf{q})$ reduces to the standard Lindhard susceptibility on a square lattice, which diverges at the nesting vector $\mathbf Q = (\pi,\pi)$. Introducing a finite next-nearest-neighbor hopping reduces the overall magnitude of the Lindhard susceptibility, as shown in Fig.~\ref{fig:peierlssus}(a).

The elastic energy term also contributes to the overall lattice-lattice interaction energy. This interaction consists of two components: (i) the harmonic oscillation of individual lattice sites and (ii) the quadratic coupling between neighboring sites. After performing a Fourier transform of the displacement field, the elastic energy takes the form  
\begin{equation}\label{eqn:phononhamiltonian}
\mathcal{V}_L  = \frac{1}{2}\sum_{\mathbf{q}} \left[k + 2\kappa (\cos q_x + \cos q_y)\right] |\mathbf{u}({\mathbf{q}})|^2 .
\end{equation}
The first term originates from the harmonic restoring energy and contributes uniformly across all $\mathbf{q}$ values within the first Brillouin zone. In contrast, the second term arising from the quadratic lattice coupling minimizes the energy at $(q_x, q_y) = (\pi, \pi)$. As a result, the elastic energy favors a displacement field $\mathbf{u}$ with $(\pi, \pi)$ modulation, thereby stabilizing the $Z_4$ CDW ground states.

The anisotropy of the electronic lattice susceptibility in momentum space determines the preferred direction of domain walls (DWs). The contribution of DW states to the energy can be expressed as
\begin{equation}
\begin{split}
    E_{\mathrm{DW}} = -4g^2 \big(\chi^P(\mathbf Q + d\mathbf{q})|\mathbf{u}(\mathbf Q + d\mathbf{q})|^2 \\+ \chi^P(\mathbf Q - d\mathbf{q})|\mathbf{u}(\mathbf Q - d\mathbf{q})|^2\big).
\end{split}
\end{equation}
Expanding for small $d\mathbf{q}$ makes
\begin{equation}\label{eqn:FDW}
    E_{\mathrm{DW}} = -4g^2 \bigg( \chi^P(\mathbf Q) + \frac{1}{2}\frac{d^2 \chi^P(\mathbf{q})}{d \mathbf{q}^2} \bigg|_{\mathbf{q}=\mathbf Q} d\mathbf{q}^2 + \mathit{O}(d\mathbf{q}^3)\bigg).
\end{equation}
The first term stabilizes the CDW ground state without DWs at $\mathbf{q}=\mathbf{Q}$. Since creating the DWs always requires an energy cost, the second derivative of susceptibility, $d^2\chi^P/d\mathbf{q}^2$, at the peak $\mathbf Q$ is always negative. The additional energy cost is minimized when $d\mathbf{q}$ is chosen along the direction where $|d^2\chi^P/d\mathbf{q}^2|$ is smallest, i.e., along the direction where the peak of $\chi^P$ decays most slowly. This means DW has a preferred direction along the peak of $\chi^P$ decays most slowly. 



\bibliography{ref}

\begin{thebibliography}{104}%
\makeatletter
\providecommand \@ifxundefined [1]{%
 \@ifx{#1\undefined}
}%
\providecommand \@ifnum [1]{%
 \ifnum #1\expandafter \@firstoftwo
 \else \expandafter \@secondoftwo
 \fi
}%
\providecommand \@ifx [1]{%
 \ifx #1\expandafter \@firstoftwo
 \else \expandafter \@secondoftwo
 \fi
}%
\providecommand \natexlab [1]{#1}%
\providecommand \enquote  [1]{``#1''}%
\providecommand \bibnamefont  [1]{#1}%
\providecommand \bibfnamefont [1]{#1}%
\providecommand \citenamefont [1]{#1}%
\providecommand \href@noop [0]{\@secondoftwo}%
\providecommand \href [0]{\begingroup \@sanitize@url \@href}%
\providecommand \@href[1]{\@@startlink{#1}\@@href}%
\providecommand \@@href[1]{\endgroup#1\@@endlink}%
\providecommand \@sanitize@url [0]{\catcode `\\12\catcode `\$12\catcode
  `\&12\catcode `\#12\catcode `\^12\catcode `\_12\catcode `\%12\relax}%
\providecommand \@@startlink[1]{}%
\providecommand \@@endlink[0]{}%
\providecommand \url  [0]{\begingroup\@sanitize@url \@url }%
\providecommand \@url [1]{\endgroup\@href {#1}{\urlprefix }}%
\providecommand \urlprefix  [0]{URL }%
\providecommand \Eprint [0]{\href }%
\providecommand \doibase [0]{https://doi.org/}%
\providecommand \selectlanguage [0]{\@gobble}%
\providecommand \bibinfo  [0]{\@secondoftwo}%
\providecommand \bibfield  [0]{\@secondoftwo}%
\providecommand \translation [1]{[#1]}%
\providecommand \BibitemOpen [0]{}%
\providecommand \bibitemStop [0]{}%
\providecommand \bibitemNoStop [0]{.\EOS\space}%
\providecommand \EOS [0]{\spacefactor3000\relax}%
\providecommand \BibitemShut  [1]{\csname bibitem#1\endcsname}%
\let\auto@bib@innerbib\@empty
\bibitem [{\citenamefont {{Tranquada}}\ \emph {et~al.}(1995)\citenamefont
  {{Tranquada}}, \citenamefont {{Sternlieb}}, \citenamefont {{Axe}},
  \citenamefont {{Nakamura}},\ and\ \citenamefont {{Uchida}}}]{tranquada95}%
  \BibitemOpen
  \bibfield  {author} {\bibinfo {author} {\bibfnamefont {J.~M.}\ \bibnamefont
  {{Tranquada}}}, \bibinfo {author} {\bibfnamefont {B.~J.}\ \bibnamefont
  {{Sternlieb}}}, \bibinfo {author} {\bibfnamefont {J.~D.}\ \bibnamefont
  {{Axe}}}, \bibinfo {author} {\bibfnamefont {Y.}~\bibnamefont {{Nakamura}}},\
  and\ \bibinfo {author} {\bibfnamefont {S.}~\bibnamefont {{Uchida}}},\
  }\bibfield  {title} {\bibinfo {title} {{Evidence for stripe correlations of
  spins and holes in copper oxide superconductors}},\ }\href
  {https://doi.org/10.1038/375561a0} {\bibfield  {journal} {\bibinfo  {journal}
  {Nature}\ }\textbf {\bibinfo {volume} {375}},\ \bibinfo {pages} {561}
  (\bibinfo {year} {1995})}\BibitemShut {NoStop}%
\bibitem [{\citenamefont {Fradkin}\ \emph {et~al.}(2015)\citenamefont
  {Fradkin}, \citenamefont {Kivelson},\ and\ \citenamefont
  {Tranquada}}]{fradkin15}%
  \BibitemOpen
  \bibfield  {author} {\bibinfo {author} {\bibfnamefont {E.}~\bibnamefont
  {Fradkin}}, \bibinfo {author} {\bibfnamefont {S.~A.}\ \bibnamefont
  {Kivelson}},\ and\ \bibinfo {author} {\bibfnamefont {J.~M.}\ \bibnamefont
  {Tranquada}},\ }\bibfield  {title} {\bibinfo {title} {Colloquium: Theory of
  intertwined orders in high temperature superconductors},\ }\href
  {https://doi.org/10.1103/RevModPhys.87.457} {\bibfield  {journal} {\bibinfo
  {journal} {Rev. Mod. Phys.}\ }\textbf {\bibinfo {volume} {87}},\ \bibinfo
  {pages} {457} (\bibinfo {year} {2015})}\BibitemShut {NoStop}%
\bibitem [{\citenamefont {Kivelson}\ \emph {et~al.}(2003)\citenamefont
  {Kivelson}, \citenamefont {Bindloss}, \citenamefont {Fradkin}, \citenamefont
  {Oganesyan}, \citenamefont {Tranquada}, \citenamefont {Kapitulnik},\ and\
  \citenamefont {Howald}}]{kivelson03}%
  \BibitemOpen
  \bibfield  {author} {\bibinfo {author} {\bibfnamefont {S.~A.}\ \bibnamefont
  {Kivelson}}, \bibinfo {author} {\bibfnamefont {I.~P.}\ \bibnamefont
  {Bindloss}}, \bibinfo {author} {\bibfnamefont {E.}~\bibnamefont {Fradkin}},
  \bibinfo {author} {\bibfnamefont {V.}~\bibnamefont {Oganesyan}}, \bibinfo
  {author} {\bibfnamefont {J.~M.}\ \bibnamefont {Tranquada}}, \bibinfo {author}
  {\bibfnamefont {A.}~\bibnamefont {Kapitulnik}},\ and\ \bibinfo {author}
  {\bibfnamefont {C.}~\bibnamefont {Howald}},\ }\bibfield  {title} {\bibinfo
  {title} {How to detect fluctuating stripes in the high-temperature
  superconductors},\ }\href {https://doi.org/10.1103/RevModPhys.75.1201}
  {\bibfield  {journal} {\bibinfo  {journal} {Rev. Mod. Phys.}\ }\textbf
  {\bibinfo {volume} {75}},\ \bibinfo {pages} {1201} (\bibinfo {year}
  {2003})}\BibitemShut {NoStop}%
\bibitem [{\citenamefont {Gr\"uner}(1988)}]{gruner1988}%
  \BibitemOpen
  \bibfield  {author} {\bibinfo {author} {\bibfnamefont {G.}~\bibnamefont
  {Gr\"uner}},\ }\bibfield  {title} {\bibinfo {title} {The dynamics of
  charge-density waves},\ }\href {https://doi.org/10.1103/RevModPhys.60.1129}
  {\bibfield  {journal} {\bibinfo  {journal} {Rev. Mod. Phys.}\ }\textbf
  {\bibinfo {volume} {60}},\ \bibinfo {pages} {1129} (\bibinfo {year}
  {1988})}\BibitemShut {NoStop}%
\bibitem [{\citenamefont {Gr{\"u}ner}(1994)}]{Grner1994DensityWI}%
  \BibitemOpen
  \bibfield  {author} {\bibinfo {author} {\bibfnamefont {G.}~\bibnamefont
  {Gr{\"u}ner}},\ }\bibfield  {title} {\bibinfo {title} {Density waves in
  solids}\ }(\bibinfo {year} {1994})\BibitemShut {NoStop}%
\bibitem [{\citenamefont {Thorne}(1996)}]{thorne_charge-density-wave_1996}%
  \BibitemOpen
  \bibfield  {author} {\bibinfo {author} {\bibfnamefont {R.~E.}\ \bibnamefont
  {Thorne}},\ }\bibfield  {title} {\bibinfo {title} {Charge-{Density}-{Wave}
  {Conductors}},\ }\href {https://doi.org/10.1063/1.881498} {\bibfield
  {journal} {\bibinfo  {journal} {Physics Today}\ }\textbf {\bibinfo {volume}
  {49}},\ \bibinfo {pages} {42} (\bibinfo {year} {1996})}\BibitemShut {NoStop}%
\bibitem [{\citenamefont {{Pierre Monceau}}(2012)}]{Monceau2012}%
  \BibitemOpen
  \bibfield  {author} {\bibinfo {author} {\bibnamefont {{Pierre Monceau}}},\
  }\bibfield  {title} {\bibinfo {title} {{Electronic crystals : an experimental
  overview}},\ }\href {https://doi.org/10.1080/00018732.2012.719674} {\bibfield
   {journal} {\bibinfo  {journal} {Advances in Physics}\ }\textbf {\bibinfo
  {volume} {61}},\ \bibinfo {pages} {325} (\bibinfo {year} {2012})}\BibitemShut
  {NoStop}%
\bibitem [{\citenamefont {Anderson}\ \emph {et~al.}(1973)\citenamefont
  {Anderson}, \citenamefont {Lee},\ and\ \citenamefont
  {Saitoh}}]{anderson1973}%
  \BibitemOpen
  \bibfield  {author} {\bibinfo {author} {\bibfnamefont {P.}~\bibnamefont
  {Anderson}}, \bibinfo {author} {\bibfnamefont {P.}~\bibnamefont {Lee}},\ and\
  \bibinfo {author} {\bibfnamefont {M.}~\bibnamefont {Saitoh}},\ }\bibfield
  {title} {\bibinfo {title} {Remarks on giant conductivity in {TTF}-{TCNQ}},\
  }\href {https://doi.org/10.1016/S0038-1098(73)80020-1} {\bibfield  {journal}
  {\bibinfo  {journal} {Solid State Communications}\ }\textbf {\bibinfo
  {volume} {13}},\ \bibinfo {pages} {595} (\bibinfo {year} {1973})}\BibitemShut
  {NoStop}%
\bibitem [{\citenamefont {{Balandin, Alexander A. and Zaitsev-Zotov, Sergei V.
  and Grüner, George}}(2021)}]{balandin2021}%
  \BibitemOpen
  \bibfield  {author} {\bibinfo {author} {\bibnamefont {{Balandin, Alexander A.
  and Zaitsev-Zotov, Sergei V. and Grüner, George}}},\ }\bibfield  {title}
  {\bibinfo {title} {{Charge-density-wave quantum materials and devices—{New}
  developments and future prospects}},\ }\href
  {https://doi.org/10.1063/5.0074613} {\bibfield  {journal} {\bibinfo
  {journal} {Applied Physics Letters}\ }\textbf {\bibinfo {volume} {119}},\
  \bibinfo {pages} {170401} (\bibinfo {year} {2021})}\BibitemShut {NoStop}%
\bibitem [{\citenamefont {Porer}\ \emph {et~al.}(2014)\citenamefont {Porer},
  \citenamefont {Leierseder}, \citenamefont {Ménard}, \citenamefont
  {Dachraoui}, \citenamefont {Mouchliadis}, \citenamefont {Perakis},
  \citenamefont {Heinzmann}, \citenamefont {Demsar}, \citenamefont
  {Rossnagel},\ and\ \citenamefont {Huber}}]{porer2014}%
  \BibitemOpen
  \bibfield  {author} {\bibinfo {author} {\bibfnamefont {M.}~\bibnamefont
  {Porer}}, \bibinfo {author} {\bibfnamefont {U.}~\bibnamefont {Leierseder}},
  \bibinfo {author} {\bibfnamefont {J.-M.}\ \bibnamefont {Ménard}}, \bibinfo
  {author} {\bibfnamefont {H.}~\bibnamefont {Dachraoui}}, \bibinfo {author}
  {\bibfnamefont {L.}~\bibnamefont {Mouchliadis}}, \bibinfo {author}
  {\bibfnamefont {I.~E.}\ \bibnamefont {Perakis}}, \bibinfo {author}
  {\bibfnamefont {U.}~\bibnamefont {Heinzmann}}, \bibinfo {author}
  {\bibfnamefont {J.}~\bibnamefont {Demsar}}, \bibinfo {author} {\bibfnamefont
  {K.}~\bibnamefont {Rossnagel}},\ and\ \bibinfo {author} {\bibfnamefont
  {R.}~\bibnamefont {Huber}},\ }\bibfield  {title} {\bibinfo {title}
  {Non-thermal separation of electronic and structural orders in a persisting
  charge density wave},\ }\href {https://doi.org/10.1038/nmat4042} {\bibfield
  {journal} {\bibinfo  {journal} {Nature Materials}\ }\textbf {\bibinfo
  {volume} {13}},\ \bibinfo {pages} {857} (\bibinfo {year} {2014})}\BibitemShut
  {NoStop}%
\bibitem [{\citenamefont {{Samnakay, R. and Wickramaratne, D. and Pope, T. R.
  and Lake, R. K. and Salguero, T. T. and Balandin, A.
  A.}}(2015)}]{samnakay2015}%
  \BibitemOpen
  \bibfield  {author} {\bibinfo {author} {\bibnamefont {{Samnakay, R. and
  Wickramaratne, D. and Pope, T. R. and Lake, R. K. and Salguero, T. T. and
  Balandin, A. A.}}},\ }\bibfield  {title} {\bibinfo {title} {Zone-{Folded}
  {Phonons} and the {Commensurate}–{Incommensurate} {Charge}-{Density}-{Wave}
  {Transition} in 1 \textit{{T}} -{TaSe}$_{\textrm{2}}$ {Thin} {Films}},\
  }\href {https://doi.org/10.1021/nl504811s} {\bibfield  {journal} {\bibinfo
  {journal} {Nano Letters}\ }\textbf {\bibinfo {volume} {15}},\ \bibinfo
  {pages} {2965} (\bibinfo {year} {2015})}\BibitemShut {NoStop}%
\bibitem [{\citenamefont {Cho}\ \emph {et~al.}(2016)\citenamefont {Cho},
  \citenamefont {Cheon}, \citenamefont {Kim}, \citenamefont {Lee},
  \citenamefont {Cho}, \citenamefont {Cheong},\ and\ \citenamefont
  {Yeom}}]{cho2016}%
  \BibitemOpen
  \bibfield  {author} {\bibinfo {author} {\bibfnamefont {D.}~\bibnamefont
  {Cho}}, \bibinfo {author} {\bibfnamefont {S.}~\bibnamefont {Cheon}}, \bibinfo
  {author} {\bibfnamefont {K.-S.}\ \bibnamefont {Kim}}, \bibinfo {author}
  {\bibfnamefont {S.-H.}\ \bibnamefont {Lee}}, \bibinfo {author} {\bibfnamefont
  {Y.-H.}\ \bibnamefont {Cho}}, \bibinfo {author} {\bibfnamefont {S.-W.}\
  \bibnamefont {Cheong}},\ and\ \bibinfo {author} {\bibfnamefont {H.~W.}\
  \bibnamefont {Yeom}},\ }\bibfield  {title} {\bibinfo {title} {Nanoscale
  manipulation of the {Mott} insulating state coupled to charge order in
  {1T}-{TaS2}},\ }\href {https://doi.org/10.1038/ncomms10453} {\bibfield
  {journal} {\bibinfo  {journal} {Nature Communications}\ }\textbf {\bibinfo
  {volume} {7}},\ \bibinfo {pages} {10453} (\bibinfo {year}
  {2016})}\BibitemShut {NoStop}%
\bibitem [{\citenamefont {Chen}\ \emph {et~al.}(2015)\citenamefont {Chen},
  \citenamefont {Chan}, \citenamefont {Fang}, \citenamefont {Zhang},
  \citenamefont {Chou}, \citenamefont {Mo}, \citenamefont {Hussain},
  \citenamefont {Fedorov},\ and\ \citenamefont {Chiang}}]{chen_charge_2015}%
  \BibitemOpen
  \bibfield  {author} {\bibinfo {author} {\bibfnamefont {P.}~\bibnamefont
  {Chen}}, \bibinfo {author} {\bibfnamefont {Y.~H.}\ \bibnamefont {Chan}},
  \bibinfo {author} {\bibfnamefont {X.~Y.}\ \bibnamefont {Fang}}, \bibinfo
  {author} {\bibfnamefont {Y.}~\bibnamefont {Zhang}}, \bibinfo {author}
  {\bibfnamefont {M.~Y.}\ \bibnamefont {Chou}}, \bibinfo {author}
  {\bibfnamefont {S.~K.}\ \bibnamefont {Mo}}, \bibinfo {author} {\bibfnamefont
  {Z.}~\bibnamefont {Hussain}}, \bibinfo {author} {\bibfnamefont {A.~V.}\
  \bibnamefont {Fedorov}},\ and\ \bibinfo {author} {\bibfnamefont {T.~C.}\
  \bibnamefont {Chiang}},\ }\bibfield  {title} {\bibinfo {title} {Charge
  density wave transition in single-layer titanium diselenide},\ }\href
  {https://doi.org/10.1038/ncomms9943} {\bibfield  {journal} {\bibinfo
  {journal} {Nature Communications}\ }\textbf {\bibinfo {volume} {6}},\
  \bibinfo {pages} {8943} (\bibinfo {year} {2015})}\BibitemShut {NoStop}%
\bibitem [{\citenamefont {Zhao}\ \emph {et~al.}(2022)\citenamefont {Zhao},
  \citenamefont {Zhu}, \citenamefont {Nie}, \citenamefont {Li}, \citenamefont
  {Wang}, \citenamefont {Dou}, \citenamefont {Hu}, \citenamefont {Xian},
  \citenamefont {Meng},\ and\ \citenamefont {Li}}]{zhao_moire_2022}%
  \BibitemOpen
  \bibfield  {author} {\bibinfo {author} {\bibfnamefont {W.-M.}\ \bibnamefont
  {Zhao}}, \bibinfo {author} {\bibfnamefont {L.}~\bibnamefont {Zhu}}, \bibinfo
  {author} {\bibfnamefont {Z.}~\bibnamefont {Nie}}, \bibinfo {author}
  {\bibfnamefont {Q.-Y.}\ \bibnamefont {Li}}, \bibinfo {author} {\bibfnamefont
  {Q.-W.}\ \bibnamefont {Wang}}, \bibinfo {author} {\bibfnamefont {L.-G.}\
  \bibnamefont {Dou}}, \bibinfo {author} {\bibfnamefont {J.-G.}\ \bibnamefont
  {Hu}}, \bibinfo {author} {\bibfnamefont {L.}~\bibnamefont {Xian}}, \bibinfo
  {author} {\bibfnamefont {S.}~\bibnamefont {Meng}},\ and\ \bibinfo {author}
  {\bibfnamefont {S.-C.}\ \bibnamefont {Li}},\ }\bibfield  {title} {\bibinfo
  {title} {Moiré enhanced charge density wave state in twisted
  {1T}-{TiTe2}/{1T}-{TiSe2} heterostructures},\ }\href
  {https://doi.org/10.1038/s41563-021-01167-0} {\bibfield  {journal} {\bibinfo
  {journal} {Nature Materials}\ }\textbf {\bibinfo {volume} {21}},\ \bibinfo
  {pages} {284} (\bibinfo {year} {2022})}\BibitemShut {NoStop}%
\bibitem [{\citenamefont {Holstein}(1959)}]{holstein1959}%
  \BibitemOpen
  \bibfield  {author} {\bibinfo {author} {\bibfnamefont {T.}~\bibnamefont
  {Holstein}},\ }\bibfield  {title} {\bibinfo {title} {Studies of polaron
  motion: {Part} {I}. {The} molecular-crystal model},\ }\href
  {https://doi.org/https://doi.org/10.1016/0003-4916(59)90002-8} {\bibfield
  {journal} {\bibinfo  {journal} {Annals of Physics}\ }\textbf {\bibinfo
  {volume} {8}},\ \bibinfo {pages} {325} (\bibinfo {year} {1959})}\BibitemShut
  {NoStop}%
\bibitem [{\citenamefont {Scalettar}\ \emph {et~al.}(1989)\citenamefont
  {Scalettar}, \citenamefont {Bickers},\ and\ \citenamefont
  {Scalapino}}]{scalettar1989}%
  \BibitemOpen
  \bibfield  {author} {\bibinfo {author} {\bibfnamefont {R.~T.}\ \bibnamefont
  {Scalettar}}, \bibinfo {author} {\bibfnamefont {N.~E.}\ \bibnamefont
  {Bickers}},\ and\ \bibinfo {author} {\bibfnamefont {D.~J.}\ \bibnamefont
  {Scalapino}},\ }\bibfield  {title} {\bibinfo {title} {Scalettar1989},\ }\href
  {https://doi.org/10.1103/PhysRevB.40.197} {\bibfield  {journal} {\bibinfo
  {journal} {Phys. Rev. B}\ }\textbf {\bibinfo {volume} {40}},\ \bibinfo
  {pages} {197} (\bibinfo {year} {1989})}\BibitemShut {NoStop}%
\bibitem [{\citenamefont {Noack}\ \emph {et~al.}(1991)\citenamefont {Noack},
  \citenamefont {Scalapino},\ and\ \citenamefont {Scalettar}}]{noack91}%
  \BibitemOpen
  \bibfield  {author} {\bibinfo {author} {\bibfnamefont {R.~M.}\ \bibnamefont
  {Noack}}, \bibinfo {author} {\bibfnamefont {D.~J.}\ \bibnamefont
  {Scalapino}},\ and\ \bibinfo {author} {\bibfnamefont {R.~T.}\ \bibnamefont
  {Scalettar}},\ }\bibfield  {title} {\bibinfo {title} {Charge-density-wave and
  pairing susceptibilities in a two-dimensional electron-phonon model},\ }\href
  {https://doi.org/10.1103/PhysRevLett.66.778} {\bibfield  {journal} {\bibinfo
  {journal} {Phys. Rev. Lett.}\ }\textbf {\bibinfo {volume} {66}},\ \bibinfo
  {pages} {778} (\bibinfo {year} {1991})}\BibitemShut {NoStop}%
\bibitem [{\citenamefont {Loh}\ \emph {et~al.}(1990)\citenamefont {Loh},
  \citenamefont {Gubernatis}, \citenamefont {Scalettar}, \citenamefont {White},
  \citenamefont {Scalapino},\ and\ \citenamefont {Sugar}}]{EYLoh1990}%
  \BibitemOpen
  \bibfield  {author} {\bibinfo {author} {\bibfnamefont {E.~Y.}\ \bibnamefont
  {Loh}}, \bibinfo {author} {\bibfnamefont {J.~E.}\ \bibnamefont {Gubernatis}},
  \bibinfo {author} {\bibfnamefont {R.~T.}\ \bibnamefont {Scalettar}}, \bibinfo
  {author} {\bibfnamefont {S.~R.}\ \bibnamefont {White}}, \bibinfo {author}
  {\bibfnamefont {D.~J.}\ \bibnamefont {Scalapino}},\ and\ \bibinfo {author}
  {\bibfnamefont {R.~L.}\ \bibnamefont {Sugar}},\ }\bibfield  {title} {\bibinfo
  {title} {Sign problem in the numerical simulation of many-electron systems},\
  }\href {https://doi.org/10.1103/PhysRevB.41.9301} {\bibfield  {journal}
  {\bibinfo  {journal} {Phys. Rev. B}\ }\textbf {\bibinfo {volume} {41}},\
  \bibinfo {pages} {9301} (\bibinfo {year} {1990})}\BibitemShut {NoStop}%
\bibitem [{\citenamefont {Troyer}\ and\ \citenamefont
  {Wiese}(2005)}]{Troyer2005}%
  \BibitemOpen
  \bibfield  {author} {\bibinfo {author} {\bibfnamefont {M.}~\bibnamefont
  {Troyer}}\ and\ \bibinfo {author} {\bibfnamefont {U.-J.}\ \bibnamefont
  {Wiese}},\ }\bibfield  {title} {\bibinfo {title} {Computational complexity
  and fundamental limitations to fermionic quantum monte carlo simulations},\
  }\href {https://doi.org/10.1103/PhysRevLett.94.170201} {\bibfield  {journal}
  {\bibinfo  {journal} {Phys. Rev. Lett.}\ }\textbf {\bibinfo {volume} {94}},\
  \bibinfo {pages} {170201} (\bibinfo {year} {2005})}\BibitemShut {NoStop}%
\bibitem [{\citenamefont {Peierls}(1955)}]{peierls1955}%
  \BibitemOpen
  \bibfield  {author} {\bibinfo {author} {\bibfnamefont {R.~E.}\ \bibnamefont
  {Peierls}},\ }\href@noop {} {\emph {\bibinfo {title} {Quantum Theory of
  Solids}}}\ (\bibinfo  {publisher} {Oxford University Press},\ \bibinfo
  {address} {Oxford},\ \bibinfo {year} {1955})\BibitemShut {NoStop}%
\bibitem [{\citenamefont {Su}\ \emph {et~al.}(1980)\citenamefont {Su},
  \citenamefont {Schrieffer},\ and\ \citenamefont {Heeger}}]{SSH}%
  \BibitemOpen
  \bibfield  {author} {\bibinfo {author} {\bibfnamefont {W.~P.}\ \bibnamefont
  {Su}}, \bibinfo {author} {\bibfnamefont {J.~R.}\ \bibnamefont {Schrieffer}},\
  and\ \bibinfo {author} {\bibfnamefont {A.~J.}\ \bibnamefont {Heeger}},\
  }\bibfield  {title} {\bibinfo {title} {Soliton excitations in
  polyacetylene},\ }\href {https://doi.org/10.1103/PhysRevB.22.2099} {\bibfield
   {journal} {\bibinfo  {journal} {Phys. Rev. B}\ }\textbf {\bibinfo {volume}
  {22}},\ \bibinfo {pages} {2099} (\bibinfo {year} {1980})}\BibitemShut
  {NoStop}%
\bibitem [{\citenamefont {Yam}\ \emph {et~al.}(2020)\citenamefont {Yam},
  \citenamefont {Moeller}, \citenamefont {Sawatzky},\ and\ \citenamefont
  {Berciu}}]{Yam2020}%
  \BibitemOpen
  \bibfield  {author} {\bibinfo {author} {\bibfnamefont {Y.-C.}\ \bibnamefont
  {Yam}}, \bibinfo {author} {\bibfnamefont {M.~M.}\ \bibnamefont {Moeller}},
  \bibinfo {author} {\bibfnamefont {G.~A.}\ \bibnamefont {Sawatzky}},\ and\
  \bibinfo {author} {\bibfnamefont {M.}~\bibnamefont {Berciu}},\ }\bibfield
  {title} {\bibinfo {title} {Peierls versus holstein models for describing
  electron-phonon coupling in perovskites},\ }\href
  {https://doi.org/10.1103/PhysRevB.102.235145} {\bibfield  {journal} {\bibinfo
   {journal} {Phys. Rev. B}\ }\textbf {\bibinfo {volume} {102}},\ \bibinfo
  {pages} {235145} (\bibinfo {year} {2020})}\BibitemShut {NoStop}%
\bibitem [{\citenamefont {Xing}\ \emph {et~al.}(2021)\citenamefont {Xing},
  \citenamefont {Chiu}, \citenamefont {Poletti}, \citenamefont {Scalettar},\
  and\ \citenamefont {Batrouni}}]{Xing2021}%
  \BibitemOpen
  \bibfield  {author} {\bibinfo {author} {\bibfnamefont {B.}~\bibnamefont
  {Xing}}, \bibinfo {author} {\bibfnamefont {W.-T.}\ \bibnamefont {Chiu}},
  \bibinfo {author} {\bibfnamefont {D.}~\bibnamefont {Poletti}}, \bibinfo
  {author} {\bibfnamefont {R.~T.}\ \bibnamefont {Scalettar}},\ and\ \bibinfo
  {author} {\bibfnamefont {G.}~\bibnamefont {Batrouni}},\ }\bibfield  {title}
  {\bibinfo {title} {Quantum monte carlo simulations of the 2d
  su-schrieffer-heeger model},\ }\href
  {https://doi.org/10.1103/PhysRevLett.126.017601} {\bibfield  {journal}
  {\bibinfo  {journal} {Phys. Rev. Lett.}\ }\textbf {\bibinfo {volume} {126}},\
  \bibinfo {pages} {017601} (\bibinfo {year} {2021})}\BibitemShut {NoStop}%
\bibitem [{\citenamefont {Tsen}\ \emph {et~al.}(2015)\citenamefont {Tsen},
  \citenamefont {Hovden}, \citenamefont {Wang}, \citenamefont {Kim},
  \citenamefont {Okamoto}, \citenamefont {Spoth}, \citenamefont {Liu},
  \citenamefont {Lu}, \citenamefont {Sun}, \citenamefont {Hone}, \citenamefont
  {Kourkoutis}, \citenamefont {Kim},\ and\ \citenamefont
  {Pasupathy}}]{tsen2015}%
  \BibitemOpen
  \bibfield  {author} {\bibinfo {author} {\bibfnamefont {A.~W.}\ \bibnamefont
  {Tsen}}, \bibinfo {author} {\bibfnamefont {R.}~\bibnamefont {Hovden}},
  \bibinfo {author} {\bibfnamefont {D.}~\bibnamefont {Wang}}, \bibinfo {author}
  {\bibfnamefont {Y.~D.}\ \bibnamefont {Kim}}, \bibinfo {author} {\bibfnamefont
  {J.}~\bibnamefont {Okamoto}}, \bibinfo {author} {\bibfnamefont {K.~A.}\
  \bibnamefont {Spoth}}, \bibinfo {author} {\bibfnamefont {Y.}~\bibnamefont
  {Liu}}, \bibinfo {author} {\bibfnamefont {W.}~\bibnamefont {Lu}}, \bibinfo
  {author} {\bibfnamefont {Y.}~\bibnamefont {Sun}}, \bibinfo {author}
  {\bibfnamefont {J.~C.}\ \bibnamefont {Hone}}, \bibinfo {author}
  {\bibfnamefont {L.~F.}\ \bibnamefont {Kourkoutis}}, \bibinfo {author}
  {\bibfnamefont {P.}~\bibnamefont {Kim}},\ and\ \bibinfo {author}
  {\bibfnamefont {A.~N.}\ \bibnamefont {Pasupathy}},\ }\bibfield  {title}
  {\bibinfo {title} {Structure and control of charge density waves in
  two-dimensional {1T}-{TaS}$_{\textrm{2}}$},\ }\href
  {https://doi.org/10.1073/pnas.1512092112} {\bibfield  {journal} {\bibinfo
  {journal} {Proceedings of the National Academy of Sciences}\ }\textbf
  {\bibinfo {volume} {112}},\ \bibinfo {pages} {15054} (\bibinfo {year}
  {2015})}\BibitemShut {NoStop}%
\bibitem [{\citenamefont {Behler}\ and\ \citenamefont
  {Parrinello}(2007)}]{behler07}%
  \BibitemOpen
  \bibfield  {author} {\bibinfo {author} {\bibfnamefont {J.}~\bibnamefont
  {Behler}}\ and\ \bibinfo {author} {\bibfnamefont {M.}~\bibnamefont
  {Parrinello}},\ }\bibfield  {title} {\bibinfo {title} {Generalized
  neural-network representation of high-dimensional potential-energy
  surfaces},\ }\href {https://doi.org/10.1103/PhysRevLett.98.146401} {\bibfield
   {journal} {\bibinfo  {journal} {Phys. Rev. Lett.}\ }\textbf {\bibinfo
  {volume} {98}},\ \bibinfo {pages} {146401} (\bibinfo {year}
  {2007})}\BibitemShut {NoStop}%
\bibitem [{\citenamefont {Bart\'ok}\ \emph {et~al.}(2010)\citenamefont
  {Bart\'ok}, \citenamefont {Payne}, \citenamefont {Kondor},\ and\
  \citenamefont {Cs\'anyi}}]{bartok10}%
  \BibitemOpen
  \bibfield  {author} {\bibinfo {author} {\bibfnamefont {A.~P.}\ \bibnamefont
  {Bart\'ok}}, \bibinfo {author} {\bibfnamefont {M.~C.}\ \bibnamefont {Payne}},
  \bibinfo {author} {\bibfnamefont {R.}~\bibnamefont {Kondor}},\ and\ \bibinfo
  {author} {\bibfnamefont {G.}~\bibnamefont {Cs\'anyi}},\ }\bibfield  {title}
  {\bibinfo {title} {Gaussian approximation potentials: The accuracy of quantum
  mechanics, without the electrons},\ }\href
  {https://doi.org/10.1103/PhysRevLett.104.136403} {\bibfield  {journal}
  {\bibinfo  {journal} {Phys. Rev. Lett.}\ }\textbf {\bibinfo {volume} {104}},\
  \bibinfo {pages} {136403} (\bibinfo {year} {2010})}\BibitemShut {NoStop}%
\bibitem [{\citenamefont {Li}\ \emph {et~al.}(2015)\citenamefont {Li},
  \citenamefont {Kermode},\ and\ \citenamefont {De~Vita}}]{Kermode2015}%
  \BibitemOpen
  \bibfield  {author} {\bibinfo {author} {\bibfnamefont {Z.}~\bibnamefont
  {Li}}, \bibinfo {author} {\bibfnamefont {J.~R.}\ \bibnamefont {Kermode}},\
  and\ \bibinfo {author} {\bibfnamefont {A.}~\bibnamefont {De~Vita}},\
  }\bibfield  {title} {\bibinfo {title} {Molecular dynamics with on-the-fly
  machine learning of quantum-mechanical forces},\ }\href
  {https://doi.org/10.1103/PhysRevLett.114.096405} {\bibfield  {journal}
  {\bibinfo  {journal} {Phys. Rev. Lett.}\ }\textbf {\bibinfo {volume} {114}},\
  \bibinfo {pages} {096405} (\bibinfo {year} {2015})}\BibitemShut {NoStop}%
\bibitem [{\citenamefont {Smith}\ \emph {et~al.}(2017)\citenamefont {Smith},
  \citenamefont {Isayev},\ and\ \citenamefont {Roitberg}}]{smith17}%
  \BibitemOpen
  \bibfield  {author} {\bibinfo {author} {\bibfnamefont {J.~S.}\ \bibnamefont
  {Smith}}, \bibinfo {author} {\bibfnamefont {O.}~\bibnamefont {Isayev}},\ and\
  \bibinfo {author} {\bibfnamefont {A.~E.}\ \bibnamefont {Roitberg}},\
  }\bibfield  {title} {\bibinfo {title} {Ani-1: an extensible neural network
  potential with dft accuracy at force field computational cost},\ }\href
  {https://doi.org/10.1039/C6SC05720A} {\bibfield  {journal} {\bibinfo
  {journal} {Chem. Sci.}\ }\textbf {\bibinfo {volume} {8}},\ \bibinfo {pages}
  {3192} (\bibinfo {year} {2017})}\BibitemShut {NoStop}%
\bibitem [{\citenamefont {Zhang}\ \emph {et~al.}(2018)\citenamefont {Zhang},
  \citenamefont {Han}, \citenamefont {Wang}, \citenamefont {Car},\ and\
  \citenamefont {E}}]{zhang18}%
  \BibitemOpen
  \bibfield  {author} {\bibinfo {author} {\bibfnamefont {L.}~\bibnamefont
  {Zhang}}, \bibinfo {author} {\bibfnamefont {J.}~\bibnamefont {Han}}, \bibinfo
  {author} {\bibfnamefont {H.}~\bibnamefont {Wang}}, \bibinfo {author}
  {\bibfnamefont {R.}~\bibnamefont {Car}},\ and\ \bibinfo {author}
  {\bibfnamefont {W.}~\bibnamefont {E}},\ }\bibfield  {title} {\bibinfo {title}
  {Deep potential molecular dynamics: A scalable model with the accuracy of
  quantum mechanics},\ }\href {https://doi.org/10.1103/PhysRevLett.120.143001}
  {\bibfield  {journal} {\bibinfo  {journal} {Phys. Rev. Lett.}\ }\textbf
  {\bibinfo {volume} {120}},\ \bibinfo {pages} {143001} (\bibinfo {year}
  {2018})}\BibitemShut {NoStop}%
\bibitem [{\citenamefont {Behler}(2016)}]{behler16}%
  \BibitemOpen
  \bibfield  {author} {\bibinfo {author} {\bibfnamefont {J.}~\bibnamefont
  {Behler}},\ }\bibfield  {title} {\bibinfo {title} {{Perspective: Machine
  learning potentials for atomistic simulations}},\ }\href
  {https://doi.org/10.1063/1.4966192} {\bibfield  {journal} {\bibinfo
  {journal} {The Journal of Chemical Physics}\ }\textbf {\bibinfo {volume}
  {145}},\ \bibinfo {pages} {170901} (\bibinfo {year} {2016})}\BibitemShut
  {NoStop}%
\bibitem [{\citenamefont {Deringer}\ \emph {et~al.}(2019)\citenamefont
  {Deringer}, \citenamefont {Caro},\ and\ \citenamefont {Csanyi}}]{deringer19}%
  \BibitemOpen
  \bibfield  {author} {\bibinfo {author} {\bibfnamefont {V.~L.}\ \bibnamefont
  {Deringer}}, \bibinfo {author} {\bibfnamefont {M.~A.}\ \bibnamefont {Caro}},\
  and\ \bibinfo {author} {\bibfnamefont {G.}~\bibnamefont {Csanyi}},\
  }\bibfield  {title} {\bibinfo {title} {Machine learning interatomic
  potentials as emerging tools for materials science},\ }\href
  {https://doi.org/https://doi.org/10.1002/adma.201902765} {\bibfield
  {journal} {\bibinfo  {journal} {Advanced Materials}\ }\textbf {\bibinfo
  {volume} {31}},\ \bibinfo {pages} {1902765} (\bibinfo {year}
  {2019})}\BibitemShut {NoStop}%
\bibitem [{\citenamefont {Mueller}\ \emph {et~al.}(2020)\citenamefont
  {Mueller}, \citenamefont {Hernandez},\ and\ \citenamefont
  {Wang}}]{Mueller2020}%
  \BibitemOpen
  \bibfield  {author} {\bibinfo {author} {\bibfnamefont {T.}~\bibnamefont
  {Mueller}}, \bibinfo {author} {\bibfnamefont {A.}~\bibnamefont {Hernandez}},\
  and\ \bibinfo {author} {\bibfnamefont {C.}~\bibnamefont {Wang}},\ }\bibfield
  {title} {\bibinfo {title} {Machine learning for interatomic potential
  models},\ }\href {https://doi.org/10.1063/1.5126336} {\bibfield  {journal}
  {\bibinfo  {journal} {The Journal of Chemical Physics}\ }\textbf {\bibinfo
  {volume} {152}},\ \bibinfo {pages} {050902} (\bibinfo {year}
  {2020})}\BibitemShut {NoStop}%
\bibitem [{\citenamefont {McGibbon}\ \emph {et~al.}(2017)\citenamefont
  {McGibbon}, \citenamefont {Taube}, \citenamefont {Donchev}, \citenamefont
  {Siva}, \citenamefont {Hernandez}, \citenamefont {Hargus}, \citenamefont
  {Law}, \citenamefont {Klepeis},\ and\ \citenamefont {Shaw}}]{mcgibbon17}%
  \BibitemOpen
  \bibfield  {author} {\bibinfo {author} {\bibfnamefont {R.~T.}\ \bibnamefont
  {McGibbon}}, \bibinfo {author} {\bibfnamefont {A.~G.}\ \bibnamefont {Taube}},
  \bibinfo {author} {\bibfnamefont {A.~G.}\ \bibnamefont {Donchev}}, \bibinfo
  {author} {\bibfnamefont {K.}~\bibnamefont {Siva}}, \bibinfo {author}
  {\bibfnamefont {F.}~\bibnamefont {Hernandez}}, \bibinfo {author}
  {\bibfnamefont {C.}~\bibnamefont {Hargus}}, \bibinfo {author} {\bibfnamefont
  {K.-H.}\ \bibnamefont {Law}}, \bibinfo {author} {\bibfnamefont {J.~L.}\
  \bibnamefont {Klepeis}},\ and\ \bibinfo {author} {\bibfnamefont {D.~E.}\
  \bibnamefont {Shaw}},\ }\bibfield  {title} {\bibinfo {title} {Improving the
  accuracy of m\"oller-plesset perturbation theory with neural networks},\
  }\href {https://doi.org/10.1063/1.4986081} {\bibfield  {journal} {\bibinfo
  {journal} {The Journal of Chemical Physics}\ }\textbf {\bibinfo {volume}
  {147}},\ \bibinfo {pages} {161725} (\bibinfo {year} {2017})}\BibitemShut
  {NoStop}%
\bibitem [{\citenamefont {Suwa}\ \emph {et~al.}(2019)\citenamefont {Suwa},
  \citenamefont {Smith}, \citenamefont {Lubbers}, \citenamefont {Batista},
  \citenamefont {Chern},\ and\ \citenamefont {Barros}}]{suwa19}%
  \BibitemOpen
  \bibfield  {author} {\bibinfo {author} {\bibfnamefont {H.}~\bibnamefont
  {Suwa}}, \bibinfo {author} {\bibfnamefont {J.~S.}\ \bibnamefont {Smith}},
  \bibinfo {author} {\bibfnamefont {N.}~\bibnamefont {Lubbers}}, \bibinfo
  {author} {\bibfnamefont {C.~D.}\ \bibnamefont {Batista}}, \bibinfo {author}
  {\bibfnamefont {G.-W.}\ \bibnamefont {Chern}},\ and\ \bibinfo {author}
  {\bibfnamefont {K.}~\bibnamefont {Barros}},\ }\bibfield  {title} {\bibinfo
  {title} {Machine learning for molecular dynamics with strongly correlated
  electrons},\ }\href {https://doi.org/10.1103/PhysRevB.99.161107} {\bibfield
  {journal} {\bibinfo  {journal} {Phys. Rev. B}\ }\textbf {\bibinfo {volume}
  {99}},\ \bibinfo {pages} {161107} (\bibinfo {year} {2019})}\BibitemShut
  {NoStop}%
\bibitem [{\citenamefont {Chmiela}\ \emph {et~al.}(2017)\citenamefont
  {Chmiela}, \citenamefont {Tkatchenko}, \citenamefont {Sauceda}, \citenamefont
  {Poltavsky}, \citenamefont {Schütt},\ and\ \citenamefont
  {Müller}}]{chmiela17}%
  \BibitemOpen
  \bibfield  {author} {\bibinfo {author} {\bibfnamefont {S.}~\bibnamefont
  {Chmiela}}, \bibinfo {author} {\bibfnamefont {A.}~\bibnamefont {Tkatchenko}},
  \bibinfo {author} {\bibfnamefont {H.~E.}\ \bibnamefont {Sauceda}}, \bibinfo
  {author} {\bibfnamefont {I.}~\bibnamefont {Poltavsky}}, \bibinfo {author}
  {\bibfnamefont {K.~T.}\ \bibnamefont {Schütt}},\ and\ \bibinfo {author}
  {\bibfnamefont {K.-R.}\ \bibnamefont {Müller}},\ }\bibfield  {title}
  {\bibinfo {title} {Machine learning of accurate energy-conserving molecular
  force fields},\ }\href {https://doi.org/10.1126/sciadv.1603015} {\bibfield
  {journal} {\bibinfo  {journal} {Science Advances}\ }\textbf {\bibinfo
  {volume} {3}},\ \bibinfo {pages} {e1603015} (\bibinfo {year}
  {2017})}\BibitemShut {NoStop}%
\bibitem [{\citenamefont {Chmiela}\ \emph {et~al.}(2018)\citenamefont
  {Chmiela}, \citenamefont {Sauceda}, \citenamefont {M{\"u}ller},\ and\
  \citenamefont {Tkatchenko}}]{chmiela18}%
  \BibitemOpen
  \bibfield  {author} {\bibinfo {author} {\bibfnamefont {S.}~\bibnamefont
  {Chmiela}}, \bibinfo {author} {\bibfnamefont {H.~E.}\ \bibnamefont
  {Sauceda}}, \bibinfo {author} {\bibfnamefont {K.-R.}\ \bibnamefont
  {M{\"u}ller}},\ and\ \bibinfo {author} {\bibfnamefont {A.}~\bibnamefont
  {Tkatchenko}},\ }\bibfield  {title} {\bibinfo {title} {Towards exact
  molecular dynamics simulations with machine-learned force fields},\ }\href
  {https://doi.org/10.1038/s41467-018-06169-2} {\bibfield  {journal} {\bibinfo
  {journal} {Nature Communications}\ }\textbf {\bibinfo {volume} {9}},\
  \bibinfo {pages} {3887} (\bibinfo {year} {2018})}\BibitemShut {NoStop}%
\bibitem [{\citenamefont {Sauceda}\ \emph {et~al.}(2020)\citenamefont
  {Sauceda}, \citenamefont {Gastegger}, \citenamefont {Chmiela}, \citenamefont
  {Müller},\ and\ \citenamefont {Tkatchenko}}]{sauceda20}%
  \BibitemOpen
  \bibfield  {author} {\bibinfo {author} {\bibfnamefont {H.~E.}\ \bibnamefont
  {Sauceda}}, \bibinfo {author} {\bibfnamefont {M.}~\bibnamefont {Gastegger}},
  \bibinfo {author} {\bibfnamefont {S.}~\bibnamefont {Chmiela}}, \bibinfo
  {author} {\bibfnamefont {K.-R.}\ \bibnamefont {Müller}},\ and\ \bibinfo
  {author} {\bibfnamefont {A.}~\bibnamefont {Tkatchenko}},\ }\bibfield  {title}
  {\bibinfo {title} {{Molecular force fields with gradient-domain machine
  learning (GDML): Comparison and synergies with classical force fields}},\
  }\href {https://doi.org/10.1063/5.0023005} {\bibfield  {journal} {\bibinfo
  {journal} {The Journal of Chemical Physics}\ }\textbf {\bibinfo {volume}
  {153}},\ \bibinfo {pages} {124109} (\bibinfo {year} {2020})}\BibitemShut
  {NoStop}%
\bibitem [{\citenamefont {Iftimie}\ \emph {et~al.}(2005)\citenamefont
  {Iftimie}, \citenamefont {Minary},\ and\ \citenamefont
  {Tuckerman}}]{iftimie_ab_2005}%
  \BibitemOpen
  \bibfield  {author} {\bibinfo {author} {\bibfnamefont {R.}~\bibnamefont
  {Iftimie}}, \bibinfo {author} {\bibfnamefont {P.}~\bibnamefont {Minary}},\
  and\ \bibinfo {author} {\bibfnamefont {M.~E.}\ \bibnamefont {Tuckerman}},\
  }\bibfield  {title} {\bibinfo {title} {\textit{{Ab} initio} molecular
  dynamics: {Concepts}, recent developments, and future trends},\ }\href
  {https://doi.org/10.1073/pnas.0500193102} {\bibfield  {journal} {\bibinfo
  {journal} {Proceedings of the National Academy of Sciences}\ }\textbf
  {\bibinfo {volume} {102}},\ \bibinfo {pages} {6654} (\bibinfo {year}
  {2005})}\BibitemShut {NoStop}%
\bibitem [{\citenamefont {Kohn}(1996)}]{kohn1996}%
  \BibitemOpen
  \bibfield  {author} {\bibinfo {author} {\bibfnamefont {W.}~\bibnamefont
  {Kohn}},\ }\bibfield  {title} {\bibinfo {title} {Density functional and
  density matrix method scaling linearly with the number of atoms},\ }\href
  {https://doi.org/10.1103/PhysRevLett.76.3168} {\bibfield  {journal} {\bibinfo
   {journal} {Phys. Rev. Lett.}\ }\textbf {\bibinfo {volume} {76}},\ \bibinfo
  {pages} {3168} (\bibinfo {year} {1996})}\BibitemShut {NoStop}%
\bibitem [{\citenamefont {Prodan}\ and\ \citenamefont
  {Kohn}(2005)}]{prodan2005}%
  \BibitemOpen
  \bibfield  {author} {\bibinfo {author} {\bibfnamefont {E.}~\bibnamefont
  {Prodan}}\ and\ \bibinfo {author} {\bibfnamefont {W.}~\bibnamefont {Kohn}},\
  }\bibfield  {title} {\bibinfo {title} {Nearsightedness of electronic
  matter},\ }\href {https://doi.org/10.1073/pnas.0505436102} {\bibfield
  {journal} {\bibinfo  {journal} {Proc. Natl. Acad. Sci. USA}\ }\textbf
  {\bibinfo {volume} {102}},\ \bibinfo {pages} {11635} (\bibinfo {year}
  {2005})}\BibitemShut {NoStop}%
\bibitem [{\citenamefont {Sanjay~Puri}(2009)}]{Puri2009}%
  \BibitemOpen
  \bibfield  {author} {\bibinfo {author} {\bibfnamefont {V.~W.}\ \bibnamefont
  {Sanjay~Puri}},\ }\href@noop {} {\emph {\bibinfo {title} {Kinetics of phase
  transitions}}}\ (\bibinfo  {publisher} {CRC Press},\ \bibinfo {year}
  {2009})\BibitemShut {NoStop}%
\bibitem [{\citenamefont {Bray}(1994)}]{Bray1994}%
  \BibitemOpen
  \bibfield  {author} {\bibinfo {author} {\bibfnamefont {A.}~\bibnamefont
  {Bray}},\ }\bibfield  {title} {\bibinfo {title} {Theory of phase-ordering
  kinetics},\ }\href {https://doi.org/10.1080/00018739400101505} {\bibfield
  {journal} {\bibinfo  {journal} {Advances in Physics}\ }\textbf {\bibinfo
  {volume} {43}},\ \bibinfo {pages} {357} (\bibinfo {year} {1994})}\BibitemShut
  {NoStop}%
\bibitem [{\citenamefont {Chatterjee}\ \emph {et~al.}(2018)\citenamefont
  {Chatterjee}, \citenamefont {Puri},\ and\ \citenamefont
  {Paul}}]{chatterjee2018}%
  \BibitemOpen
  \bibfield  {author} {\bibinfo {author} {\bibfnamefont {S.}~\bibnamefont
  {Chatterjee}}, \bibinfo {author} {\bibfnamefont {S.}~\bibnamefont {Puri}},\
  and\ \bibinfo {author} {\bibfnamefont {R.}~\bibnamefont {Paul}},\ }\bibfield
  {title} {\bibinfo {title} {Ordering kinetics in the $q$-state clock model:
  Scaling properties and growth laws},\ }\href
  {https://doi.org/10.1103/PhysRevE.98.032109} {\bibfield  {journal} {\bibinfo
  {journal} {Phys. Rev. E}\ }\textbf {\bibinfo {volume} {98}},\ \bibinfo
  {pages} {032109} (\bibinfo {year} {2018})}\BibitemShut {NoStop}%
\bibitem [{\citenamefont {Chatterjee}\ \emph {et~al.}(2020)\citenamefont
  {Chatterjee}, \citenamefont {Sutradhar}, \citenamefont {Puri},\ and\
  \citenamefont {Paul}}]{chatterjee2020}%
  \BibitemOpen
  \bibfield  {author} {\bibinfo {author} {\bibfnamefont {S.}~\bibnamefont
  {Chatterjee}}, \bibinfo {author} {\bibfnamefont {S.}~\bibnamefont
  {Sutradhar}}, \bibinfo {author} {\bibfnamefont {S.}~\bibnamefont {Puri}},\
  and\ \bibinfo {author} {\bibfnamefont {R.}~\bibnamefont {Paul}},\ }\bibfield
  {title} {\bibinfo {title} {Ordering kinetics in a $q$-state random-bond clock
  model: Role of vortices and interfaces},\ }\href
  {https://doi.org/10.1103/PhysRevE.101.032128} {\bibfield  {journal} {\bibinfo
   {journal} {Phys. Rev. E}\ }\textbf {\bibinfo {volume} {101}},\ \bibinfo
  {pages} {032128} (\bibinfo {year} {2020})}\BibitemShut {NoStop}%
\bibitem [{\citenamefont {Corberi}\ \emph {et~al.}(2006)\citenamefont
  {Corberi}, \citenamefont {Lippiello},\ and\ \citenamefont
  {Zannetti}}]{Corberi06}%
  \BibitemOpen
  \bibfield  {author} {\bibinfo {author} {\bibfnamefont {F.}~\bibnamefont
  {Corberi}}, \bibinfo {author} {\bibfnamefont {E.}~\bibnamefont {Lippiello}},\
  and\ \bibinfo {author} {\bibfnamefont {M.}~\bibnamefont {Zannetti}},\
  }\bibfield  {title} {\bibinfo {title} {Scaling and universality in the aging
  kinetics of the two-dimensional clock model},\ }\href
  {https://doi.org/10.1103/PhysRevE.74.041106} {\bibfield  {journal} {\bibinfo
  {journal} {Phys. Rev. E}\ }\textbf {\bibinfo {volume} {74}},\ \bibinfo
  {pages} {041106} (\bibinfo {year} {2006})}\BibitemShut {NoStop}%
\bibitem [{\citenamefont {Jos\'e}\ \emph {et~al.}(1977)\citenamefont {Jos\'e},
  \citenamefont {Kadanoff}, \citenamefont {Kirkpatrick},\ and\ \citenamefont
  {Nelson}}]{jose1977}%
  \BibitemOpen
  \bibfield  {author} {\bibinfo {author} {\bibfnamefont {J.~V.}\ \bibnamefont
  {Jos\'e}}, \bibinfo {author} {\bibfnamefont {L.~P.}\ \bibnamefont
  {Kadanoff}}, \bibinfo {author} {\bibfnamefont {S.}~\bibnamefont
  {Kirkpatrick}},\ and\ \bibinfo {author} {\bibfnamefont {D.~R.}\ \bibnamefont
  {Nelson}},\ }\bibfield  {title} {\bibinfo {title} {Renormalization, vortices,
  and symmetry-breaking perturbations in the two-dimensional planar model},\
  }\href {https://doi.org/10.1103/PhysRevB.16.1217} {\bibfield  {journal}
  {\bibinfo  {journal} {Phys. Rev. B}\ }\textbf {\bibinfo {volume} {16}},\
  \bibinfo {pages} {1217} (\bibinfo {year} {1977})}\BibitemShut {NoStop}%
\bibitem [{\citenamefont {Tang}\ and\ \citenamefont {Hirsch}(1988)}]{Tang1988}%
  \BibitemOpen
  \bibfield  {author} {\bibinfo {author} {\bibfnamefont {S.}~\bibnamefont
  {Tang}}\ and\ \bibinfo {author} {\bibfnamefont {J.~E.}\ \bibnamefont
  {Hirsch}},\ }\bibfield  {title} {\bibinfo {title} {Peierls instability in the
  two-dimensional half-filled hubbard model},\ }\href
  {https://doi.org/10.1103/PhysRevB.37.9546} {\bibfield  {journal} {\bibinfo
  {journal} {Phys. Rev. B}\ }\textbf {\bibinfo {volume} {37}},\ \bibinfo
  {pages} {9546} (\bibinfo {year} {1988})}\BibitemShut {NoStop}%
\bibitem [{\citenamefont {Ono}\ and\ \citenamefont {Hamano}(2000)}]{Ono2000}%
  \BibitemOpen
  \bibfield  {author} {\bibinfo {author} {\bibfnamefont {Y.}~\bibnamefont
  {Ono}}\ and\ \bibinfo {author} {\bibfnamefont {T.}~\bibnamefont {Hamano}},\
  }\bibfield  {title} {\bibinfo {title} {Peierls distortion in two-dimensional
  tight-binding model},\ }\href {https://doi.org/10.1143/JPSJ.69.1769}
  {\bibfield  {journal} {\bibinfo  {journal} {Journal of the Physical Society
  of Japan}\ }\textbf {\bibinfo {volume} {69}},\ \bibinfo {pages} {1769}
  (\bibinfo {year} {2000})}\BibitemShut {NoStop}%
\bibitem [{\citenamefont {Watanabe}\ \emph {et~al.}(2007)\citenamefont
  {Watanabe}, \citenamefont {Chiba},\ and\ \citenamefont {Ono}}]{Watanabe2007}%
  \BibitemOpen
  \bibfield  {author} {\bibinfo {author} {\bibfnamefont {C.}~\bibnamefont
  {Watanabe}}, \bibinfo {author} {\bibfnamefont {S.}~\bibnamefont {Chiba}},\
  and\ \bibinfo {author} {\bibfnamefont {Y.}~\bibnamefont {Ono}},\ }\bibfield
  {title} {\bibinfo {title} {Phonon softening in two-dimensional
  electron–lattice system with anisotropy},\ }\href
  {https://doi.org/10.1143/JPSJ.76.114704} {\bibfield  {journal} {\bibinfo
  {journal} {Journal of the Physical Society of Japan}\ }\textbf {\bibinfo
  {volume} {76}},\ \bibinfo {pages} {114704} (\bibinfo {year}
  {2007})}\BibitemShut {NoStop}%
\bibitem [{\citenamefont {Yuan}\ \emph {et~al.}(2001)\citenamefont {Yuan},
  \citenamefont {Nunner},\ and\ \citenamefont {Kopp}}]{Yuan01}%
  \BibitemOpen
  \bibfield  {author} {\bibinfo {author} {\bibfnamefont {Q.}~\bibnamefont
  {Yuan}}, \bibinfo {author} {\bibfnamefont {T.}~\bibnamefont {Nunner}},\ and\
  \bibinfo {author} {\bibfnamefont {T.}~\bibnamefont {Kopp}},\ }\bibfield
  {title} {\bibinfo {title} {Imperfect nesting and peierls instability for a
  two-dimensional tight-binding model},\ }\href
  {https://doi.org/10.1007/PL00011133} {\bibfield  {journal} {\bibinfo
  {journal} {The European Physical Journal B - Condensed Matter and Complex
  Systems}\ }\textbf {\bibinfo {volume} {22}},\ \bibinfo {pages} {37} (\bibinfo
  {year} {2001})}\BibitemShut {NoStop}%
\bibitem [{\citenamefont {Yuan}(2002)}]{yuan02}%
  \BibitemOpen
  \bibfield  {author} {\bibinfo {author} {\bibfnamefont {Q.}~\bibnamefont
  {Yuan}},\ }\bibfield  {title} {\bibinfo {title} {A study of peierls
  instabilities for a two-dimensional t-t' model},\ }\href
  {https://doi.org/10.1140/epjb/e20020032} {\bibfield  {journal} {\bibinfo
  {journal} {The European Physical Journal B - Condensed Matter and Complex
  Systems}\ }\textbf {\bibinfo {volume} {25}},\ \bibinfo {pages} {281}
  (\bibinfo {year} {2002})}\BibitemShut {NoStop}%
\bibitem [{\citenamefont {Ermak}\ and\ \citenamefont
  {Buckholz}(1980)}]{ermak80}%
  \BibitemOpen
  \bibfield  {author} {\bibinfo {author} {\bibfnamefont {D.~L.}\ \bibnamefont
  {Ermak}}\ and\ \bibinfo {author} {\bibfnamefont {H.}~\bibnamefont
  {Buckholz}},\ }\bibfield  {title} {\bibinfo {title} {Numerical integration of
  the {Langevin} equation: {Monte} {Carlo} simulation},\ }\href
  {https://doi.org/https://doi.org/10.1016/0021-9991(80)90084-4} {\bibfield
  {journal} {\bibinfo  {journal} {Journal of Computational Physics}\ }\textbf
  {\bibinfo {volume} {35}},\ \bibinfo {pages} {169} (\bibinfo {year}
  {1980})}\BibitemShut {NoStop}%
\bibitem [{\citenamefont {Hohenberg}\ and\ \citenamefont
  {Halperin}(1977)}]{hohenberg77}%
  \BibitemOpen
  \bibfield  {author} {\bibinfo {author} {\bibfnamefont {P.~C.}\ \bibnamefont
  {Hohenberg}}\ and\ \bibinfo {author} {\bibfnamefont {B.~I.}\ \bibnamefont
  {Halperin}},\ }\bibfield  {title} {\bibinfo {title} {Theory of dynamic
  critical phenomena},\ }\href {https://doi.org/10.1103/RevModPhys.49.435}
  {\bibfield  {journal} {\bibinfo  {journal} {Rev. Mod. Phys.}\ }\textbf
  {\bibinfo {volume} {49}},\ \bibinfo {pages} {435} (\bibinfo {year}
  {1977})}\BibitemShut {NoStop}%
\bibitem [{\citenamefont {Glauber}(1963)}]{glauber_time-dependent_1963}%
  \BibitemOpen
  \bibfield  {author} {\bibinfo {author} {\bibfnamefont {R.~J.}\ \bibnamefont
  {Glauber}},\ }\bibfield  {title} {\bibinfo {title} {Time-{Dependent}
  {Statistics} of the {Ising} {Model}},\ }\href
  {https://doi.org/10.1063/1.1703954} {\bibfield  {journal} {\bibinfo
  {journal} {Journal of Mathematical Physics}\ }\textbf {\bibinfo {volume}
  {4}},\ \bibinfo {pages} {294} (\bibinfo {year} {1963})}\BibitemShut {NoStop}%
\bibitem [{\citenamefont {Hänggi}\ and\ \citenamefont
  {Thomas}(1982)}]{hanggi1982}%
  \BibitemOpen
  \bibfield  {author} {\bibinfo {author} {\bibfnamefont {P.}~\bibnamefont
  {Hänggi}}\ and\ \bibinfo {author} {\bibfnamefont {H.}~\bibnamefont
  {Thomas}},\ }\bibfield  {title} {\bibinfo {title} {Stochastic processes:
  {Time} evolution, symmetries and linear response},\ }\href
  {https://doi.org/https://doi.org/10.1016/0370-1573(82)90045-X} {\bibfield
  {journal} {\bibinfo  {journal} {Physics Reports}\ }\textbf {\bibinfo {volume}
  {88}},\ \bibinfo {pages} {207} (\bibinfo {year} {1982})}\BibitemShut
  {NoStop}%
\bibitem [{\citenamefont {Di~Ventra}\ and\ \citenamefont
  {Pantelides}(2000)}]{diventra00}%
  \BibitemOpen
  \bibfield  {author} {\bibinfo {author} {\bibfnamefont {M.}~\bibnamefont
  {Di~Ventra}}\ and\ \bibinfo {author} {\bibfnamefont {S.~T.}\ \bibnamefont
  {Pantelides}},\ }\bibfield  {title} {\bibinfo {title} {Hellmann-feynman
  theorem and the definition of forces in quantum time-dependent and transport
  problems},\ }\href {https://doi.org/10.1103/PhysRevB.61.16207} {\bibfield
  {journal} {\bibinfo  {journal} {Phys. Rev. B}\ }\textbf {\bibinfo {volume}
  {61}},\ \bibinfo {pages} {16207} (\bibinfo {year} {2000})}\BibitemShut
  {NoStop}%
\bibitem [{\citenamefont {Todorov}\ \emph {et~al.}(2010)\citenamefont
  {Todorov}, \citenamefont {Dundas},\ and\ \citenamefont
  {McEniry}}]{todorov10}%
  \BibitemOpen
  \bibfield  {author} {\bibinfo {author} {\bibfnamefont {T.~N.}\ \bibnamefont
  {Todorov}}, \bibinfo {author} {\bibfnamefont {D.}~\bibnamefont {Dundas}},\
  and\ \bibinfo {author} {\bibfnamefont {E.~J.}\ \bibnamefont {McEniry}},\
  }\bibfield  {title} {\bibinfo {title} {Nonconservative generalized
  current-induced forces},\ }\href {https://doi.org/10.1103/PhysRevB.81.075416}
  {\bibfield  {journal} {\bibinfo  {journal} {Phys. Rev. B}\ }\textbf {\bibinfo
  {volume} {81}},\ \bibinfo {pages} {075416} (\bibinfo {year}
  {2010})}\BibitemShut {NoStop}%
\bibitem [{\citenamefont {L\"u}\ \emph {et~al.}(2012)\citenamefont {L\"u},
  \citenamefont {Brandbyge}, \citenamefont {Hedeg\aa{}rd}, \citenamefont
  {Todorov},\ and\ \citenamefont {Dundas}}]{lu12}%
  \BibitemOpen
  \bibfield  {author} {\bibinfo {author} {\bibfnamefont {J.-T.}\ \bibnamefont
  {L\"u}}, \bibinfo {author} {\bibfnamefont {M.}~\bibnamefont {Brandbyge}},
  \bibinfo {author} {\bibfnamefont {P.}~\bibnamefont {Hedeg\aa{}rd}}, \bibinfo
  {author} {\bibfnamefont {T.~N.}\ \bibnamefont {Todorov}},\ and\ \bibinfo
  {author} {\bibfnamefont {D.}~\bibnamefont {Dundas}},\ }\bibfield  {title}
  {\bibinfo {title} {Current-induced atomic dynamics, instabilities, and raman
  signals: Quasiclassical langevin equation approach},\ }\href
  {https://doi.org/10.1103/PhysRevB.85.245444} {\bibfield  {journal} {\bibinfo
  {journal} {Phys. Rev. B}\ }\textbf {\bibinfo {volume} {85}},\ \bibinfo
  {pages} {245444} (\bibinfo {year} {2012})}\BibitemShut {NoStop}%
\bibitem [{\citenamefont {Dundas}\ \emph {et~al.}(2009)\citenamefont {Dundas},
  \citenamefont {McEniry},\ and\ \citenamefont {Todorov}}]{dundas09}%
  \BibitemOpen
  \bibfield  {author} {\bibinfo {author} {\bibfnamefont {D.}~\bibnamefont
  {Dundas}}, \bibinfo {author} {\bibfnamefont {E.~J.}\ \bibnamefont
  {McEniry}},\ and\ \bibinfo {author} {\bibfnamefont {T.~N.}\ \bibnamefont
  {Todorov}},\ }\bibfield  {title} {\bibinfo {title} {Current-driven atomic
  waterwheels},\ }\href {https://doi.org/10.1038/nnano.2008.411} {\bibfield
  {journal} {\bibinfo  {journal} {Nature Nanotechnology}\ }\textbf {\bibinfo
  {volume} {4}},\ \bibinfo {pages} {99} (\bibinfo {year} {2009})}\BibitemShut
  {NoStop}%
\bibitem [{\citenamefont {Marx}\ and\ \citenamefont {Hutter}(2009)}]{marx2009}%
  \BibitemOpen
  \bibfield  {author} {\bibinfo {author} {\bibfnamefont {D.}~\bibnamefont
  {Marx}}\ and\ \bibinfo {author} {\bibfnamefont {J.}~\bibnamefont {Hutter}},\
  }\href {https://doi.org/10.1017/CBO9780511609633} {\emph {\bibinfo {title}
  {Ab Initio Molecular Dynamics: Basic Theory and Advanced Methods}}}\
  (\bibinfo  {publisher} {Cambridge University Press},\ \bibinfo {year}
  {2009})\BibitemShut {NoStop}%
\bibitem [{\citenamefont {Weiler}\ \emph {et~al.}(2018)\citenamefont {Weiler},
  \citenamefont {Geiger}, \citenamefont {Welling}, \citenamefont {Boomsma},\
  and\ \citenamefont {Cohen}}]{weiler18}%
  \BibitemOpen
  \bibfield  {author} {\bibinfo {author} {\bibfnamefont {M.}~\bibnamefont
  {Weiler}}, \bibinfo {author} {\bibfnamefont {M.}~\bibnamefont {Geiger}},
  \bibinfo {author} {\bibfnamefont {M.}~\bibnamefont {Welling}}, \bibinfo
  {author} {\bibfnamefont {W.}~\bibnamefont {Boomsma}},\ and\ \bibinfo {author}
  {\bibfnamefont {T.}~\bibnamefont {Cohen}},\ }\bibfield  {title} {\bibinfo
  {title} {3d steerable cnns: Learning rotationally equivariant features in
  volumetric data},\ }\href {https://arxiv.org/abs/1807.02547} {\  (\bibinfo
  {year} {2018})},\ \Eprint {https://arxiv.org/abs/1807.02547}
  {arXiv:1807.02547 [cs.LG]} \BibitemShut {NoStop}%
\bibitem [{\citenamefont {Wang}\ \emph {et~al.}(2018)\citenamefont {Wang},
  \citenamefont {Chern}, \citenamefont {Batista},\ and\ \citenamefont
  {Barros}}]{wang2018}%
  \BibitemOpen
  \bibfield  {author} {\bibinfo {author} {\bibfnamefont {Z.}~\bibnamefont
  {Wang}}, \bibinfo {author} {\bibfnamefont {G.-W.}\ \bibnamefont {Chern}},
  \bibinfo {author} {\bibfnamefont {C.~D.}\ \bibnamefont {Batista}},\ and\
  \bibinfo {author} {\bibfnamefont {K.}~\bibnamefont {Barros}},\ }\bibfield
  {title} {\bibinfo {title} {Gradient-based stochastic estimation of the
  density matrix},\ }\href {https://doi.org/10.1063/1.5017741} {\bibfield
  {journal} {\bibinfo  {journal} {The Journal of Chemical Physics}\ }\textbf
  {\bibinfo {volume} {148}},\ \bibinfo {pages} {094107} (\bibinfo {year}
  {2018})}\BibitemShut {NoStop}%
\bibitem [{\citenamefont {Berger}(1996)}]{berger96}%
  \BibitemOpen
  \bibfield  {author} {\bibinfo {author} {\bibfnamefont {L.}~\bibnamefont
  {Berger}},\ }\bibfield  {title} {\bibinfo {title} {Emission of spin waves by
  a magnetic multilayer traversed by a current},\ }\href
  {https://doi.org/10.1103/PhysRevB.54.9353} {\bibfield  {journal} {\bibinfo
  {journal} {Phys. Rev. B}\ }\textbf {\bibinfo {volume} {54}},\ \bibinfo
  {pages} {9353} (\bibinfo {year} {1996})}\BibitemShut {NoStop}%
\bibitem [{\citenamefont {Johnston}\ \emph {et~al.}(2013)\citenamefont
  {Johnston}, \citenamefont {Nowadnick}, \citenamefont {Kung}, \citenamefont
  {Moritz}, \citenamefont {Scalettar},\ and\ \citenamefont
  {Devereaux}}]{johnston2013}%
  \BibitemOpen
  \bibfield  {author} {\bibinfo {author} {\bibfnamefont {S.}~\bibnamefont
  {Johnston}}, \bibinfo {author} {\bibfnamefont {E.~A.}\ \bibnamefont
  {Nowadnick}}, \bibinfo {author} {\bibfnamefont {Y.~F.}\ \bibnamefont {Kung}},
  \bibinfo {author} {\bibfnamefont {B.}~\bibnamefont {Moritz}}, \bibinfo
  {author} {\bibfnamefont {R.~T.}\ \bibnamefont {Scalettar}},\ and\ \bibinfo
  {author} {\bibfnamefont {T.~P.}\ \bibnamefont {Devereaux}},\ }\bibfield
  {title} {\bibinfo {title} {Determinant quantum monte carlo study of the
  two-dimensional single-band hubbard-holstein model},\ }\href
  {https://doi.org/10.1103/PhysRevB.87.235133} {\bibfield  {journal} {\bibinfo
  {journal} {Phys. Rev. B}\ }\textbf {\bibinfo {volume} {87}},\ \bibinfo
  {pages} {235133} (\bibinfo {year} {2013})}\BibitemShut {NoStop}%
\bibitem [{\citenamefont {Costa}\ \emph {et~al.}(2020)\citenamefont {Costa},
  \citenamefont {Seki}, \citenamefont {Yunoki},\ and\ \citenamefont
  {Sorella}}]{costa2020}%
  \BibitemOpen
  \bibfield  {author} {\bibinfo {author} {\bibfnamefont {N.~C.}\ \bibnamefont
  {Costa}}, \bibinfo {author} {\bibfnamefont {K.}~\bibnamefont {Seki}},
  \bibinfo {author} {\bibfnamefont {S.}~\bibnamefont {Yunoki}},\ and\ \bibinfo
  {author} {\bibfnamefont {S.}~\bibnamefont {Sorella}},\ }\bibfield  {title}
  {\bibinfo {title} {Phase diagram of the two-dimensional hubbard-holstein
  model},\ }\href {https://doi.org/10.1038/s42005-020-0342-2} {\bibfield
  {journal} {\bibinfo  {journal} {Communications Physics}\ }\textbf {\bibinfo
  {volume} {3}},\ \bibinfo {pages} {80} (\bibinfo {year} {2020})}\BibitemShut
  {NoStop}%
\bibitem [{\citenamefont {Cheng}\ \emph {et~al.}(2023)\citenamefont {Cheng},
  \citenamefont {Zhang},\ and\ \citenamefont {Chern}}]{chen23_CDW}%
  \BibitemOpen
  \bibfield  {author} {\bibinfo {author} {\bibfnamefont {C.}~\bibnamefont
  {Cheng}}, \bibinfo {author} {\bibfnamefont {S.}~\bibnamefont {Zhang}},\ and\
  \bibinfo {author} {\bibfnamefont {G.-W.}\ \bibnamefont {Chern}},\ }\bibfield
  {title} {\bibinfo {title} {Machine learning for phase ordering dynamics of
  charge density waves},\ }\href {https://doi.org/10.1103/PhysRevB.108.014301}
  {\bibfield  {journal} {\bibinfo  {journal} {Phys. Rev. B}\ }\textbf {\bibinfo
  {volume} {108}},\ \bibinfo {pages} {014301} (\bibinfo {year}
  {2023})}\BibitemShut {NoStop}%
\bibitem [{\citenamefont {Ghosh}\ \emph {et~al.}(2024)\citenamefont {Ghosh},
  \citenamefont {Zhang}, \citenamefont {Cheng},\ and\ \citenamefont
  {Chern}}]{Supriyo2024}%
  \BibitemOpen
  \bibfield  {author} {\bibinfo {author} {\bibfnamefont {S.}~\bibnamefont
  {Ghosh}}, \bibinfo {author} {\bibfnamefont {S.}~\bibnamefont {Zhang}},
  \bibinfo {author} {\bibfnamefont {C.}~\bibnamefont {Cheng}},\ and\ \bibinfo
  {author} {\bibfnamefont {G.-W.}\ \bibnamefont {Chern}},\ }\bibfield  {title}
  {\bibinfo {title} {Kinetics of orbital ordering in cooperative jahn-teller
  models: Machine-learning enabled large-scale simulations},\ }\href
  {https://doi.org/10.1103/PhysRevMaterials.8.123602} {\bibfield  {journal}
  {\bibinfo  {journal} {Phys. Rev. Mater.}\ }\textbf {\bibinfo {volume} {8}},\
  \bibinfo {pages} {123602} (\bibinfo {year} {2024})}\BibitemShut {NoStop}%
\bibitem [{\citenamefont {Ma}\ \emph {et~al.}(2019)\citenamefont {Ma},
  \citenamefont {Zhang}, \citenamefont {Tan}, \citenamefont {Ghosh},\ and\
  \citenamefont {Chern}}]{ma19}%
  \BibitemOpen
  \bibfield  {author} {\bibinfo {author} {\bibfnamefont {J.}~\bibnamefont
  {Ma}}, \bibinfo {author} {\bibfnamefont {P.}~\bibnamefont {Zhang}}, \bibinfo
  {author} {\bibfnamefont {Y.}~\bibnamefont {Tan}}, \bibinfo {author}
  {\bibfnamefont {A.~W.}\ \bibnamefont {Ghosh}},\ and\ \bibinfo {author}
  {\bibfnamefont {G.-W.}\ \bibnamefont {Chern}},\ }\bibfield  {title} {\bibinfo
  {title} {Machine learning electron correlation in a disordered medium},\
  }\href {https://doi.org/10.1103/PhysRevB.99.085118} {\bibfield  {journal}
  {\bibinfo  {journal} {Phys. Rev. B}\ }\textbf {\bibinfo {volume} {99}},\
  \bibinfo {pages} {085118} (\bibinfo {year} {2019})}\BibitemShut {NoStop}%
\bibitem [{\citenamefont {Liu}\ \emph {et~al.}(2022)\citenamefont {Liu},
  \citenamefont {Zhang}, \citenamefont {Zhang}, \citenamefont {Lee},\ and\
  \citenamefont {Chern}}]{liu22}%
  \BibitemOpen
  \bibfield  {author} {\bibinfo {author} {\bibfnamefont {Y.-H.}\ \bibnamefont
  {Liu}}, \bibinfo {author} {\bibfnamefont {S.}~\bibnamefont {Zhang}}, \bibinfo
  {author} {\bibfnamefont {P.}~\bibnamefont {Zhang}}, \bibinfo {author}
  {\bibfnamefont {T.-K.}\ \bibnamefont {Lee}},\ and\ \bibinfo {author}
  {\bibfnamefont {G.-W.}\ \bibnamefont {Chern}},\ }\bibfield  {title} {\bibinfo
  {title} {Machine learning predictions for local electronic properties of
  disordered correlated electron systems},\ }\href
  {https://doi.org/10.1103/PhysRevB.106.035131} {\bibfield  {journal} {\bibinfo
   {journal} {Phys. Rev. B}\ }\textbf {\bibinfo {volume} {106}},\ \bibinfo
  {pages} {035131} (\bibinfo {year} {2022})}\BibitemShut {NoStop}%
\bibitem [{\citenamefont {Zhang}\ \emph
  {et~al.}(2022{\natexlab{a}})\citenamefont {Zhang}, \citenamefont {Zhang},\
  and\ \citenamefont {Chern}}]{zhang2022}%
  \BibitemOpen
  \bibfield  {author} {\bibinfo {author} {\bibfnamefont {S.}~\bibnamefont
  {Zhang}}, \bibinfo {author} {\bibfnamefont {P.}~\bibnamefont {Zhang}},\ and\
  \bibinfo {author} {\bibfnamefont {G.-W.}\ \bibnamefont {Chern}},\ }\bibfield
  {title} {\bibinfo {title} {Anomalous phase separation in a correlated
  electron system: Machine-learning–enabled large-scale kinetic monte carlo
  simulations},\ }\href {https://doi.org/10.1073/pnas.2119957119} {\bibfield
  {journal} {\bibinfo  {journal} {Proceedings of the National Academy of
  Sciences}\ }\textbf {\bibinfo {volume} {119}},\ \bibinfo {pages}
  {e2119957119} (\bibinfo {year} {2022}{\natexlab{a}})}\BibitemShut {NoStop}%
\bibitem [{\citenamefont {Zhang}\ \emph {et~al.}(2020)\citenamefont {Zhang},
  \citenamefont {Saha},\ and\ \citenamefont {Chern}}]{zhang2020}%
  \BibitemOpen
  \bibfield  {author} {\bibinfo {author} {\bibfnamefont {P.}~\bibnamefont
  {Zhang}}, \bibinfo {author} {\bibfnamefont {P.}~\bibnamefont {Saha}},\ and\
  \bibinfo {author} {\bibfnamefont {G.-W.}\ \bibnamefont {Chern}},\ }\href
  {https://arxiv.org/abs/2006.04205} {\bibinfo {title} {Machine learning
  dynamics of phase separation in correlated electron magnets}} (\bibinfo
  {year} {2020}),\ \Eprint {https://arxiv.org/abs/2006.04205} {arXiv:2006.04205
  [cond-mat.str-el]} \BibitemShut {NoStop}%
\bibitem [{\citenamefont {Zhang}\ and\ \citenamefont
  {Chern}(2021{\natexlab{a}})}]{zhang2021}%
  \BibitemOpen
  \bibfield  {author} {\bibinfo {author} {\bibfnamefont {P.}~\bibnamefont
  {Zhang}}\ and\ \bibinfo {author} {\bibfnamefont {G.-W.}\ \bibnamefont
  {Chern}},\ }\bibfield  {title} {\bibinfo {title} {Arrested phase separation
  in double-exchange models: Large-scale simulation enabled by machine
  learning},\ }\href {https://doi.org/10.1103/PhysRevLett.127.146401}
  {\bibfield  {journal} {\bibinfo  {journal} {Phys. Rev. Lett.}\ }\textbf
  {\bibinfo {volume} {127}},\ \bibinfo {pages} {146401} (\bibinfo {year}
  {2021}{\natexlab{a}})}\BibitemShut {NoStop}%
\bibitem [{\citenamefont {Zhang}\ and\ \citenamefont
  {Chern}(2023{\natexlab{a}})}]{zhang2023}%
  \BibitemOpen
  \bibfield  {author} {\bibinfo {author} {\bibfnamefont {P.}~\bibnamefont
  {Zhang}}\ and\ \bibinfo {author} {\bibfnamefont {G.-W.}\ \bibnamefont
  {Chern}},\ }\bibfield  {title} {\bibinfo {title} {Machine learning
  nonequilibrium electron forces for spin dynamics of itinerant magnets},\
  }\href {https://doi.org/10.1038/s41524-023-00990-0} {\bibfield  {journal}
  {\bibinfo  {journal} {npj Computational Materials}\ }\textbf {\bibinfo
  {volume} {9}},\ \bibinfo {pages} {32} (\bibinfo {year}
  {2023}{\natexlab{a}})}\BibitemShut {NoStop}%
\bibitem [{\citenamefont {Fan}\ \emph {et~al.}(2024)\citenamefont {Fan},
  \citenamefont {Zhang},\ and\ \citenamefont {Chern}}]{Yunhao2024}%
  \BibitemOpen
  \bibfield  {author} {\bibinfo {author} {\bibfnamefont {Y.}~\bibnamefont
  {Fan}}, \bibinfo {author} {\bibfnamefont {S.}~\bibnamefont {Zhang}},\ and\
  \bibinfo {author} {\bibfnamefont {G.-W.}\ \bibnamefont {Chern}},\ }\bibfield
  {title} {\bibinfo {title} {Coarsening of chiral domains in itinerant electron
  magnets: A machine learning force-field approach},\ }\href
  {https://doi.org/10.1103/PhysRevB.110.245105} {\bibfield  {journal} {\bibinfo
   {journal} {Phys. Rev. B}\ }\textbf {\bibinfo {volume} {110}},\ \bibinfo
  {pages} {245105} (\bibinfo {year} {2024})}\BibitemShut {NoStop}%
\bibitem [{\citenamefont {Zhang}\ \emph
  {et~al.}(2022{\natexlab{b}})\citenamefont {Zhang}, \citenamefont {Zhang},\
  and\ \citenamefont {Chern}}]{puhan22}%
  \BibitemOpen
  \bibfield  {author} {\bibinfo {author} {\bibfnamefont {P.}~\bibnamefont
  {Zhang}}, \bibinfo {author} {\bibfnamefont {S.}~\bibnamefont {Zhang}},\ and\
  \bibinfo {author} {\bibfnamefont {G.-W.}\ \bibnamefont {Chern}},\ }\href
  {https://arxiv.org/abs/2201.00798} {\bibinfo {title} {Descriptors for machine
  learning model of generalized force field in condensed matter systems}}
  (\bibinfo {year} {2022}{\natexlab{b}}),\ \Eprint
  {https://arxiv.org/abs/2201.00798} {arXiv:2201.00798 [cond-mat.str-el]}
  \BibitemShut {NoStop}%
\bibitem [{\citenamefont {Zhang}\ \emph
  {et~al.}(2022{\natexlab{c}})\citenamefont {Zhang}, \citenamefont {Zhang},\
  and\ \citenamefont {Chern}}]{sheng22}%
  \BibitemOpen
  \bibfield  {author} {\bibinfo {author} {\bibfnamefont {S.}~\bibnamefont
  {Zhang}}, \bibinfo {author} {\bibfnamefont {P.}~\bibnamefont {Zhang}},\ and\
  \bibinfo {author} {\bibfnamefont {G.-W.}\ \bibnamefont {Chern}},\ }\bibfield
  {title} {\bibinfo {title} {Anomalous phase separation in a correlated
  electron system: Machine-learning–enabled large-scale kinetic monte carlo
  simulations},\ }\href {https://doi.org/10.1073/pnas.2119957119} {\bibfield
  {journal} {\bibinfo  {journal} {Proceedings of the National Academy of
  Sciences}\ }\textbf {\bibinfo {volume} {119}},\ \bibinfo {pages}
  {e2119957119} (\bibinfo {year} {2022}{\natexlab{c}})}\BibitemShut {NoStop}%
\bibitem [{\citenamefont {Zhang}\ and\ \citenamefont
  {Chern}(2021{\natexlab{b}})}]{puhan21}%
  \BibitemOpen
  \bibfield  {author} {\bibinfo {author} {\bibfnamefont {P.}~\bibnamefont
  {Zhang}}\ and\ \bibinfo {author} {\bibfnamefont {G.-W.}\ \bibnamefont
  {Chern}},\ }\bibfield  {title} {\bibinfo {title} {Arrested phase separation
  in double-exchange models: Large-scale simulation enabled by machine
  learning},\ }\href {https://doi.org/10.1103/PhysRevLett.127.146401}
  {\bibfield  {journal} {\bibinfo  {journal} {Phys. Rev. Lett.}\ }\textbf
  {\bibinfo {volume} {127}},\ \bibinfo {pages} {146401} (\bibinfo {year}
  {2021}{\natexlab{b}})}\BibitemShut {NoStop}%
\bibitem [{\citenamefont {Zhang}\ and\ \citenamefont
  {Chern}(2023{\natexlab{b}})}]{puhan23}%
  \BibitemOpen
  \bibfield  {author} {\bibinfo {author} {\bibfnamefont {P.}~\bibnamefont
  {Zhang}}\ and\ \bibinfo {author} {\bibfnamefont {G.-W.}\ \bibnamefont
  {Chern}},\ }\bibfield  {title} {\bibinfo {title} {Machine learning
  nonequilibrium electron forces for spin dynamics of itinerant magnets},\
  }\href {https://doi.org/10.1038/s41524-023-00990-0} {\bibfield  {journal}
  {\bibinfo  {journal} {npj Computational Materials}\ }\textbf {\bibinfo
  {volume} {9}},\ \bibinfo {pages} {32} (\bibinfo {year}
  {2023}{\natexlab{b}})}\BibitemShut {NoStop}%
\bibitem [{\citenamefont {Behler}(2011)}]{behler11}%
  \BibitemOpen
  \bibfield  {author} {\bibinfo {author} {\bibfnamefont {J.}~\bibnamefont
  {Behler}},\ }\bibfield  {title} {\bibinfo {title} {{Atom-centered symmetry
  functions for constructing high-dimensional neural network potentials}},\
  }\href {https://doi.org/10.1063/1.3553717} {\bibfield  {journal} {\bibinfo
  {journal} {The Journal of Chemical Physics}\ }\textbf {\bibinfo {volume}
  {134}},\ \bibinfo {pages} {074106} (\bibinfo {year} {2011})}\BibitemShut
  {NoStop}%
\bibitem [{\citenamefont {Bart\'ok}\ \emph {et~al.}(2013)\citenamefont
  {Bart\'ok}, \citenamefont {Kondor},\ and\ \citenamefont
  {Cs\'anyi}}]{bartok13}%
  \BibitemOpen
  \bibfield  {author} {\bibinfo {author} {\bibfnamefont {A.~P.}\ \bibnamefont
  {Bart\'ok}}, \bibinfo {author} {\bibfnamefont {R.}~\bibnamefont {Kondor}},\
  and\ \bibinfo {author} {\bibfnamefont {G.}~\bibnamefont {Cs\'anyi}},\
  }\bibfield  {title} {\bibinfo {title} {On representing chemical
  environments},\ }\href {https://doi.org/10.1103/PhysRevB.87.184115}
  {\bibfield  {journal} {\bibinfo  {journal} {Phys. Rev. B}\ }\textbf {\bibinfo
  {volume} {87}},\ \bibinfo {pages} {184115} (\bibinfo {year}
  {2013})}\BibitemShut {NoStop}%
\bibitem [{\citenamefont {Ghiringhelli}\ \emph {et~al.}(2015)\citenamefont
  {Ghiringhelli}, \citenamefont {Vybiral}, \citenamefont {Levchenko},
  \citenamefont {Draxl},\ and\ \citenamefont {Scheffler}}]{ghiringhelli15}%
  \BibitemOpen
  \bibfield  {author} {\bibinfo {author} {\bibfnamefont {L.~M.}\ \bibnamefont
  {Ghiringhelli}}, \bibinfo {author} {\bibfnamefont {J.}~\bibnamefont
  {Vybiral}}, \bibinfo {author} {\bibfnamefont {S.~V.}\ \bibnamefont
  {Levchenko}}, \bibinfo {author} {\bibfnamefont {C.}~\bibnamefont {Draxl}},\
  and\ \bibinfo {author} {\bibfnamefont {M.}~\bibnamefont {Scheffler}},\
  }\bibfield  {title} {\bibinfo {title} {Big data of materials science:
  Critical role of the descriptor},\ }\href
  {https://doi.org/10.1103/PhysRevLett.114.105503} {\bibfield  {journal}
  {\bibinfo  {journal} {Phys. Rev. Lett.}\ }\textbf {\bibinfo {volume} {114}},\
  \bibinfo {pages} {105503} (\bibinfo {year} {2015})}\BibitemShut {NoStop}%
\bibitem [{\citenamefont {Himanen}\ \emph {et~al.}(2020)\citenamefont
  {Himanen}, \citenamefont {Jäger}, \citenamefont {Morooka}, \citenamefont
  {{Federici Canova}}, \citenamefont {Ranawat}, \citenamefont {Gao},
  \citenamefont {Rinke},\ and\ \citenamefont {Foster}}]{himanen20}%
  \BibitemOpen
  \bibfield  {author} {\bibinfo {author} {\bibfnamefont {L.}~\bibnamefont
  {Himanen}}, \bibinfo {author} {\bibfnamefont {M.~O.}\ \bibnamefont {Jäger}},
  \bibinfo {author} {\bibfnamefont {E.~V.}\ \bibnamefont {Morooka}}, \bibinfo
  {author} {\bibfnamefont {F.}~\bibnamefont {{Federici Canova}}}, \bibinfo
  {author} {\bibfnamefont {Y.~S.}\ \bibnamefont {Ranawat}}, \bibinfo {author}
  {\bibfnamefont {D.~Z.}\ \bibnamefont {Gao}}, \bibinfo {author} {\bibfnamefont
  {P.}~\bibnamefont {Rinke}},\ and\ \bibinfo {author} {\bibfnamefont {A.~S.}\
  \bibnamefont {Foster}},\ }\bibfield  {title} {\bibinfo {title} {Dscribe:
  Library of descriptors for machine learning in materials science},\ }\href
  {https://doi.org/https://doi.org/10.1016/j.cpc.2019.106949} {\bibfield
  {journal} {\bibinfo  {journal} {Computer Physics Communications}\ }\textbf
  {\bibinfo {volume} {247}},\ \bibinfo {pages} {106949} (\bibinfo {year}
  {2020})}\BibitemShut {NoStop}%
\bibitem [{\citenamefont {Rupp}\ \emph {et~al.}(2012)\citenamefont {Rupp},
  \citenamefont {Tkatchenko}, \citenamefont {M\"uller},\ and\ \citenamefont
  {von Lilienfeld}}]{Rupp2012}%
  \BibitemOpen
  \bibfield  {author} {\bibinfo {author} {\bibfnamefont {M.}~\bibnamefont
  {Rupp}}, \bibinfo {author} {\bibfnamefont {A.}~\bibnamefont {Tkatchenko}},
  \bibinfo {author} {\bibfnamefont {K.-R.}\ \bibnamefont {M\"uller}},\ and\
  \bibinfo {author} {\bibfnamefont {O.~A.}\ \bibnamefont {von Lilienfeld}},\
  }\bibfield  {title} {\bibinfo {title} {Fast and accurate modeling of
  molecular atomization energies with machine learning},\ }\href
  {https://doi.org/10.1103/PhysRevLett.108.058301} {\bibfield  {journal}
  {\bibinfo  {journal} {Phys. Rev. Lett.}\ }\textbf {\bibinfo {volume} {108}},\
  \bibinfo {pages} {058301} (\bibinfo {year} {2012})}\BibitemShut {NoStop}%
\bibitem [{\citenamefont {Shapeev}(2016)}]{shapeev16}%
  \BibitemOpen
  \bibfield  {author} {\bibinfo {author} {\bibfnamefont {A.~V.}\ \bibnamefont
  {Shapeev}},\ }\bibfield  {title} {\bibinfo {title} {Moment tensor potentials:
  A class of systematically improvable interatomic potentials},\ }\href
  {https://doi.org/10.1137/15M1054183} {\bibfield  {journal} {\bibinfo
  {journal} {Multiscale Modeling \& Simulation}\ }\textbf {\bibinfo {volume}
  {14}},\ \bibinfo {pages} {1153} (\bibinfo {year} {2016})}\BibitemShut
  {NoStop}%
\bibitem [{\citenamefont {Drautz}(2019)}]{drautz19}%
  \BibitemOpen
  \bibfield  {author} {\bibinfo {author} {\bibfnamefont {R.}~\bibnamefont
  {Drautz}},\ }\bibfield  {title} {\bibinfo {title} {Atomic cluster expansion
  for accurate and transferable interatomic potentials},\ }\href
  {https://doi.org/10.1103/PhysRevB.99.014104} {\bibfield  {journal} {\bibinfo
  {journal} {Phys. Rev. B}\ }\textbf {\bibinfo {volume} {99}},\ \bibinfo
  {pages} {014104} (\bibinfo {year} {2019})}\BibitemShut {NoStop}%
\bibitem [{\citenamefont {Hansen}\ \emph {et~al.}(2015)\citenamefont {Hansen},
  \citenamefont {Biegler}, \citenamefont {Ramakrishnan}, \citenamefont
  {Pronobis}, \citenamefont {von Lilienfeld}, \citenamefont {M{\"u}ller},\ and\
  \citenamefont {Tkatchenko}}]{Hansen2015}%
  \BibitemOpen
  \bibfield  {author} {\bibinfo {author} {\bibfnamefont {K.}~\bibnamefont
  {Hansen}}, \bibinfo {author} {\bibfnamefont {F.}~\bibnamefont {Biegler}},
  \bibinfo {author} {\bibfnamefont {R.}~\bibnamefont {Ramakrishnan}}, \bibinfo
  {author} {\bibfnamefont {W.}~\bibnamefont {Pronobis}}, \bibinfo {author}
  {\bibfnamefont {O.~A.}\ \bibnamefont {von Lilienfeld}}, \bibinfo {author}
  {\bibfnamefont {K.-R.}\ \bibnamefont {M{\"u}ller}},\ and\ \bibinfo {author}
  {\bibfnamefont {A.}~\bibnamefont {Tkatchenko}},\ }\bibfield  {title}
  {\bibinfo {title} {Machine learning predictions of molecular properties:
  Accurate many-body potentials and nonlocality in chemical space},\ }\href
  {https://doi.org/10.1021/acs.jpclett.5b00831} {\bibfield  {journal} {\bibinfo
   {journal} {The Journal of Physical Chemistry Letters}\ }\textbf {\bibinfo
  {volume} {6}},\ \bibinfo {pages} {2326} (\bibinfo {year} {2015})}\BibitemShut
  {NoStop}%
\bibitem [{\citenamefont {Faber}\ \emph {et~al.}(2015)\citenamefont {Faber},
  \citenamefont {Lindmaa}, \citenamefont {von Lilienfeld},\ and\ \citenamefont
  {Armiento}}]{Faber2015}%
  \BibitemOpen
  \bibfield  {author} {\bibinfo {author} {\bibfnamefont {F.}~\bibnamefont
  {Faber}}, \bibinfo {author} {\bibfnamefont {A.}~\bibnamefont {Lindmaa}},
  \bibinfo {author} {\bibfnamefont {O.~A.}\ \bibnamefont {von Lilienfeld}},\
  and\ \bibinfo {author} {\bibfnamefont {R.}~\bibnamefont {Armiento}},\
  }\bibfield  {title} {\bibinfo {title} {Crystal structure representations for
  machine learning models of formation energies},\ }\href
  {https://doi.org/https://doi.org/10.1002/qua.24917} {\bibfield  {journal}
  {\bibinfo  {journal} {International Journal of Quantum Chemistry}\ }\textbf
  {\bibinfo {volume} {115}},\ \bibinfo {pages} {1094} (\bibinfo {year}
  {2015})}\BibitemShut {NoStop}%
\bibitem [{\citenamefont {Huo}\ and\ \citenamefont {Rupp}(2022)}]{huo22}%
  \BibitemOpen
  \bibfield  {author} {\bibinfo {author} {\bibfnamefont {H.}~\bibnamefont
  {Huo}}\ and\ \bibinfo {author} {\bibfnamefont {M.}~\bibnamefont {Rupp}},\
  }\bibfield  {title} {\bibinfo {title} {Unified representation of molecules
  and crystals for machine learning},\ }\href
  {https://doi.org/10.1088/2632-2153/aca005} {\bibfield  {journal} {\bibinfo
  {journal} {Machine Learning: Science and Technology}\ }\textbf {\bibinfo
  {volume} {3}},\ \bibinfo {pages} {045017} (\bibinfo {year}
  {2022})}\BibitemShut {NoStop}%
\bibitem [{\citenamefont {Kondor}(2007)}]{kondor07}%
  \BibitemOpen
  \bibfield  {author} {\bibinfo {author} {\bibfnamefont {R.}~\bibnamefont
  {Kondor}},\ }\href {https://arxiv.org/abs/cs/0701127} {\bibinfo {title} {A
  novel set of rotationally and translationally invariant features for images
  based on the non-commutative bispectrum}} (\bibinfo {year} {2007}),\ \Eprint
  {https://arxiv.org/abs/cs/0701127} {arXiv:cs/0701127 [cs.CV]} \BibitemShut
  {NoStop}%
\bibitem [{\citenamefont {Hamermesh}()}]{hamermesh_group_1989}%
  \BibitemOpen
  \bibfield  {author} {\bibinfo {author} {\bibfnamefont {M.}~\bibnamefont
  {Hamermesh}},\ }\href {https://books.google.com/books?id=c0o9_wlCzgcC} {\emph
  {\bibinfo {title} {Group Theory and Its Application to Physical Problems}}},\
  Addison Wesley Series in Physics\ (\bibinfo  {publisher} {Dover
  Publications})\BibitemShut {NoStop}%
\bibitem [{\citenamefont {Paszke}\ \emph {et~al.}(2019)\citenamefont {Paszke},
  \citenamefont {Gross}, \citenamefont {Massa}, \citenamefont {Lerer},
  \citenamefont {Bradbury}, \citenamefont {Chanan}, \citenamefont {Killeen},
  \citenamefont {Lin}, \citenamefont {Gimelshein}, \citenamefont {Antiga},
  \citenamefont {Desmaison}, \citenamefont {Köpf}, \citenamefont {Yang},
  \citenamefont {DeVito}, \citenamefont {Raison}, \citenamefont {Tejani},
  \citenamefont {Chilamkurthy}, \citenamefont {Steiner}, \citenamefont {Fang},
  \citenamefont {Bai},\ and\ \citenamefont {Chintala}}]{paszke19}%
  \BibitemOpen
  \bibfield  {author} {\bibinfo {author} {\bibfnamefont {A.}~\bibnamefont
  {Paszke}}, \bibinfo {author} {\bibfnamefont {S.}~\bibnamefont {Gross}},
  \bibinfo {author} {\bibfnamefont {F.}~\bibnamefont {Massa}}, \bibinfo
  {author} {\bibfnamefont {A.}~\bibnamefont {Lerer}}, \bibinfo {author}
  {\bibfnamefont {J.}~\bibnamefont {Bradbury}}, \bibinfo {author}
  {\bibfnamefont {G.}~\bibnamefont {Chanan}}, \bibinfo {author} {\bibfnamefont
  {T.}~\bibnamefont {Killeen}}, \bibinfo {author} {\bibfnamefont
  {Z.}~\bibnamefont {Lin}}, \bibinfo {author} {\bibfnamefont {N.}~\bibnamefont
  {Gimelshein}}, \bibinfo {author} {\bibfnamefont {L.}~\bibnamefont {Antiga}},
  \bibinfo {author} {\bibfnamefont {A.}~\bibnamefont {Desmaison}}, \bibinfo
  {author} {\bibfnamefont {A.}~\bibnamefont {Köpf}}, \bibinfo {author}
  {\bibfnamefont {E.}~\bibnamefont {Yang}}, \bibinfo {author} {\bibfnamefont
  {Z.}~\bibnamefont {DeVito}}, \bibinfo {author} {\bibfnamefont
  {M.}~\bibnamefont {Raison}}, \bibinfo {author} {\bibfnamefont
  {A.}~\bibnamefont {Tejani}}, \bibinfo {author} {\bibfnamefont
  {S.}~\bibnamefont {Chilamkurthy}}, \bibinfo {author} {\bibfnamefont
  {B.}~\bibnamefont {Steiner}}, \bibinfo {author} {\bibfnamefont
  {L.}~\bibnamefont {Fang}}, \bibinfo {author} {\bibfnamefont {J.}~\bibnamefont
  {Bai}},\ and\ \bibinfo {author} {\bibfnamefont {S.}~\bibnamefont
  {Chintala}},\ }\href {https://arxiv.org/abs/1912.01703} {\bibinfo {title}
  {Pytorch: An imperative style, high-performance deep learning library}}
  (\bibinfo {year} {2019}),\ \Eprint {https://arxiv.org/abs/1912.01703}
  {arXiv:1912.01703 [cs.LG]} \BibitemShut {NoStop}%
\bibitem [{\citenamefont {Nair}\ and\ \citenamefont {Hinton}(2010)}]{Nair10}%
  \BibitemOpen
  \bibfield  {author} {\bibinfo {author} {\bibfnamefont {V.}~\bibnamefont
  {Nair}}\ and\ \bibinfo {author} {\bibfnamefont {G.~E.}\ \bibnamefont
  {Hinton}},\ }\bibfield  {title} {\bibinfo {title} {Rectified linear units
  improve restricted boltzmann machines},\ }in\ \href
  {https://api.semanticscholar.org/CorpusID:15539264} {\emph {\bibinfo
  {booktitle} {International Conference on Machine Learning}}}\ (\bibinfo
  {year} {2010})\BibitemShut {NoStop}%
\bibitem [{\citenamefont {Barron}(2017)}]{barron2017}%
  \BibitemOpen
  \bibfield  {author} {\bibinfo {author} {\bibfnamefont {J.~T.}\ \bibnamefont
  {Barron}},\ }\href {https://arxiv.org/abs/1704.07483} {\bibinfo {title}
  {Continuously differentiable exponential linear units}} (\bibinfo {year}
  {2017}),\ \Eprint {https://arxiv.org/abs/1704.07483} {arXiv:1704.07483
  [cs.LG]} \BibitemShut {NoStop}%
\bibitem [{\citenamefont {Paszke}\ \emph {et~al.}(2017)\citenamefont {Paszke},
  \citenamefont {Gross}, \citenamefont {Chintala}, \citenamefont {Chanan},
  \citenamefont {Yang}, \citenamefont {DeVito}, \citenamefont {Lin},
  \citenamefont {Desmaison}, \citenamefont {Antiga},\ and\ \citenamefont
  {Lerer}}]{Paszke2017}%
  \BibitemOpen
  \bibfield  {author} {\bibinfo {author} {\bibfnamefont {A.}~\bibnamefont
  {Paszke}}, \bibinfo {author} {\bibfnamefont {S.}~\bibnamefont {Gross}},
  \bibinfo {author} {\bibfnamefont {S.}~\bibnamefont {Chintala}}, \bibinfo
  {author} {\bibfnamefont {G.}~\bibnamefont {Chanan}}, \bibinfo {author}
  {\bibfnamefont {E.}~\bibnamefont {Yang}}, \bibinfo {author} {\bibfnamefont
  {Z.}~\bibnamefont {DeVito}}, \bibinfo {author} {\bibfnamefont
  {Z.}~\bibnamefont {Lin}}, \bibinfo {author} {\bibfnamefont {A.}~\bibnamefont
  {Desmaison}}, \bibinfo {author} {\bibfnamefont {L.}~\bibnamefont {Antiga}},\
  and\ \bibinfo {author} {\bibfnamefont {A.}~\bibnamefont {Lerer}},\ }\bibfield
   {title} {\bibinfo {title} {Automatic differentiation in pytorch},\ }in\
  \href {https://openreview.net/forum?id=BJJsrmfCZ} {\emph {\bibinfo
  {booktitle} {NIPS 2017 Workshop on Autodiff}}}\ (\bibinfo {year}
  {2017})\BibitemShut {NoStop}%
\bibitem [{\citenamefont {Kingma}\ and\ \citenamefont {Ba}(2017)}]{kingma2017}%
  \BibitemOpen
  \bibfield  {author} {\bibinfo {author} {\bibfnamefont {D.~P.}\ \bibnamefont
  {Kingma}}\ and\ \bibinfo {author} {\bibfnamefont {J.}~\bibnamefont {Ba}},\
  }\href {https://arxiv.org/abs/1412.6980} {\bibinfo {title} {Adam: A method
  for stochastic optimization}} (\bibinfo {year} {2017}),\ \Eprint
  {https://arxiv.org/abs/1412.6980} {arXiv:1412.6980 [cs.LG]} \BibitemShut
  {NoStop}%
\bibitem [{\citenamefont {Verlet}(1967)}]{verlet67}%
  \BibitemOpen
  \bibfield  {author} {\bibinfo {author} {\bibfnamefont {L.}~\bibnamefont
  {Verlet}},\ }\bibfield  {title} {\bibinfo {title} {Computer "experiments" on
  classical fluids. i. thermodynamical properties of lennard-jones molecules},\
  }\href {https://doi.org/10.1103/PhysRev.159.98} {\bibfield  {journal}
  {\bibinfo  {journal} {Phys. Rev.}\ }\textbf {\bibinfo {volume} {159}},\
  \bibinfo {pages} {98} (\bibinfo {year} {1967})}\BibitemShut {NoStop}%
\bibitem [{\citenamefont {Vyas}\ \emph {et~al.}(2025)\citenamefont {Vyas},
  \citenamefont {Ottino}, \citenamefont {Lueptow},\ and\ \citenamefont
  {Umbanhowar}}]{vyas2010}%
  \BibitemOpen
  \bibfield  {author} {\bibinfo {author} {\bibfnamefont {D.~R.}\ \bibnamefont
  {Vyas}}, \bibinfo {author} {\bibfnamefont {J.~M.}\ \bibnamefont {Ottino}},
  \bibinfo {author} {\bibfnamefont {R.~M.}\ \bibnamefont {Lueptow}},\ and\
  \bibinfo {author} {\bibfnamefont {P.~B.}\ \bibnamefont {Umbanhowar}},\
  }\bibfield  {title} {\bibinfo {title} {Improved velocity-{Verlet} algorithm
  for the discrete element method},\ }\href
  {https://doi.org/https://doi.org/10.1016/j.cpc.2025.109524} {\bibfield
  {journal} {\bibinfo  {journal} {Computer Physics Communications}\ }\textbf
  {\bibinfo {volume} {310}},\ \bibinfo {pages} {109524} (\bibinfo {year}
  {2025})}\BibitemShut {NoStop}%
\bibitem [{\citenamefont {Yurke}\ \emph {et~al.}(1993)\citenamefont {Yurke},
  \citenamefont {Pargellis}, \citenamefont {Kovacs},\ and\ \citenamefont
  {Huse}}]{Yurke93}%
  \BibitemOpen
  \bibfield  {author} {\bibinfo {author} {\bibfnamefont {B.}~\bibnamefont
  {Yurke}}, \bibinfo {author} {\bibfnamefont {A.~N.}\ \bibnamefont
  {Pargellis}}, \bibinfo {author} {\bibfnamefont {T.}~\bibnamefont {Kovacs}},\
  and\ \bibinfo {author} {\bibfnamefont {D.~A.}\ \bibnamefont {Huse}},\
  }\bibfield  {title} {\bibinfo {title} {Coarsening dynamics of the xy model},\
  }\href {https://doi.org/10.1103/PhysRevE.47.1525} {\bibfield  {journal}
  {\bibinfo  {journal} {Phys. Rev. E}\ }\textbf {\bibinfo {volume} {47}},\
  \bibinfo {pages} {1525} (\bibinfo {year} {1993})}\BibitemShut {NoStop}%
\bibitem [{\citenamefont {Bray}\ \emph {et~al.}(2000)\citenamefont {Bray},
  \citenamefont {Briant},\ and\ \citenamefont {Jervis}}]{Bray00}%
  \BibitemOpen
  \bibfield  {author} {\bibinfo {author} {\bibfnamefont {A.~J.}\ \bibnamefont
  {Bray}}, \bibinfo {author} {\bibfnamefont {A.~J.}\ \bibnamefont {Briant}},\
  and\ \bibinfo {author} {\bibfnamefont {D.~K.}\ \bibnamefont {Jervis}},\
  }\bibfield  {title} {\bibinfo {title} {Breakdown of scaling in the
  nonequilibrium critical dynamics of the two-dimensional $\mathit{XY}$
  model},\ }\href {https://doi.org/10.1103/PhysRevLett.84.1503} {\bibfield
  {journal} {\bibinfo  {journal} {Phys. Rev. Lett.}\ }\textbf {\bibinfo
  {volume} {84}},\ \bibinfo {pages} {1503} (\bibinfo {year}
  {2000})}\BibitemShut {NoStop}%
\bibitem [{\citenamefont {Kaski}\ and\ \citenamefont {Gunton}(1983)}]{Kaski83}%
  \BibitemOpen
  \bibfield  {author} {\bibinfo {author} {\bibfnamefont {K.}~\bibnamefont
  {Kaski}}\ and\ \bibinfo {author} {\bibfnamefont {J.~D.}\ \bibnamefont
  {Gunton}},\ }\bibfield  {title} {\bibinfo {title} {Universal dynamical
  scaling in the clock model},\ }\href
  {https://doi.org/10.1103/PhysRevB.28.5371} {\bibfield  {journal} {\bibinfo
  {journal} {Phys. Rev. B}\ }\textbf {\bibinfo {volume} {28}},\ \bibinfo
  {pages} {5371} (\bibinfo {year} {1983})}\BibitemShut {NoStop}%
\bibitem [{\citenamefont {Kaski}\ \emph
  {et~al.}(1985{\natexlab{a}})\citenamefont {Kaski}, \citenamefont {Grant},\
  and\ \citenamefont {Gunton}}]{Kaski85}%
  \BibitemOpen
  \bibfield  {author} {\bibinfo {author} {\bibfnamefont {K.}~\bibnamefont
  {Kaski}}, \bibinfo {author} {\bibfnamefont {M.}~\bibnamefont {Grant}},\ and\
  \bibinfo {author} {\bibfnamefont {J.~D.}\ \bibnamefont {Gunton}},\ }\bibfield
   {title} {\bibinfo {title} {Domain growth in the clock model},\ }\href
  {https://doi.org/10.1103/PhysRevB.31.3040} {\bibfield  {journal} {\bibinfo
  {journal} {Phys. Rev. B}\ }\textbf {\bibinfo {volume} {31}},\ \bibinfo
  {pages} {3040} (\bibinfo {year} {1985}{\natexlab{a}})}\BibitemShut {NoStop}%
\bibitem [{\citenamefont {Kaski}\ \emph
  {et~al.}(1985{\natexlab{b}})\citenamefont {Kaski}, \citenamefont {Nieminen},\
  and\ \citenamefont {Gunton}}]{Kaski85b}%
  \BibitemOpen
  \bibfield  {author} {\bibinfo {author} {\bibfnamefont {K.}~\bibnamefont
  {Kaski}}, \bibinfo {author} {\bibfnamefont {J.}~\bibnamefont {Nieminen}},\
  and\ \bibinfo {author} {\bibfnamefont {J.~D.}\ \bibnamefont {Gunton}},\
  }\bibfield  {title} {\bibinfo {title} {Domain growth and scaling in the
  q-state potts model},\ }\href {https://doi.org/10.1103/PhysRevB.31.2998}
  {\bibfield  {journal} {\bibinfo  {journal} {Phys. Rev. B}\ }\textbf {\bibinfo
  {volume} {31}},\ \bibinfo {pages} {2998} (\bibinfo {year}
  {1985}{\natexlab{b}})}\BibitemShut {NoStop}%
\bibitem [{\citenamefont {Grest}\ and\ \citenamefont
  {Srolovitz}(1984)}]{Grest84}%
  \BibitemOpen
  \bibfield  {author} {\bibinfo {author} {\bibfnamefont {G.~S.}\ \bibnamefont
  {Grest}}\ and\ \bibinfo {author} {\bibfnamefont {D.~J.}\ \bibnamefont
  {Srolovitz}},\ }\bibfield  {title} {\bibinfo {title} {Vortex effects on
  domain growth: The clock model},\ }\href
  {https://doi.org/10.1103/PhysRevB.30.6535} {\bibfield  {journal} {\bibinfo
  {journal} {Phys. Rev. B}\ }\textbf {\bibinfo {volume} {30}},\ \bibinfo
  {pages} {6535} (\bibinfo {year} {1984})}\BibitemShut {NoStop}%
\bibitem [{\citenamefont {Enomoto}\ and\ \citenamefont
  {Kato}(1990)}]{Enomoto90}%
  \BibitemOpen
  \bibfield  {author} {\bibinfo {author} {\bibfnamefont {Y.}~\bibnamefont
  {Enomoto}}\ and\ \bibinfo {author} {\bibfnamefont {R.}~\bibnamefont {Kato}},\
  }\bibfield  {title} {\bibinfo {title} {A model simulation study of domain
  growth in a system with multiply degenerate ordered states},\ }\href
  {https://doi.org/10.1088/0953-8984/2/46/021} {\bibfield  {journal} {\bibinfo
  {journal} {Journal of Physics: Condensed Matter}\ }\textbf {\bibinfo {volume}
  {2}},\ \bibinfo {pages} {9215} (\bibinfo {year} {1990})}\BibitemShut
  {NoStop}%
\end{thebibliography}%
\end{document}